\documentclass[aps,10pt,prd,notitlepage,nofootinbib]{revtex4-1}

\usepackage[utf8]{inputenc}
\usepackage{newtxtext,newtxmath}
\usepackage{bbold}
\usepackage{bm}
\usepackage{graphicx}
\usepackage[usenames,dvipsnames]{xcolor}
\usepackage{color}
\usepackage{tikz}
\usepackage[colorlinks=true,linkcolor=blue,urlcolor=blue,citecolor=blue]{hyperref}
\usepackage{amsmath,amssymb,amsfonts}
\usepackage{slashed}
\usepackage[english]{babel}
\usepackage{titlesec}
\usepackage{microtype}
\usepackage{dcolumn}
\usepackage{blindtext}
\usepackage{epsfig}
\usepackage{pifont}
\usepackage{dsfont,mathrsfs}
\usepackage{cancel}
\usepackage{bigints}

\newcommand{\be}{\begin{equation}}
\newcommand{\ee}{\end{equation}}
\newcommand{\ba}{\begin{eqnarray}}
\newcommand{\ea}{\end{eqnarray}}

\newcommand{\nn}{\nonumber}

\newcommand{\MB}[1]{\left|#1\right|}
\newcommand{\FB}[1]{\left(#1\right)}
\newcommand{\SB}[1]{\left\{#1\right\}}
\newcommand{\TB}[1]{\left[#1\right]}
\newcommand{\AB}[1]{\left<#1\right>}

\newcommand{\scrL}{\mathscr{L}}

\newcommand{\sign}[1]{\text{sign}\left(#1\right)}

\newcommand{\munu}{{\mu\nu}}

\newcommand{\RE}{\text{Re}}

\newcommand{\unit}{\mathds{1}}

\newcommand{\D}{\mathscr{D}}

\newcommand{\fsl}{\slashed}
\newcommand{\kfourint}[1]{\int \dfrac{d^4#1}{(2\pi)^4}}

\newcommand{\Tr}[1]{ \text{Tr}\left[#1\right]}
\newcommand{\intzinf}{\int_{0}^{\infty}}
\newcommand{\intinf}{\int_{-\infty}^{\infty}}
\newcommand{\half}{\dfrac{1}{2}}

\newcommand{\vparasq}[1]{{#1}_\parallel^2 }
\newcommand{\vperpsq}[1]{{#1}_\perp^2 }

\newcommand{\pbarpasi}{\AB{\overline{\psi}\psi}}

\begin{document}
\title{Effect of anomalous magnetic moment of quarks on the phase structure and mesonic properties in the NJL model}
\author{Nilanjan Chaudhuri$^{a}$}
\email{sovon.nilanjan@gmail.com}
\author{Snigdha Ghosh$^{b}$}
\email{snigdha.physics@gmail.com, snigdha.ghosh@saha.ac.in}
\author{Sourav Sarkar$^{a,c}$}
\email{sourav@vecc.gov.in}
\author{ Pradip Roy$^{b,c}$}
\email{pradipk.roy@saha.ac.in}
\affiliation{$^a$Variable Energy Cyclotron Centre, 1/AF Bidhannagar, Kolkata - 700064, India}
\affiliation{$^b$Saha Institute of Nuclear Physics, 1/AF Bidhannagar, Kolkata - 700064, India}
\affiliation{$^c$Homi Bhabha National Institute, Training School Complex, Anushaktinagar, Mumbai - 400085, India}

\begin{abstract}
Employing a field dependent three-momentum cut-off regularization technique, we study the phase structure and mesonic masses using the $ 2 $-flavour Nambu–Jona Lasinio model at finite temperature and density in presence of arbitrary external magnetic field. This approach is then applied to incorporate the effects of the anomalous magnetic moment(AMM) of quarks on constituent quark mass and thermodynamic observables as a function of temperature/baryonic density. The critical temperature for transition from chiral symmetry broken to the restored phase is observed to decrease with the external magnetic field, which can be classified as inverse magnetic catalysis, while an opposite behaviour is realized in the case of a vanishing magnetic moment, implying magnetic catalysis. These essential features are also reflected in the phase diagram. Furthermore, the properties of the low lying scalar and neutral pseudoscalar mesons are also studied in presence of a hot and dense magnetized medium including AMM of the quarks using random phase approximation. For non-zero values of magnetic field, we notice a sudden jump in the mass of the Goldstone mode at and above the Mott transition temperature which is found to decrease substantially with the increase in magnetic field  when the AMM of the quarks are taken into consideration.
\end{abstract}
\maketitle

\section{Introduction} \label{Sec.Intro}
	Study of strongly interacting matter under the influence of high temperature and finite baryonic density in a  magnetized medium is a subject of great interest~\cite{Lect_note}. Specifically,  the presence of  a background magnetic field results in a large number of interesting physical effects in quark matter and understanding them brings 	us closer to our main objective of understanding quantum chromodynamics (QCD). Some of the most important ones among these phenomena are Chiral Magnetic Effects (CME)~\cite{Fukushima,Kharzeev,Kharzeev2,Bali}; Magnetic Catalysis (MC)~\cite{Shovkovy,Gusynin1,Gusynin2,Gusynin3}  of dynamical chiral symmetry breaking as well as Inverse Magnetic Catalysis (IMC)~\cite{Preis,Preis2} which may lead to significant modification of the nature of electro-weak~\cite{Elmfors,Skalozub,Sadooghi_ew,Navarro}, chiral and superconducting phase transitions~\cite{Fayazbakhsh1,Fayazbakhsh2,Skokov,Fukushima2}, electromagnetically induced superconductivity and superfluidity~\cite{Chernodub1,Chernodub2} and many more. Now, it is remarkable that strong magnetic fields of the order	of $ \approx 10^{18} $ G~\cite{Kharzeev,Skokov2} or larger can be generated in non-central heavy-ion collisions, at RHIC and LHC, two of the most 	important laboratories for the study of strongly interacting matter. Since this 	is comparable to the QCD scale i.e. $ eB\approx m_\pi^2 $ (note that in natural units, 	$ 10^{18} {\rm ~G} \approx m_\pi^2 \approx 0.02~ {\rm GeV}^2  $), the magnetic field in these 	laboratories is sufficient to make  noticeable influence in the properties of QCD matter. 	The charge separation in heavy ion collisions may be a consequence of such magnetic field which has been attributed to the so-called CME mentioned earlier. It is worth mentioning that, though the fields created during heavy ion collision is short lived~\cite{Kharzeev}, the presence of 	finite electrical conductivity of the hot and dense medium may lead to  substantial delay in the decay of these time-dependent magnetic fields~\cite{Gursoy}. This justifies the use of a uniform background field in most of the calculations in literature. Besides heavy ion collisions, strong magnetic field can be realized in several physical systems, such as: (i) in the early universe during electroweak transition where the magnetic field, as high as $ \approx 10^{23}$ G~\cite{Vachaspati,Campanelli} might have been produced, (ii) at the surface of certain compact stars called \textit{magnetars}, magnetic field is of the order of $ \approx 10^{15}$ G~\cite{Duncan,Duncan2}, while in the interior it might reach $\approx 10^{18} $ G~\cite{Lai}, (iii) in quasi-relativistic condensed matter systems like  graphene~\cite{Novoselov,Zhang_cond} etc.  Thus, apart from its theoretical intricacies,  the possibility of an experimental verification has attracted a large number of researchers in this domain of physics in recent years.
		
	However, the detailed analysis of the above mentioned properties involves a great deal of  complexities while evaluating quantities of interest from first principles since the large coupling  strength of QCD in low energy regime restricts the use of perturbative approach. Lattice QCD simulations provide one of the best procedures to tackle the problem at intermediate temperatures (comparable to the QCD scale) and low baryonic density  which is relevant for highly relativistic heavy ion collisions~\cite{Forcrand1,Forcrand2,Forcrand3,Bali_lat,Luschevskaya,Aoki,Aoki2}. However, for compact stars one has to deal with the low temperatures and high values of baryonic chemical potential. CBM experiment at FAIR is also expected to explore high baryonic density matter. These areas of the phase diagram are not accessible via the lattice simulation due to the so-called sign problem in Monte Carlo sampling~\cite{Preis}. Available alternative is to work with effective models which possess some of the essential features of QCD and mathematically tractable so that the basic mechanisms remain illustrative. Nambu Jona-Lasinio (NJL) model~\cite{Nambu1,Nambu2} is one such model,  which presents a useful scheme to examine the vacuum structure of QCD at arbitrary temperatures and baryonic density. This model has been extensively used  to study some of the non-perturbative properties of the QCD vacuum as it was constructed respecting the global symmetries of QCD, most importantly the chiral symmetry (see~\cite{Klevansky,Hatsuda1,Vogl,Buballa} for reviews). As mentioned in~\cite{Klevansky} the point like interaction between quarks makes the NJL  model non-renormalizable. Thus a proper regularization scheme has to be chosen to deal with the divergent integrals and the parameters associated with the model are fixed to reproduce some well known phenomenological quantities, for example pion-decay constant $ f_\pi  $, condensate etc~\cite{Reinder}.

	The problem of chiral phase transition using NJL model in presence of a uniform background  magnetic field is an extensively studied topic in the  literature~\cite{Klevansky2,Gusynin1,Gusynin2,Gusynin3,Mao,Sadooghi1,Ruggieri}. In most of the cases it was found that the magnetic field is likely to strengthen  chirally broken phase leading to the MC. But in few lattice data~\cite{Bali_lat,Bali_lat2,Bali_lat3,Bornyakov}, a contradictory behaviour of the transition temperature was observed which supports IMC. A significant amount of effort has been made  to explain this discrepancy by adopting appropriate modifications of effective models, primarily considering a field dependent coupling constant~\cite{Ruggieri2,Andersen,Ayala2}. In~\cite{Mao} it was shown that $eB$-dependence of transition temperature can vary depending upon the regularization scheme. In a recent study~\cite{Strickland}, it has been indicated that finite value of anomalous magnetic moment(AMM) of nucleons increases the level of pressure anisotropies for a system of proton and neutrons. Furthermore, in ~\cite{Arghya} it was demonstrated that, as a manifestation of inclusion of AMM of protons and neutrons, critical temperature for vacuum to nuclear medium decreases with increasing magnetic field which can be identified as IMC.  Since the chiral symmetry breaking leads to the generation of an AMM for the quarks~\cite{Pedro,Chang}, it is justified to include their contributions during evaluation of dynamics of chiral symmetry breaking and restoration at high temperatures and chemical potentials. A rigorous study of the effect of AMM in phase diagram  in a magnetized medium can be found in~\cite{Sadooghi}, where they have used zeta function regularization technique to regularise the $ eB $ dependent thermodynamic potential for non zero values of AMM of quarks and used a field dependent smoothing function.
		
	In absence of magnetic field NJL model has also been used extensively to describe the physical properties of light scalar and pseudoscalar mesons~\cite{HatsudaPRL,Klevansky,Shuryak,Ayala,Inagaki,Hansen}, which have a direct relevance with the dynamics of chiral phase transition. Progress have also been made in studying these lightest hadrons in a hot and dense magnetized medium~\cite{Sadooghi1,Zhang, Mao_pi,Avancini}. It was found that the minimum temperature for which the overlap interval starts in the crossover region increases with increasing magnetic field. As it was already pointed out in the previous paragraph that inclusion of AMM in different cases leads to an opposite behaviour in transition temperature so it will be interesting to evaluate mesonic masses considering the AMM of quarks. It is worth mentioning that there are also other methods available to study mesonic properties, such as, lattice simulations~\cite{Forcrand1,Forcrand2,Forcrand3,Pushkina,Wissel}, dimensional reduction~\cite{Hansson,Laine}, hard thermal loop approximation~\cite{Alberico,Alberico2,Karsch} etc. As mentioned in Ref.~\cite{Hansen}, the latter approaches rely on a separation of momentum scales which, strictly speaking, holds only in the weak coupling regime $ g \ll 1 $ and hence may not be justified near the vicinity of phase transition. 
		
	In this work, we aim to study the effect of AMM of the quarks on both the phase structure as well as mesonic properties (namely of scalar meson $\sigma$ and pseudoscalar neutral meson $\pi^0$) in 2-flavour NJL model. 
	We have used a magnetic field dependent three-momentum cut-off as our regularization scheme~\cite{Morimoto:2018pzk} which has been shown to be a generalization of the usual zero field three-momentum cut-off regularization. 
	We have also shown that, in the limit $B\rightarrow0$, the analytic expression reduces to the corresponding one 
	in absence of external magnetic field.
	An extensive study of the phase structure of quark matter at finite density under arbitrary external magnetic field has been made by evaluating the chiral susceptibility in the neighbourhood of transition temperature or chemical potential of quarks. MC is observed when AMM of the quarks is switched off while an opposite behaviour (IMC) is obtained considering the AMM of the quarks. Moreover, it is seen that, the nature of the (pseudo-chiral) phase transition largely depends on the external magnetic field as well as on the consideration of the AMM of the quarks. We then calculated the masses of $\sigma$ and $\pi^0$ at finite temperature, density and arbitrary external magnetic field including AMM of the quarks, of which we have not come across any prior work investigations in the literature. The $\pi^0$ mass has been seen to suffer sudden jump at and above Mott transition ($T_\text{Mott}$) temperature. Finally, the variation of $T_\text{Mott}$ with external magnetic field is studied and significant decrease of $T_\text{Mott}$ is observed with the increase in $B$ when the AMM of the quarks are taken into account. It is worth mentioning that, all the calculation presented in the work have been performed by considering all the Landau levels of the quarks and thus the results are valid for arbitrary magnetic field.
	
	The paper is organized as follows. In Sec.~\ref{deri_gap_eqn}, the gap equation is derived for the calculation of the constituent quark mass. Next in Sec.~\ref{TD_obs}, various thermodynamic quantities are obtained. Sec.~\ref{sec.meson} is devoted for the calculation of mesonic properties. We then present numerical results in Sec.~\ref{sec.results} followed by a summary and conclusion of our work in Sec.~\ref{sec.summary}.


\section{Constituent Quark Mass}
\label{deri_gap_eqn}
 The Lagrangian of the two-flavour NJL-model considering the anomalous magnetic moment of free quarks 
 in presence of constant background magnetic field is given by~\cite{Nambu1,Nambu2,Volkov,Klevansky,HatsudaPRL,Buballa}
\begin{equation}
\scrL = \overline{\psi}(x)\FB{i\fsl{D}-m +\half\hat{a} \sigma^\munu F_\munu  }\psi(x)  
+ G\SB{ \FB{\overline{\psi}(x) \psi (x)}^2 + \FB{ \overline{\psi}(x) i\gamma_5\tau \psi(x)}^2} \label{njl_lagrangian}
\end{equation}
 where we have dropped the flavour ($ f=u,d $) and color ($ c=r,g,b $) indices from the Dirac field $ \FB{ \psi^{fc}} $ for a convenient representation. In Eq.~\eqref{njl_lagrangian}, $m$ is current quark mass representing the explicit chiral  symmetry breaking (we will take $ m_u= m_d = m $ to ensure isospin symmetry of the theory at vanishing magnetic field), $ D_\mu = \partial_\mu + i QA_\mu  $ is covariant derivative which couples quark charge $ \hat{Q}=\text{diag}(2e/3,-e/3)$ with  the external magnetic field represented by the four potential $ A_\mu $, the factor $\hat{a}= \hat{Q}\hat{\kappa }$,  where $ \hat{\kappa}=\text{diag}(\kappa_u,\kappa_d) $ is a  $2\times 2 $ matrix in the flavour space  (see Ref.~\cite{Sadooghi} for details), $ F^\munu= \partial^\mu A^\nu- \partial^\nu A^\mu  $ and  $ \sigma^\munu=\dfrac{i}{2}[\gamma^\mu,\gamma^\nu]$. The metric tensor used in this work is mostly negative, $g^\munu = \text{diag}\FB{1,-1,-1,-1}$. Now expanding $ \bar{\psi } \psi  $ around the quark-condensate $ \pbarpasi $ and dropping the quadratic 
 term of the fluctuation one can write
 \begin{equation}
 \FB{\overline{\psi} \psi }^2 = \FB{\overline{\psi} \psi -\pbarpasi +\pbarpasi}^2 \approx 2\pbarpasi \FB{\overline{\psi} \psi } - \pbarpasi^2.
 \end{equation}
 In this mean field approximation (MFA), the Lagrangian becomes 
 \begin{equation}
 \scrL = \overline{\psi}(x)\FB{i\fsl{D}-M +\half\hat{a} \sigma^\munu F_\munu  }\psi(x)  - { \dfrac{(M-m)^2}{4G}}
 \end{equation}
where, $ M $ is the constituent quark mass given by
\begin{equation}
M=m - 2G\pbarpasi.
\label{gap_eqn}
\end{equation} 
Eq.~\eqref{gap_eqn} is known as `\textit{gap equation}'. Now following Ref.~\cite{Sadooghi1, Sadooghi}, the one-loop effective potential ($\Omega$) for a two-flavour NJL model considering the AMM of quarks at finite temperature ($ T $) and chemical potential ($ \mu $) in presence of a uniform background magnetic field ($ B $) fixed along $ z $-direction is expressed as
\begin{eqnarray}
\Omega = { \dfrac{(M-m)^2}{4G}} - N_c\sum_f \dfrac{\MB{e_fB }}{\beta}\sum_{n=0}^{\infty }~ \sum_{s\in\SB{\pm1}}  
\int_{-\infty}^{\infty} \dfrac{dp_z}{4\pi^2} \SB{\beta E_{nfs} - \ln \FB{1-n^+}- \ln \FB{1-n^-}}
\label{eff_pot}
\end{eqnarray}
where $ N_c = 3 $ is the number of colors, $ e_u = 2e/3, e_d =-e/3 $, $ \beta=\dfrac{1}{T} $ is the inverse temperature and $n^\pm$ being the thermal distribution functions of the quarks/antiquarks given by
\begin{equation}
n^\pm = \dfrac{1}{\exp\TB{\beta(E_{nfs} \mp \mu_q)} +1}
\label{distn_fn}
\end{equation} 
with $\mu_q$ being the chemical potential of the quark.
In Eqs.~\eqref{eff_pot} and \eqref{distn_fn}, $E_{nfs}$ are the energy eigenvalues of the quarks in presence of external magnetic field (which is a consequence of the Landau quantization of the transverse momenta of the quarks due to the external magnetic field) 
and given by 
 \begin{equation} 
E_{nfs}  = \TB{p_z^2 + \SB{ \FB{\sqrt{\MB{e_f B} (2n+1-s\xi_f) +M^2} - s\kappa_fe_fB }^2 }}^{\half}
\label{energy}
\end{equation}
where, $n$ is the Landau level index, $ s \in\SB{\pm1} $ is the spin index and $ \xi_f = \sign{e_f} $.

Following Ref.~\cite{Sadooghi}, in this article we will assume the values of $ \kappa_f $ appearing in Eq.~\eqref{energy} as constant and independent of ($ T,\mu, B, M $). Now the constituent quark mass can be obtained self-consistently by minimizing the thermodynamic potential w.r.t. $ M $. For this, we differentiate 
$\Omega$ w.r.t. $M$ and equate it to zero to get
\begin{eqnarray}
\dfrac{\partial \Omega}{\partial M } &=& \dfrac{M-m}{2G} -N_c \sum_f \MB{e_fB }\sum_{n=0}^{\infty } \sum_{\{s\}} \intinf \dfrac{dp_z}{4\pi^2} \dfrac{M}{E_{nfs}} \FB{1- \dfrac{s\kappa_fe_f B}{M_{nf}} }\FB{1 -n^+-n^-} = 0~.
\end{eqnarray} 
The solution of the above equation gives the constituent quarks mass as
\begin{equation}
M= m + 2GN_c \sum_f \MB{e_fB }\sum_{n=0}^{\infty } \sum_{\{s\}} \int_{-\infty}^\infty \dfrac{dp_z}{4\pi^2} 
\dfrac{M}{E_{nfs}} \FB{1- \dfrac{s\kappa_fe_f B}{M_{nf}} }\FB{1 -n^+-n^-}
\label{mass_constituen}
\end{equation}
where $ M_{nf} = \sqrt{\MB{e_f B} (2n+1-s\xi_f) +M^2} $. Note that in Eq.~\eqref{mass_constituen}, the medium independent integral is ultraviolet divergent. This feature is due to the fact that quarks are assumed to interact via point-like interaction~\cite{Klevansky}; as a consequence the theory becomes non-renormalizable. So it is necessary to provide an appropriate regularization scheme to make a sense of the divergent integrals. There is no unique way to introduce this regulator~\cite{Klevansky,Buballa,Mao,Menenzes,Sadooghi1,Avancini,Avancini:2019wed}. Discussions about the implementation of different regularization schemes and their outcomes in absence as well as in presence of background magnetic field can be found in these articles. In the case of non-zero background magnetic field, regularization has been done via extracting the pure vacuum part using \textit{Hurwitz zeta} function and applying a three momentum cut-off on the pure vacuum part~\cite{Menenzes,Sadooghi1,Avancini}. 
In this article, we will demonstrate the application of a field dependent 
cut-off~\cite{Morimoto:2018pzk} on the divergent $ p_z $-integral directly 
without going into the trouble of extracting the pure vacuum part which may become cumbersome.
To discuss this procedure we will start from the following integral which appeared in Eq.~\eqref{mass_constituen}:
\begin{eqnarray}
I_{\rm div }=\int_{-\infty}^\infty \dfrac{dp_z}{4\pi^2} \dfrac{M}{E_{nfs}} \FB{1- \dfrac{s\kappa_fe_f B}{M_{nf}} }.
\label{idiv}
\end{eqnarray}
First, we note that the integrands in Eqs.~\eqref{mass_constituen} and \eqref{idiv} are both even functions of 
$ p_z $; introducing the field dependent cut-off parameter $ \Lambda_z $ we get,
\begin{equation}
I_{\rm reg }=2\int_{0}^{\Lambda_z }\dfrac{dp_z}{4\pi^2} \dfrac{M}{E_{nfs}} \FB{1- \dfrac{s\kappa_fe_f B}{M_{nf}} }
\label{iregular}
\end{equation}
where,
\begin{equation}
\Lambda_z = \sqrt{\Lambda^2 - \vec{p }_\perp^2}
\end{equation}
while $ \Lambda $ being the usual  three-momentum cut-off. 
The quantity $ \vec{p}_\perp^2 $ inside the square root can be identified from Eq.~\eqref{energy} in the following manner: 
\begin{equation}
\vec{p }_\perp^2 = \FB{\sqrt{\MB{e_f B} (2n+1-s\xi_f) +M^2} - s\kappa_fe_fB }^2- M^2
= \MB{e_f B} (2n+1-s\xi_f) + \FB{\kappa_fe_fB}^2 - 2sM_{nf} \kappa_fe_fB.
\label{p_perp}
\end{equation}
As a cross check, one can see that as $ \kappa_f\rightarrow 0 $, we get $ \vec{p }_\perp^2\rightarrow (2n+1-s\xi_f)\MB{e_f B} $ which is the usual expression of the Landau quantized transverse momentum expressing the fact that the lowest landau level is non-degenerate. Now the regularization shown in Eq.~\eqref{iregular} will be valid iff $\Lambda^2 - \vec{p }_\perp^2 \ge 0$ and $ \vec{p }_\perp^2 \ge 0 $ as $ p_z, \vec{p}_\perp$  are real quantities. Note that the first condition will always be there for any finite values of $ eB $~\cite{Strickland} but the second condition is only due to the non-zero values of AMM of quarks. These conditions will constrain the contributing $ n $-vlaues in the sum. From now on we will call these two conditions as `\textit{UV-blocking'} and `\textit{AMM-blocking}' respectively. A discussion on these issues is provided in Appendix~\ref{AMM_blocking}. Putting these condition back in Eq.~\eqref{mass_constituen}, we get
\begin{eqnarray}
M&=& m + 4GN_c \sum_f \MB{e_fB }\sum_{n=0}^{\infty } \sum_{\{s\}} \int_{0}^{\Lambda_z} \dfrac{dp_z}{4\pi^2} 
\Theta\FB{ \Lambda^2 - \vec{p }_\perp^2} \ \Theta\FB{\vec{p }_\perp^2}\dfrac{M}{E_{nfs}} \FB{1- \dfrac{s\kappa_fe_f B}{M_{nf}} } \nn \\ 
&& - 4GN_c \sum_f \MB{e_fB }\sum_{n=0}^{\infty } \sum_{\{s\}} \int_{0}^\infty \dfrac{dp_z}{4\pi^2} \dfrac{M}{E_{nfs}} 
\FB{1- \dfrac{s\kappa_fe_f B}{M_{nf}} }\FB{n^++n^-}~.
\label{mass_constituen_reg}
\end{eqnarray}
where $ \Theta(x) $ represents \textit{Heaviside theta function } with $ \Theta(x)=1 $ for $ x> 0 $ and zero otherwise. As a consistency check, we take the limits $ \kappa_f\rightarrow 0 $ and $ e_fB\rightarrow0 $ of the above equation (see Appendix~\ref{Btendstozero} for details) and get,
\begin{equation}
M=m + \dfrac{GMN_fN_c }{\pi^2}\TB{  \Lambda \sqrt{\Lambda^2 + M^2 }- M^2\sinh^{-1} \FB{{\dfrac{\Lambda}{ M}}}}
\label{gap_zero.1}
\end{equation}
which is the well known expression of the gap equation in absence of external magnetic field and agrees with Ref.~\cite{Klevansky}.


\section{Thermodynamic Quantities}\label{TD_obs}
In this section we will calculate some of the thermodynamic observables which can be evaluated from the 
effective potential ($ \Omega $). For example, the entropy density is given
\begin{eqnarray}
s &= &- \dfrac{\partial \Omega }{\partial T} = - N_c \sum_f \MB{e_fB }\sum_{n=0}^{\infty } \sum_{\{s\}} \int_{-\infty}^\infty \dfrac{dp_z}{4\pi^2} \TB{ \ln \FB{1- n^+ } + \ln(1-n^-) -\dfrac{E_{nfs}}{T}\FB{n^+ + n^-} +\dfrac{\mu}{T}\FB{n^+ - n^-} }~.
\label{entropy_density}
\end{eqnarray}

To study different mechanism involving the phase transition from symmetry broken to restored phase, 
we will use the chiral susceptibility which is defined as
 \begin{eqnarray}
 \chi_{mm}&=&  \dfrac{\partial^2 \Omega}{\partial m^2} = \dfrac{1}{2G}\FB{\dfrac{\partial M}{\partial m} -1}~.
 \label{chiral_sus_eq}
 \end{eqnarray}
Evaluation of Eqs.~\eqref{entropy_density} and \eqref{chiral_sus_eq} are straight forward and can be done by 
differentiating $M$ from Eq.~\eqref{mass_constituen_reg} w.r.t. $T$ and $m$ in the following way:
\begin{eqnarray}
\dfrac{\partial M}{\partial T} &=& 
-\dfrac{GN_c}{\pi^2 D_T}  \sum_f \MB{e_fB }\sum_{n=0}^{\infty } \sum_{\{s\}}\int_{0}^{\infty}dp_z \dfrac{\beta^2 M}{E_{nfs}   }\FB{1- \dfrac{s\kappa_fe_f B}{M_{nf}}   } \nn \\ 
&& \hspace{2cm}\TB{n^+ (1- n^+) (E_{nfs} -\mu) +n^- (1- n^-) (E_{nfs} +\mu)\dfrac{}{} }  \\
\dfrac{\partial M}{\partial m} &=& \frac{1}{D_{m}}
\end{eqnarray}
where $D_T$ and $D_{m}$ appearing in the above equations are respectively given by:
\begin{eqnarray}
D_T &=& 1 - \dfrac{GN_c}{\pi^2} \sum_f \MB{e_fB }\sum_{n=0}^{\infty } \sum_{\{s\}}\int_{0}^{\Lambda_z}dp_z \TB{ \dfrac{s\kappa_fe_fBM^2}{E_{nfs} M_{nf} ^3  } +\dfrac{1}{E_{nfs}} \FB{1-\dfrac{s\kappa_fe_fB}{M_{nf}}}- \dfrac{M^2}{E_{nfs}^3} \FB{1-\dfrac{s\kappa_fe_fB}{M_{nf}} }^2} \nn \\
 && -\dfrac{GN_c}{\pi^2} \sum_f \MB{e_fB }\sum_{n=0}^{\infty } \sum_{\{s\}}\dfrac{M^2}{\FB{\Lambda^2 + M^2 }^{1/2}} \FB{1 - \dfrac{s\kappa_fe_f B}{M_{nf}}} \dfrac{s\kappa_f e_f B }{\Lambda_zM_{nf}}\nn \\
  && +  \dfrac{GN_c}{\pi^2} \sum_f \MB{e_fB }\sum_{n=0}^{\infty } \sum_{\{s\}}\int_{0}^{\infty}dp_z \TB{ \dfrac{s\kappa_fe_fBM^2}{E_{nfs} M_{nf} ^3  } +\dfrac{1}{E_{nfs}} \FB{1-\dfrac{s\kappa_fe_fB}{M_{nf}}}- \dfrac{M^2}{E_{nfs}^3} \FB{1-\dfrac{s\kappa_fe_fB}{M_{nf}} }^2} \FB{n^+ +n^-}\nn \\ 
  &&  -  \dfrac{GN_c}{\pi^2} \sum_f \MB{e_fB }\sum_{n=0}^{\infty } \sum_{\{s\}}\int_{0}^{\infty}dp_z\beta \FB{\dfrac{M}{E_{nfs}}}^2\FB{1- \dfrac{s\kappa_fe_fB}{M_{nf}}}^2 \TB{n^+(1-n^+) +n^- ( 1+ n^-)}~,
\end{eqnarray}
\begin{eqnarray}
D_{m} &=& 1 - \dfrac{GN_c}{\pi^2} \sum_f \MB{e_fB }\sum_{n=0}^{\infty } \sum_{\{s\}}\int_{0}^{\Lambda_z}dp_z \TB{ \SB{\dfrac{1}{E_{nfs}    }  - \dfrac{M^2}{E_{nfs}^3} \FB{1-  \dfrac{s\kappa_fe_fB}{M_{nf}}   } } \FB{ 1-\dfrac{s\kappa_fe_fB}{M_{nf}} } + 
	\dfrac{s\kappa_fe_fBM^2}{ E_{nfs} M_{nf }^3  } } \nn \\ 
&& -\dfrac{GN_c}{\pi^2} \sum_f \MB{e_fB }\sum_{n=0}^{\infty } \sum_{\{s\}}\dfrac{M^2}{\FB{\Lambda^2 + M^2 }^{1/2}} \FB{1 - \dfrac{s\kappa_fe_f B}{M_{nf}}}\dfrac{s\kappa_f e_f B }{\Lambda_zM_{nf}} \nn 
\\ && +  \dfrac{GN_c}{\pi^2} \sum_f \MB{e_fB }\sum_{n=0}^{\infty } \sum_{\{s\}}\int_{0}^{\infty}dp_z \TB{ \SB{\dfrac{1}{E_{nfs}    }  - \dfrac{M^2}{E_{nfs}^3} \FB{1-  \dfrac{s\kappa_fe_fB}{M_{nf}}   } } \FB{ 1-\dfrac{s\kappa_fe_fB}{M_{nf}} } + 
	\dfrac{s\kappa_fe_fBM^2}{ E_{nfs} M_{nf }^3  } }\FB{n^++ n^- }\nn \\ &&  -  \dfrac{GN_c}{\pi^2} \sum_f \MB{e_fB }\sum_{n=0}^{\infty } \sum_{\{s\}}\int_{0}^{\infty}dp_z\beta \FB{\dfrac{M}{E_{nfs}}}^2\FB{1- \dfrac{s\kappa_fe_fB}{M_{nf}}}^2 \TB{   \dfrac{}{}n^+(1-n^+) +n^- ( 1+ n^-)}.
\end{eqnarray}
The fact that $ \Lambda_z = \Lambda_z(M) $ is also a function $ M $ have been taken care of while taking the derivatives by means of using Leibniz rule.


\section{Meson properties in the NJL-Model} \label{sec.meson}
Since mesons are the bound states of quark and antiquark, the meson propagators are expressed as~\cite{Klevansky} (in RPA)
\begin{eqnarray}
D_a(q)  &=& \dfrac{2G}{  1 - 2G \Pi_a(q) }
\end{eqnarray}
where $ a \in \SB{\sigma ,\pi}  $ denotes the scalar ($ \Gamma_a =1 $) and 
pseudoscalar ($ \Gamma_a =i\gamma_5\tau $) channels respectively. 
In the above equation, $ \Pi_a(q)  $'s are the `\textit{polarization functions}' and given by
\begin{eqnarray}
\Pi_a(q) &=& i \kfourint{k}\Tr{{S(k)} \Gamma_a {S(p=k+q)} \Gamma_a     }
\end{eqnarray}
where 
\begin{eqnarray}
S(k) = \FB{\slashed{k} +M}\TB{ \dfrac{-1}{k^2 - M^2 +i\epsilon }}
\end{eqnarray}
is the \textit{`dressed'} quark propagator at zero temperature containing the vacuum constituent quark mass $M$. From the pole of the propagators, the mesonic masses can be obtained which essentially requires solving the following 
set of transcendental equations
\begin{eqnarray}
1 - 2G \Pi_a(q^0=m_a,\vec{q}=\vec{0}) &=& 0 ~,~ a \in \SB{\sigma,\pi}~.
\label{tran_meson_mass}
\end{eqnarray}

To include the effect of finite temperature and density in the mesonic excitations, we use the Real Time Formalism (RTF) of Thermal Field Theory (TFT)~\cite{Sourav,Bellac:2011kqa} in which all the two point correlation functions become $2\times2$ matrices in thermal space. However, these matrices can be put in diagonal forms in terms of analytic functions (will be denoted by bars) and thus the knowledge of any one components (say the $11$ component) of these matrices are sufficient to know the complete one. The real part of the analytic thermal polarization function is thus given by~\cite{Sourav}
\begin{eqnarray}
\RE\overline{\Pi}_a (q^0,\vec{q})&=&  \RE ~ i \kfourint{k}\Tr{{S^{11}(k)}\Gamma_a {S^{11}(p=k+q)} \Gamma_a }
\label{eq.repi.1}
\end{eqnarray}
where
\begin{equation}
S^{11}(k) = S(k) - \eta(k\cdot u) \TB{S(k)-\gamma^0S^\dagger(k)\gamma^0} 
\end{equation}
is the $11$ component of the real time \textit{`dressed'} quark propagator containing the temperature and/or density dependent constituent quark mass $M=M(T,\mu_q)$. In the above equation, 
\begin{equation}
\eta(k\cdot u) = \Theta (k\cdot u) f^+(k\cdot u) +\Theta(-k\cdot u) f^-(-k\cdot u) 
\end{equation}
where, $u^\mu$ is the medium four velocity which becomes $u^\mu_\text{LRF}\equiv(1,\vec{0})$ in the Local Rest Frame (LRF) of the medium; $ f^\pm(x) = \TB{e^{(x\mp\mu_q)/T}+1}^{-1} $ are the Fermi-Dirac distribution functions for the quark/antiquark. Performing the $dk^0$ and the angular integrals in Eq.~\eqref{eq.repi.1}, we get after some simplifications,
\begin{eqnarray}
\RE\overline{\Pi}_a (q^0=m_a ,\vec{q}=\vec{0}) &=&  \dfrac{1}{4\pi^2}\int_0^\Lambda \vec{k}^2 d|\vec{k}| 
\FB{\dfrac{1}{\omega_k q_0} } \mathcal{P} \TB{ \dfrac{\mathcal{N}_a(k^0= -q^0 +\omega_k)}{q^0-2\omega_k} 
+ \dfrac{\mathcal{N}_a(k^0= \omega_k)}{q^0+2\omega_k}} \nn \\ 
&& - \dfrac{1}{4\pi^2} \intzinf \vec{k}^2 d|\vec{k}|  \FB{\dfrac{1}{\omega_k q_0}   } \mathcal{P} 
\TB{ \dfrac{\mathcal{N}_a(k^0= -\omega_k) f^- (\omega_k)}{q^0-2\omega_k} 
+ \dfrac{\mathcal{N}_a(k^0= \omega_k)f^+ (\omega_k)}{q^0+2\omega_k}  \nn \right. \\ 
&& \hspace{2cm} \left. +  \dfrac{\mathcal{N}_a(k^0= -q^0 -\omega_k) f^- (\omega_k)}{q^0-2\omega_k} 
+ \dfrac{\mathcal{N}_a(k^0= -q^0+\omega_k)f^+ (\omega_k)}{q^0+2\omega_k}}
\label{pola_0} 
\end{eqnarray} 
where $\mathcal{P}$ denotes the Cauchy principal value integral and $\mathcal{N}_a(k,q)$ contains the factors coming from interaction vertices as well as the numerator of the fermion propagators. For the scalar and pseudoscalar channels, they are given by
\begin{eqnarray}
\mathcal{N}_\sigma(k,q) &=& N_cN_f\Tr{(\cancel{k}+\cancel{q}+M)(\cancel{k}+M)} = 4N_cN_f(M^2 +k^2 +k\cdot q ) \\
\mathcal{N}_\pi (k,q) &=& -N_cN_f\Tr{\gamma^5(\cancel{k}+\cancel{q}+M)\gamma^5(\cancel{k}+M)} = -4N_cN_f(M^2 -k^2 -k\cdot q )~. 
\end{eqnarray}
It can be noticed that, a sharp three-momentum cutoff has been used to regulate the temperature independent part in Eq.~\eqref{pola_0}. We have checked that, the substitution of the above equations into Eq.~\eqref{pola_0} followed by some simplification agrees with the expressions of the polarization functions as given in Ref.~\cite{Klevansky}.

The effect of external magnetic field in the polarization function is included by means of Schwinger proper time formalism~\cite{Schwinger:1951nm}. 
Thus, under both the external magnetic field ($B\hat{z}$) and finite temperature, the real part of the analytic thermo-magnetic polarization function (denoted by a double bar) becomes~\cite{Olivio,Ghosh:2019fet}
\begin{eqnarray}
\RE\overline{\overline{\Pi}}_a (q^0,\vec{q})&=&  \RE ~ i \kfourint{k}\Tr{{S_B^{11}(k)}\Gamma_a {S_B^{11}(p=k+q)} \Gamma_a }
\label{eq.repi.2}
\end{eqnarray}
where, $S_B^{11}(k)$ is the $11$-component of the real time thermo-magnetic quark propagator given by,
\begin{equation}
S_B^{11}(k) = S_B(k) - \eta(k\cdot u) \TB{S_B(k)-\gamma^0S_B^\dagger(k)\gamma^0} 
\label{eq.prop.11B}
\end{equation}
with $S_B(k)$ being the momentum space Schwinger proper time propagator for charged Fermions. It is worth noting that, the corresponding coordinate space Schwinger propagator contains a gauge dependent phase factor which gets canceled for the calculation of loop graphs containing equally charged particles. The expression for $S_B(k)$ for a particular quark flavour $f$ including the AMM of quarks is given by~\cite{Aguirre}
\begin{eqnarray}
S_B(k) &=& \sum_{s\in \SB{\pm 1}}\sum_{n=0}^{\infty } \FB{ 1-\delta^0_n\delta^{-1}_s} \mathscr{D}_{nfs} (k)\TB{ \dfrac{-1}{\vparasq{k} - \FB{M_n - s\kappa_fe_fB }^2 +i\epsilon }} 
\label{eq.prop.SB}
\end{eqnarray}
where,
\begin{eqnarray}
\mathscr{D}_{nfs} (k) = \dfrac{(-1)^n e^{-\alpha_k}}{2M_n} && 
\TB{(M_n + sM)(\slashed{k}_\parallel - \kappa_f e_f B + sM_n) \SB{\unit +\sign{e_f} i\gamma^1\gamma^2} L_n(2\alpha_k)  \nn \right. \\ 
&& ~~~ \left. - (M_n-sM) (\slashed{k}_\parallel + \kappa_f e_f B- sM_n) \SB{\unit -\sign{e_f} i\gamma^1\gamma^2}L_{n-1}(2\alpha_k) 
\nn  \right.\\ && \left. -4 \SB{\slashed{k}_\parallel + \sign{e_f} i\gamma^1\gamma^2s (M_n -s\kappa_fe_f B)   }  
\sign{e_f} i\gamma^1\gamma^2 s\slashed{p}_\perp L^1_{n-1} \FB{2\alpha_k} }  
\label{eq.D}
\end{eqnarray}
with $ \alpha_k =-\vperpsq{k}/e_fB $, $ M_n = \sqrt{M^2 + 2n\MB{e_fB}} $ and $L^a_n(z)$ being the generalized Laguerre polynomial (with the condition $L^a_{-1}(z) = 0$). Due to the external magnetic field in the positive $ z $-direction, the decomposition of a four vector is done as $k=(k_\parallel+k_\perp)$ where $k_\parallel^\mu=g_\parallel^\munu k_\nu$ and $k_\perp^\mu=g_\perp^\munu k_\nu$ with the corresponding metric tensors $g_\parallel^\munu=\text{diag}(1,0,0,-1)$ and $g_\perp^\munu=\text{diag}(0,-1,-1,0)$.

We now substitute Eq.~\eqref{eq.prop.11B} into Eq.~\eqref{eq.repi.2} and perform the $dk^0d^2k_\perp$ integrals. Some relevant steps for this calculation are provided in Appendix~\ref{app.pola} and we get from Eq.~\eqref{eq.pola.4}
\begin{eqnarray}
\RE\overline{\overline{\Pi}}_a (q) &=& \sum_f \sum_{s_k,s_p} \sum_{l=0}^{\infty} \int_{0}^{\sqrt{\Lambda^2-\vec{k}_{\perp l}^2}} \dfrac{dk_z}{\pi}  
\Theta\FB{\vec{k}_{\perp l}^2} \Theta\FB{\vec{p}_{\perp l}^2} \Theta\FB{\Lambda^2-\vec{k}_{\perp l}^2} \Theta\FB{\Lambda^2-\vec{p}_{\perp l}^2} \nn \\ 
&& \mathcal{P} \TB{\dfrac{\tilde{\mathcal{N}}^a_{lls_ks_p}(k^0=-q^0+\omega^{ls_p}_{k})}{2 \omega^{ls_p}_{k} 
		\SB{\FB{q^0 - \omega^{ls_p}_{k}}^2 - \FB{\omega^{ls_k}_{k}}^2 }} 
	+\dfrac{\tilde{\mathcal{N}}^a_{lls_ks_p}(k^0=\omega^{ls_k}_{k})}{2 \omega^{ls_k}_{k} \SB{\FB{q^0 + \omega^{ls_k}_{k}}^2 - \FB{\omega^{ls_p}_{k}}^2 }} }
+ \sum_f \sum_{s_k,s_p} \sum_{l=0}^{\infty} \int_{-\infty}^{+\infty} \dfrac{dk_z}{(2\pi)}  \nn \\ && 
\Theta\FB{\vec{k}_{\perp l}^2} \Theta\FB{\vec{p}_{\perp l}^2}
\mathcal{P} \TB{
	- \dfrac{\tilde{\mathcal{N}}^a_{lls_ks_p}(k^0=-\omega^{ls_k}_{k}) f^-(\omega^{ls_k}_{k}) }{2 \omega^{ls_k}_{k} 
		\SB{\FB{q^0 - \omega^{ls_k}_{k}}^2 - \FB{\omega^{ls_p}_{k}}^2  }  } - \dfrac{\tilde{\mathcal{N}}^a_{lls_ks_p}(k^0=\omega^{ls_k}_{k}) f^+(\omega^{ls_k}_{k})}
	{2 \omega^{ls_k}_{k} \SB{\FB{q^0 + \omega^{ls_k}_{k}}^2 - \FB{\omega^{ls_p}_{k}}^2  }  }   \right. \nn  \\ &&  \hspace{0cm} \left.
	-\dfrac{\tilde{\mathcal{N}}^a_{lls_ks_p}(k^0=-q^0-\omega^{ls_p}_{k}) f^-(\omega^{ls_p}_{k})   }{2 \omega^{ls_p}_{k} 
		\SB{\FB{q^0 + \omega^{ls_p}_{k}}^2 - \FB{\omega^{ls_k}_{k}}^2  }  } - \dfrac{\tilde{\mathcal{N}}^a_{lls_ks_p}(k^0=-q^0 +\omega^{ls_p}_{k}) f^+(\omega^{ls_p}_{k})}
	{2 \omega^{ls_p}_{k} \SB{\FB{q^0 - \omega^{ls_p}_{k}}^2 - \FB{\omega^{ls_k}_{k}}^2  }  } 
}
\label{eq.pola.5}
\end{eqnarray}
where 
\begin{eqnarray}
\omega^{ls_k}_k &=& \sqrt{k_z^2+(M_l-s_k\kappa eB)^2}~, \\
\vec{k}_{\perp l}^2 &=& 2 leB +\FB{\kappa e B}^2 - 2s_kM_{l}(\kappa eB)~,  \\
\vec{p}_{\perp l}^2 &=& 2 leB +\FB{\kappa e B}^2 - 2s_pM_{l}(\kappa eB)~.
\end{eqnarray}
and the expression for $\tilde{\mathcal{N}}^a_{lls_ks_p}$ is given in Eq.~\eqref{eq.Ntilde}.
%


\section{Numerical Results} \label{sec.results}
In this section, we present numerical results for constituent quark mass, several thermodynamic quantities 
and properties of scalar and neutral pseudoscalar mesons in a hot and dense medium in presence of uniform magnetic field. 
As already discussed in Sec.~\ref{deri_gap_eqn}, due to the four-fermion contact interaction among the quarks, 
NJL model becomes non-renormalizable and we have described a field dependent regularization technique where three-momentum 
cut-off parameter($ \Lambda $) was used to get rid of divergent integrals. Now, following Refs.~\cite{Klevansky,Buballa,Sasaki}, 
we have chosen $ \Lambda=587.9$ MeV, coupling constant $ G=2.44/\Lambda^2 $ and bare mass of quarks $ m=5.6$ MeV. 
These parameters have been fixed by fitting the empirical values of pion mass $ m_\pi=135.0 $ MeV and pion decay 
constant $ f_\pi=92.4$ MeV at zero temperature and baryon density in absence of background magnetic field. 
We have considered constant values of AMM of quarks $ \kappa_u= 0.29 \ {\rm GeV^{-1}}$  and $ \kappa_d= 0.36 \ {\rm GeV^{-1}}$ 
following Ref.~\cite{Sadooghi}.
\begin{figure}[h]
	\begin{center}
		\includegraphics[angle=-90, scale=0.30]{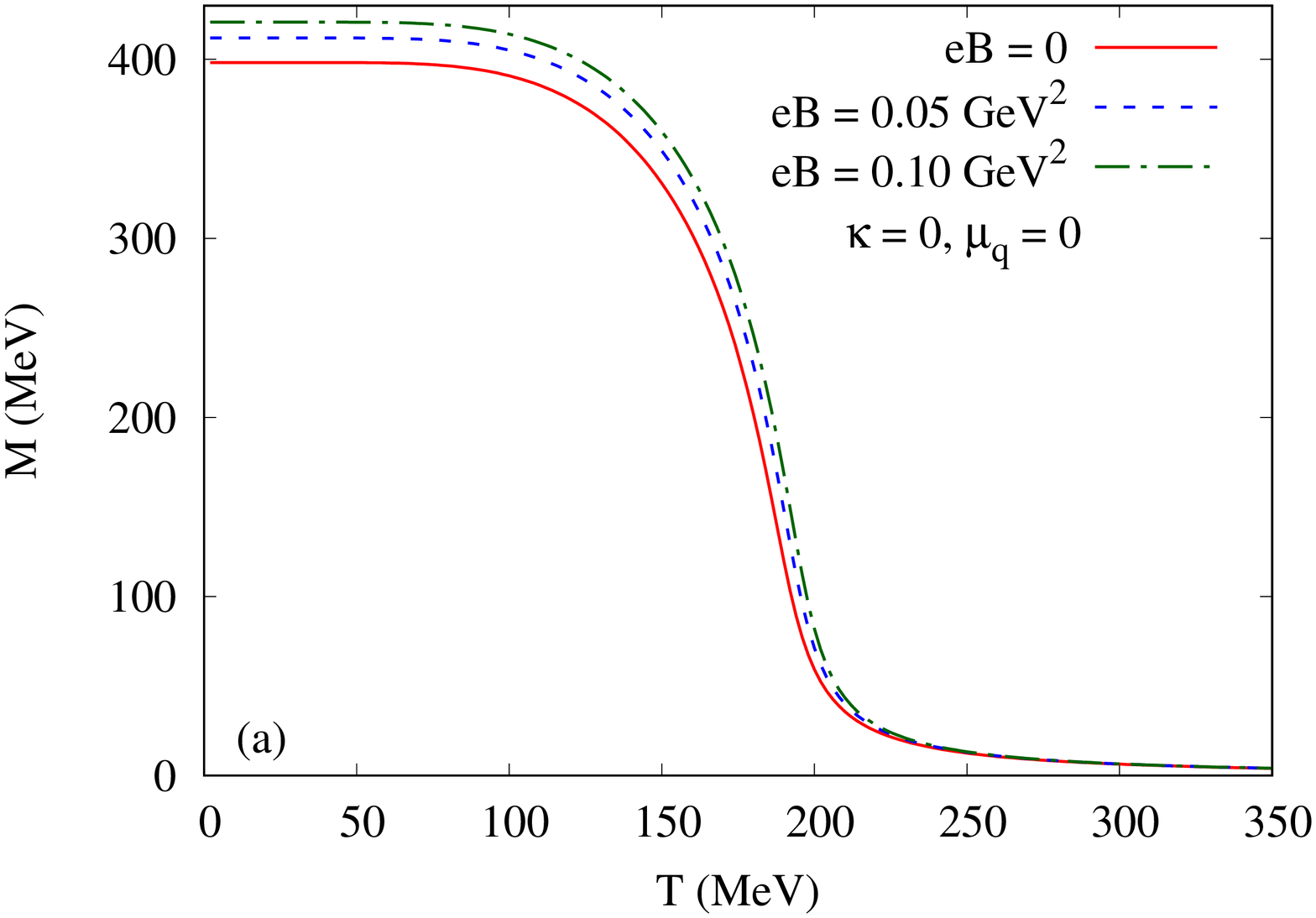} \includegraphics[angle=-90, scale=0.30]{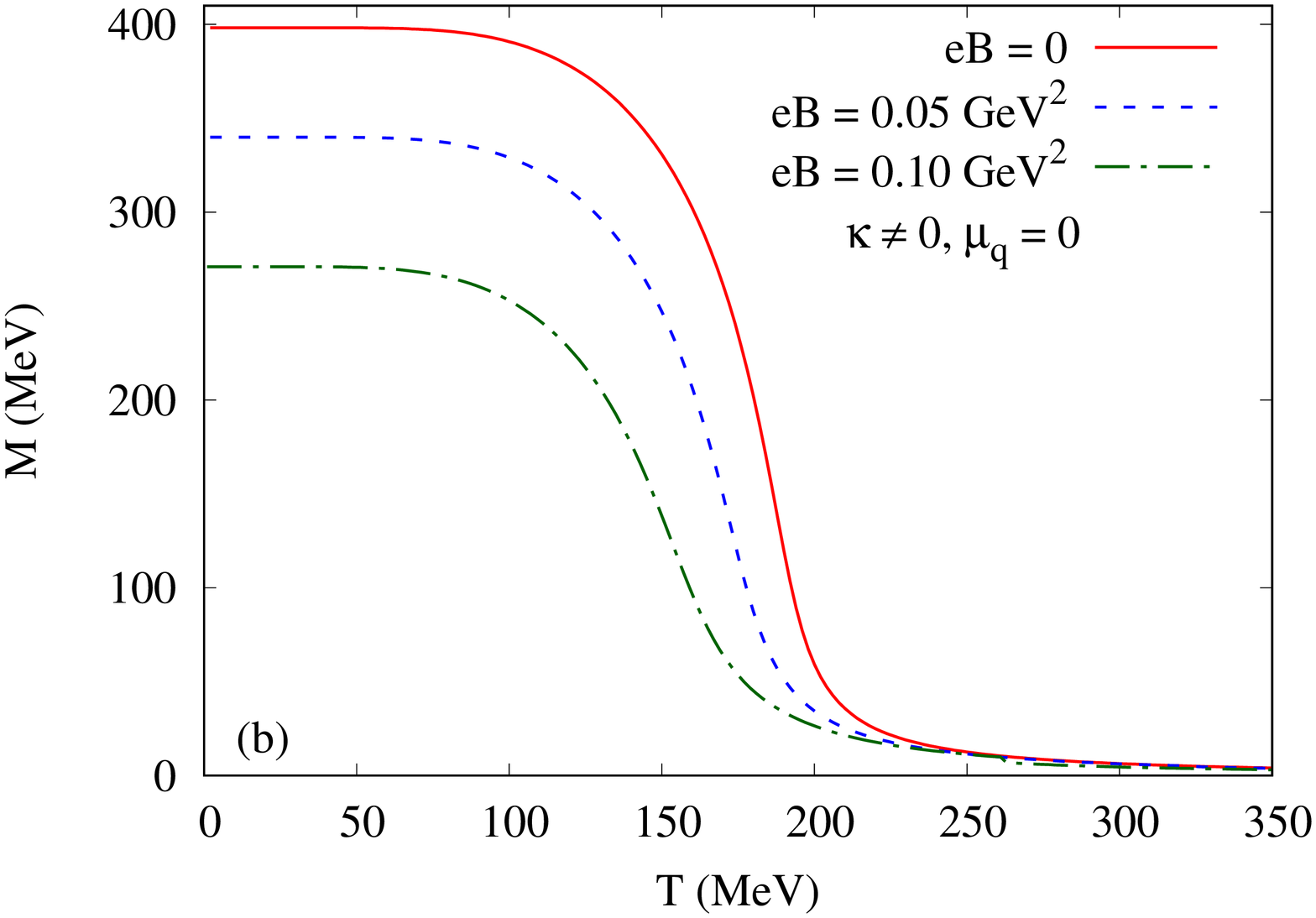} \\
		\includegraphics[angle=-90, scale=0.30]{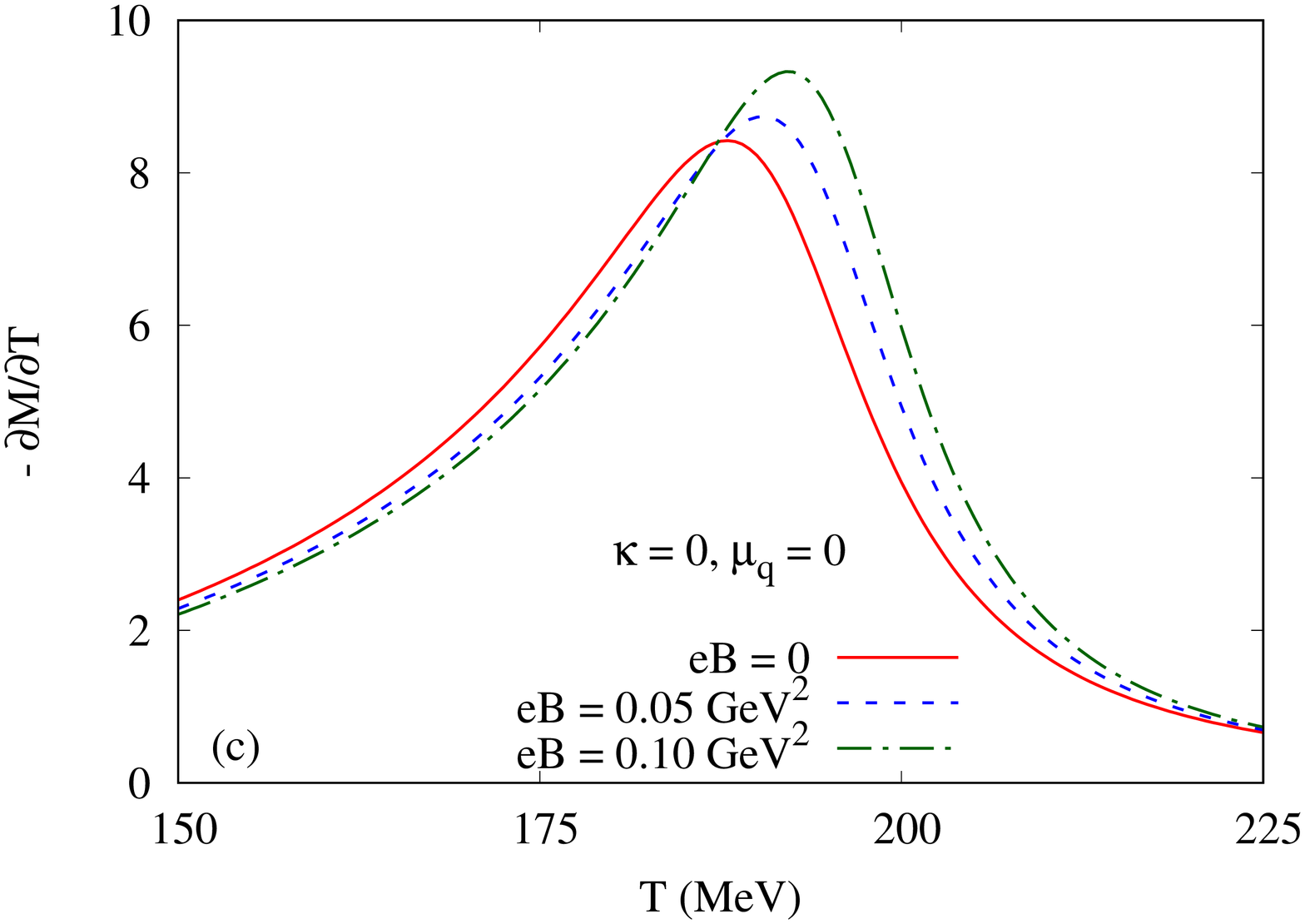} \includegraphics[angle=-90, scale=0.30]{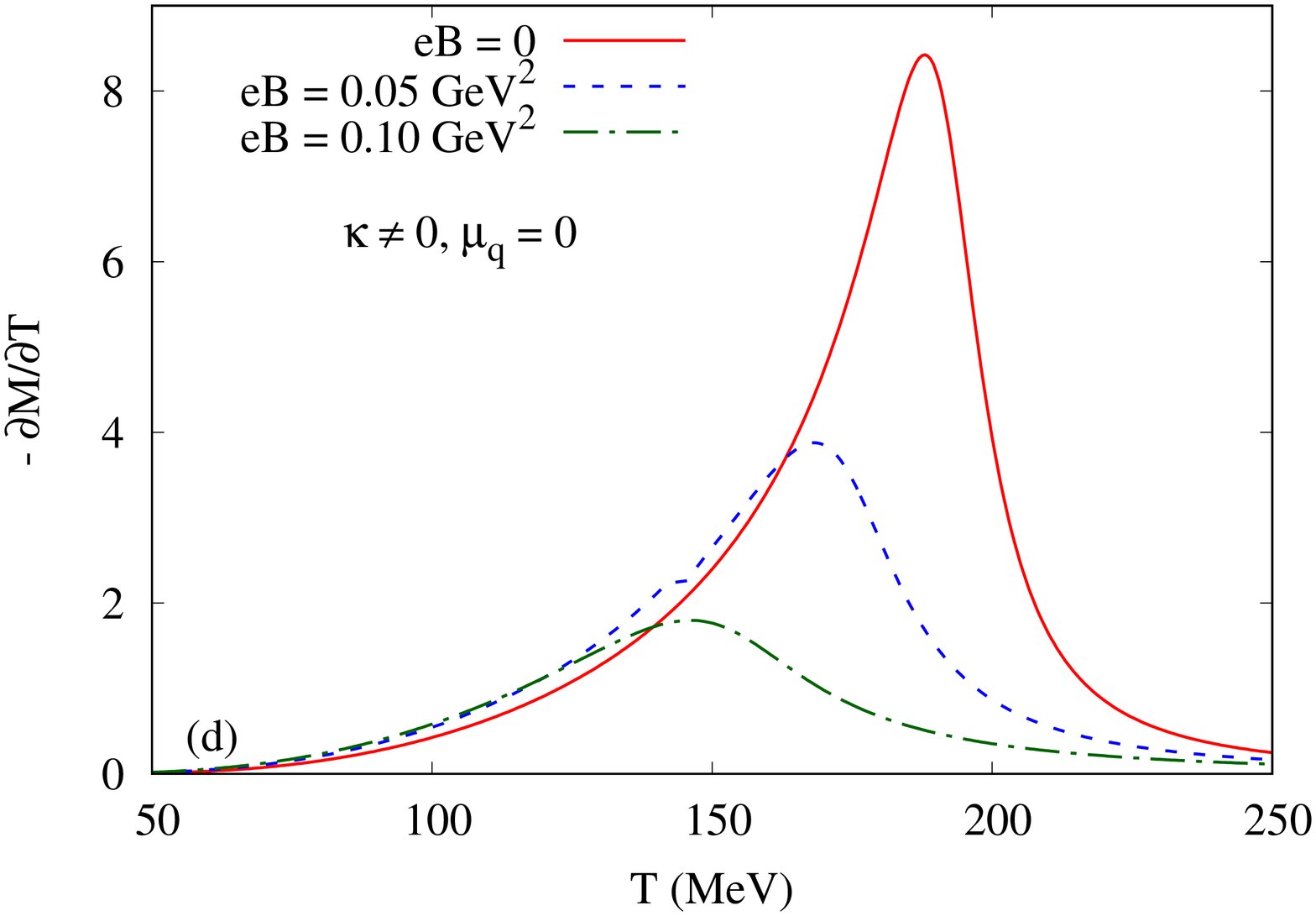} 
	\end{center}
	\caption{Variation of constituent quark mass ($ M $) with temperature ($ T $) at zero quark chemical potential 
		$ (\mu_q) $ for three different values of external magnetic field ($ eB= 0.0$ , 0.05 and 0.10 GeV$^2$) for 
		(a) $\kappa_{f} =0$ and (b) $\kappa_{f} \ne 0$. 
		Variation of $ \frac{\partial M}{\partial T}$ with temperature ($ T $) at zero quark chemical potential 
		$ (\mu_q) $ for three different values of external magnetic field ($ eB= 0.0$ , 0.05 and 0.10 GeV$^2$) for 
		(c) $\kappa_{f} =0$ and (d) $\kappa_{f} \ne 0$.}
	\label{dep_cons_qmass}
\end{figure}

In Fig.~\ref{dep_cons_qmass}(a) we have demonstrated temperature dependence of constituent quark mass at $ \mu_q=0 $ 
for $ eB= 0.0, 0.05$ and $0.10 {\rm \ GeV^2}$ respectively without considering AMM of quarks. 
In all the cases, $ M $ almost remains constant up to $ T \approx 100 $ MeV and the transition from chiral 
symmetry broken (with $ M\ne 0 $) to the restored phase (i.e. $ M\approx m\approx 0 $), is a smooth crossover. 
Since we have considered non vanishing current quark mass, $m = 5.6$ MeV, the chiral symmetry is never restored fully. 
So, the temperature for which $ M $ has the highest change can be identified as critical temperature($ T_c $). 
Note that, for stronger values of magnetic field,  $ M $ increases as $ T\rightarrow 0 $ and the transitions to the 
symmetry restored phase take place at the larger values of temperature, which is immediately evident from Fig.~\ref{dep_cons_qmass}(c) 
where we have plotted the variation of $ -\partial M/ \partial T $ as a function $ T $. It can be seen that, as $ eB $ 
increases, the peak representing $T_c $, shifts towards the higher values of temperature. 
This phenomenon is known as \textit{Magnetic Catalysis} (MC)~\cite{Shovkovy,Gusynin1,Gusynin2,Gusynin3,Kharzeev}, which 
explains that the magnetic field has a strong tendency to enhance (or `\textit{catalyze}') spin-zero 
fermion-antifermion $(\pbarpasi)$ condensates. Moreover, height of the peaks also increases with increase 
in $ eB $ implying the fact that $M$ decreases more rapidly around a certain value of temperature as we keep on 
increasing the magnetic field. Now in Fig.~\ref{dep_cons_qmass}(b) we have included AMM of quarks without altering 
the other external factors, such as, $ eB$ and $\mu_q$. In this case we observe an opposite effect. 
The value of $M$ deceases significantly with the increase of $ eB $ at low temperature values and the transitions 
occur at smaller values of temperature with increasing $ eB $. Thus, in  contrast to MC, chirality broken phase becomes 
unfavourable with increasing magnetic field. This is known as \textit{inverse magnetic catalysis}(IMC)~\cite{Inagaki2, Preis, Kharzeev,Bali_lat,Bali_lat2,Bali_lat3,Bornyakov}. The corresponding curve of $ -\dfrac{\partial M}{\partial T} $ vs $ T $ shows s broadening  of the peak in Fig.~\ref{dep_cons_qmass}(d) which implies that, for finite values of AMM of quarks, the transition occurs for larger range of temperature about $T_C$.  
  \begin{figure}[h]
  	\begin{center}
  		\includegraphics[angle=-90, scale=0.3]{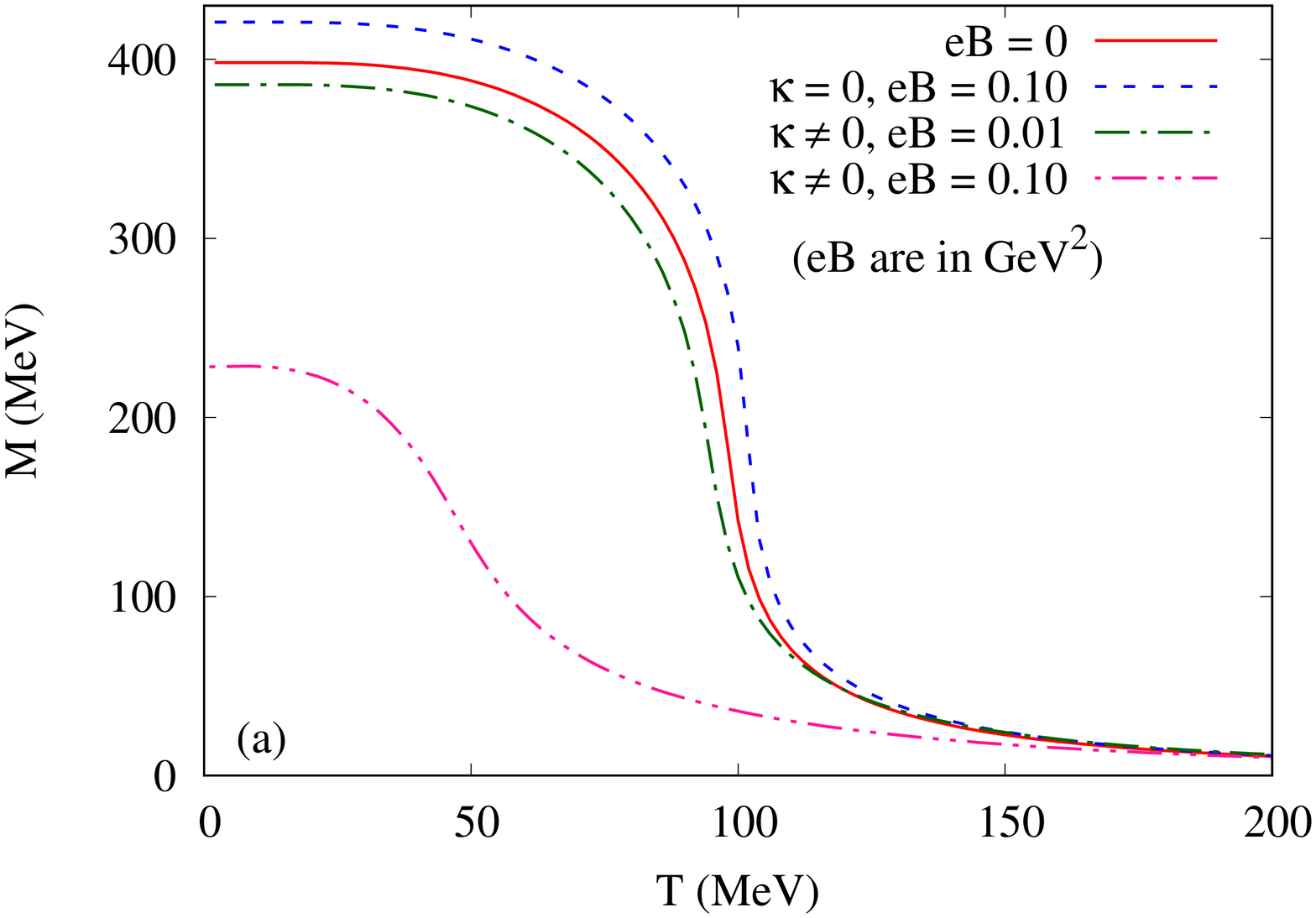}
  		\includegraphics[angle=-90, scale=0.3]{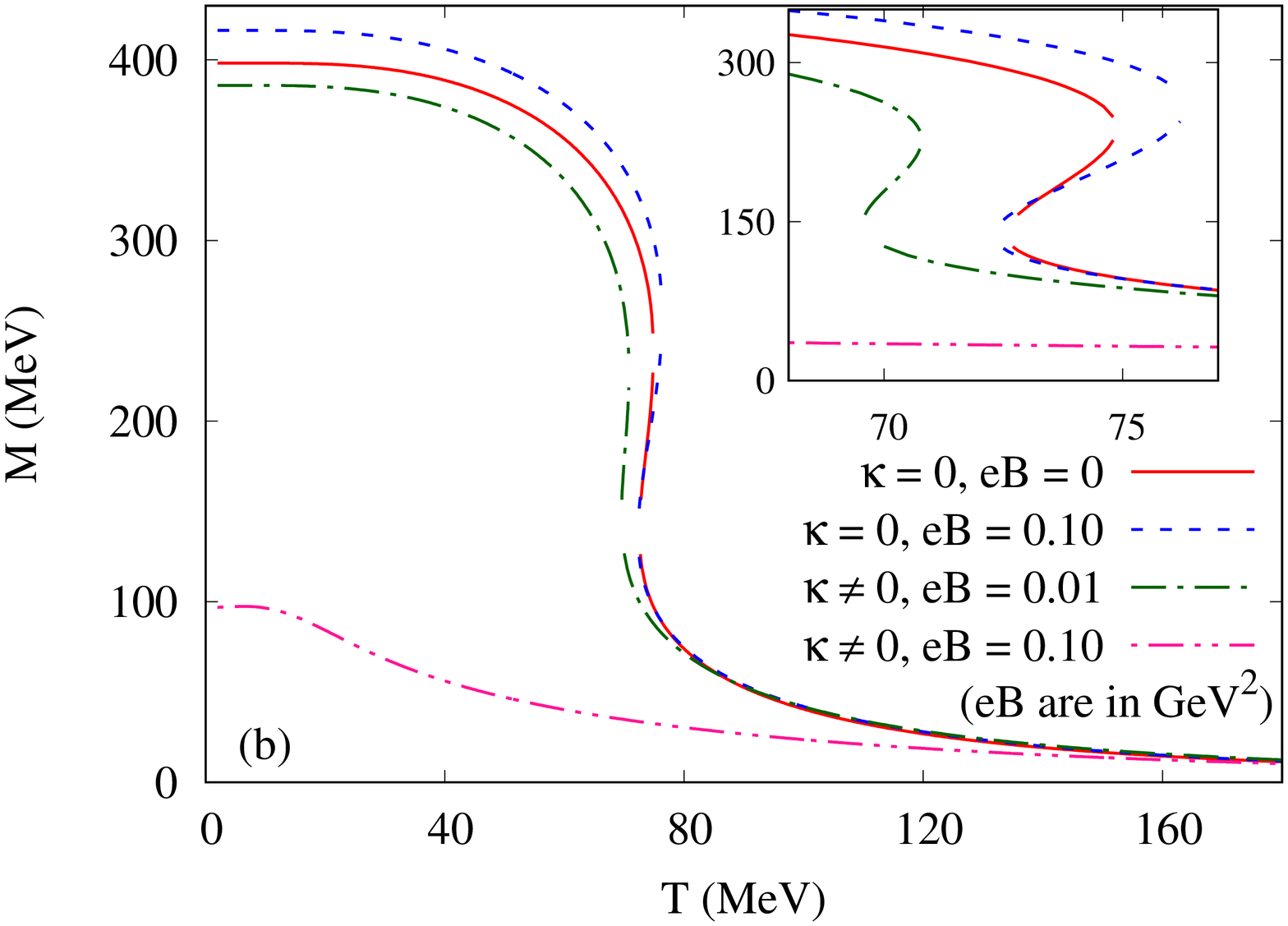}
  	\end{center}
  	\caption{ $ T $ dependence of $ M $ at (a) $ \mu_q = 300$  MeV  and (b) at $ \mu_q= 330 $ MeV for different values of $ eB $ with and without considering AMM of quarks. The inset plot in (b) shows the multi-valued nature of $M$ by focusing on the relevant temperature range.}
  	\label{M_vs_T_diff_mu}
  \end{figure}

In Figs.~\ref{M_vs_T_diff_mu}(a) and (b), we have shown $M$ as function of $T$ for two different values of 
quark chemical potential ($ \mu_q =300 $ and $ 330 $ MeV).  It is important to note that in the first case, 
for finite and vanishing values of $ eB $  as well as $ \kappa_f $,  $ M $ decreases continuously as a function of $ T $. 
But for higher value of $ \mu_q $ we obtained several possible solutions of $  M $ from the gap equation for a range of temperature. 
These solutions corresponds to the absolute (stable) minima, local minima (metastable) and maxima (unstable) of the effective potential. Existence of three different solutions can be understood from considering the gap equation in zero field scenario 
as expressed in Eq.~\eqref{gap_zero}. Since this is a cubic equation in variable $ M $, in principle, there will 
always be three roots for $ M $ (three real or one real and two imaginary roots). For lower values of $ \mu_q $ there exists only one real root and as we increase $ \mu_q $, after a certain value  the imaginary roots become real. 
This multivaluedness of $ M $  is a signature of first order transition. 
Comparing the solid-red and dashed-blue curves in both the Figs.~\ref{M_vs_T_diff_mu}(a) and (b), 
we notice that, in the absence of the AMM of quarks the nature of 
pseudo-chiral phase transition changes from crossover to first order with the increase in quark chemical potential. 
Similar behaviour is also seen even for non-zero AMM of the quarks (the dash-dot-green curves) for lower values 
of $eB$ ($\sim$ 0.01 GeV$^2$). 
However, at higher values of external magnetic field ($\sim$ 0.10 GeV$^2$), the transition remains always cross over 
(the dash-dot-dot-pink curves) for $\kappa\ne0$.
%
%
%
 \begin{figure}[h]
	\begin{center}
		\includegraphics[angle=-90, scale=0.3]{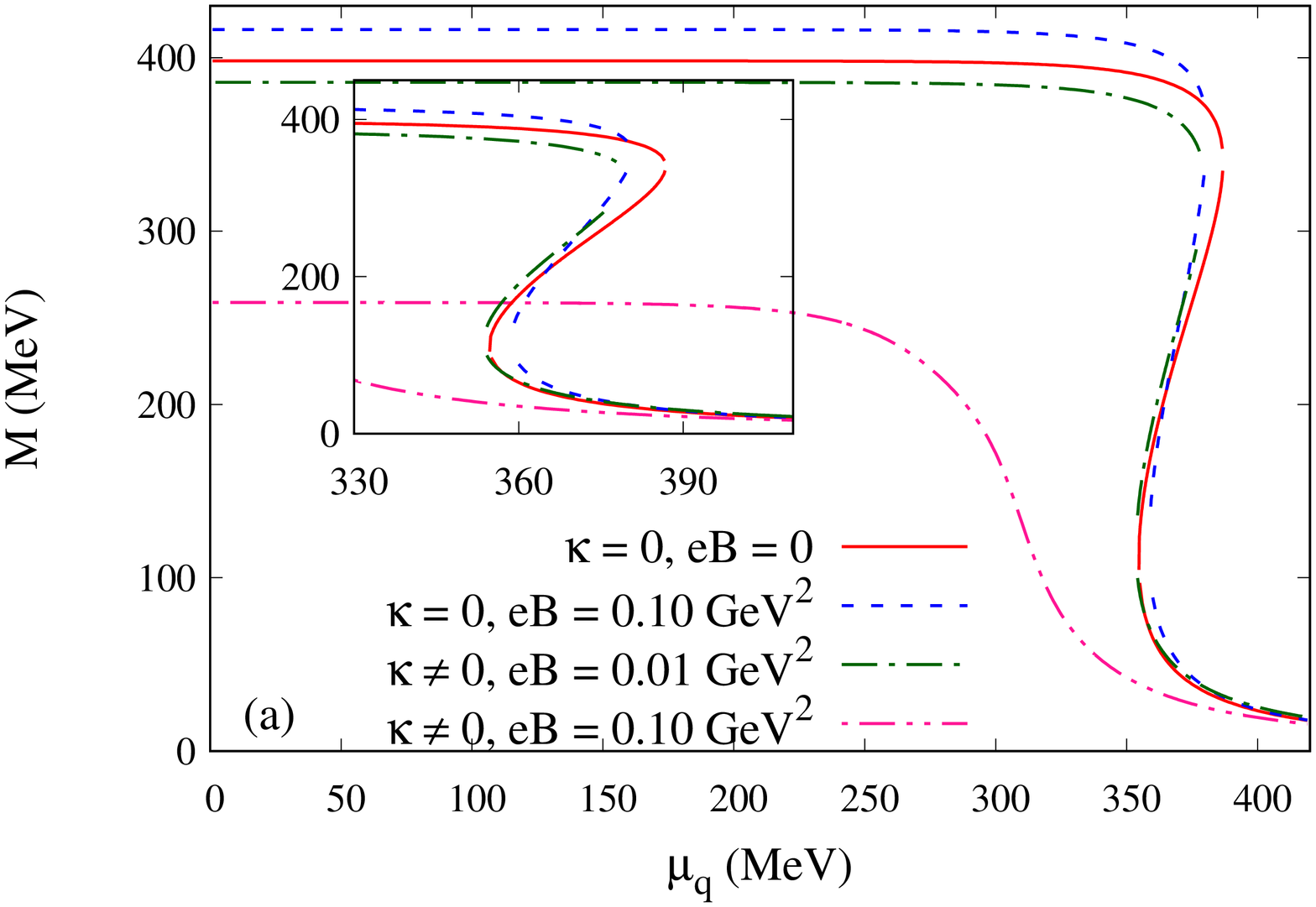}
		\includegraphics[angle=-90, scale=0.3]{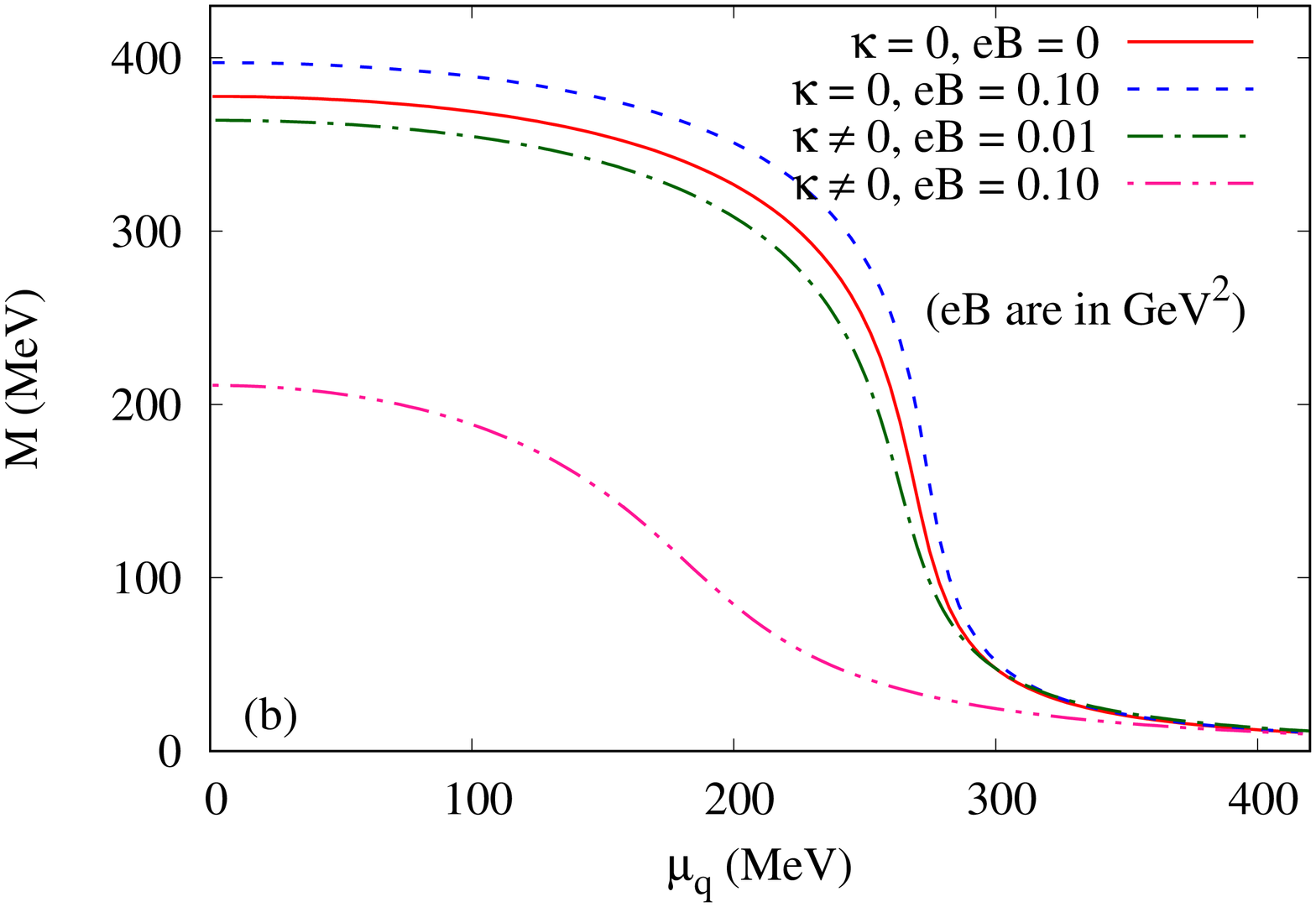}
	\end{center}
	\caption{ $ \mu_q $ dependence of Constituent quark mass ($ M $) at (a) $ T = 30$  MeV  and (b) at $ T= 120 $ MeV for different values of $ eB  $  and $  \kappa $. The inset plot in (a) shows the multi-valued nature of $M$ by focusing on the relevant $\mu_q$ range.}  
	\label{M_vs_mu}
\end{figure}

Next, in Figs.~\ref{M_vs_mu}(a) and (b), $M$ is plotted as a function of $ \mu_q $ at two different temperature ($ T= 30 $ 
and $ 120 $ MeV) for four different cases as discussed in the previous paragraph. 
Here also we get multiple solutions of $ M $ but at lower 
values of temperature for similar external conditions as discussed above. This interplay of $ T $ and $ \mu_q $ on the 
mechanism of phase transition will be revisited while discussing the phase diagram. It is to be noted that, with the increase in 
temperature, $\mu_c $ decreases. 
 \begin{figure}[h]
 	\begin{center}
 		\includegraphics[angle=-90, scale=0.3]{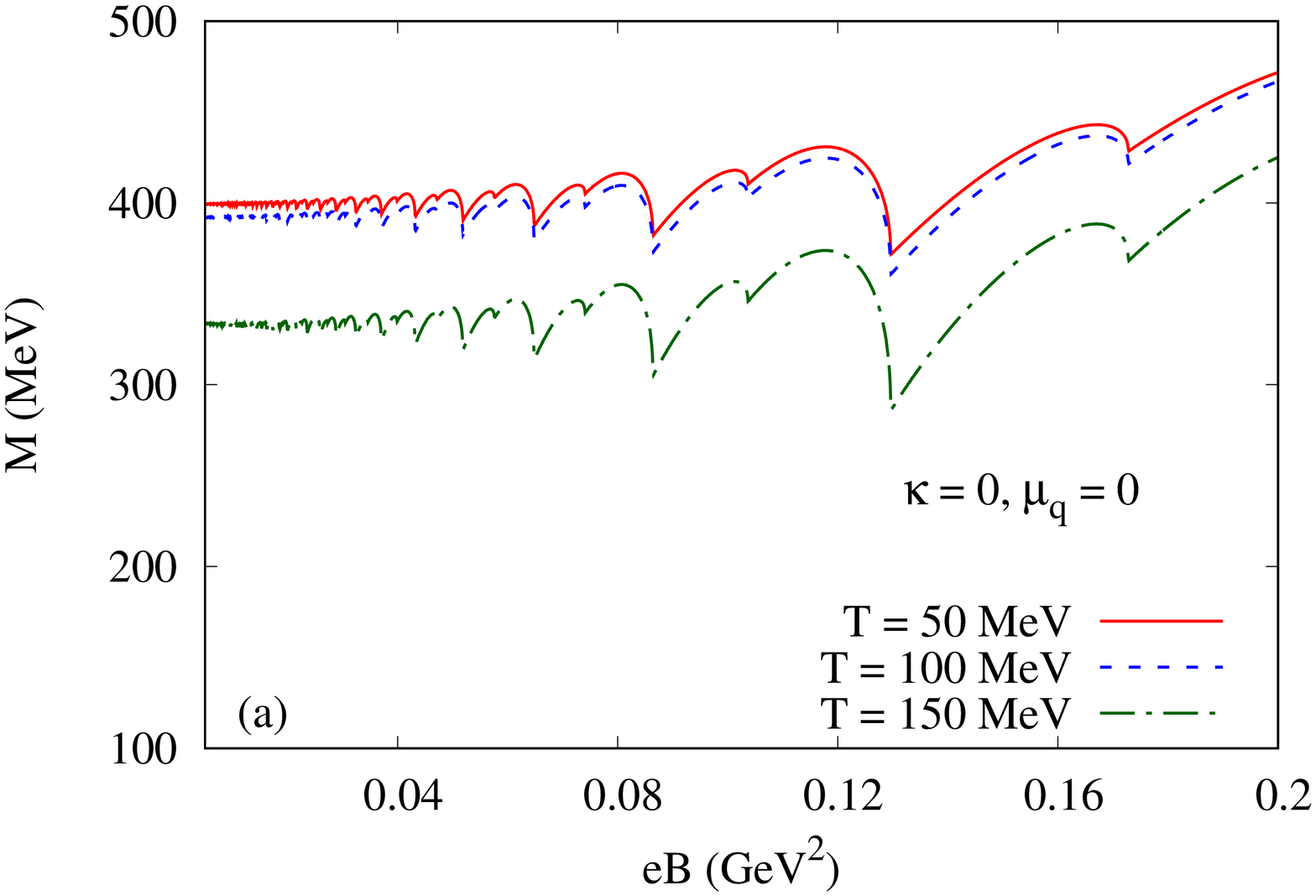}
 		\includegraphics[angle=-90, scale=0.3]{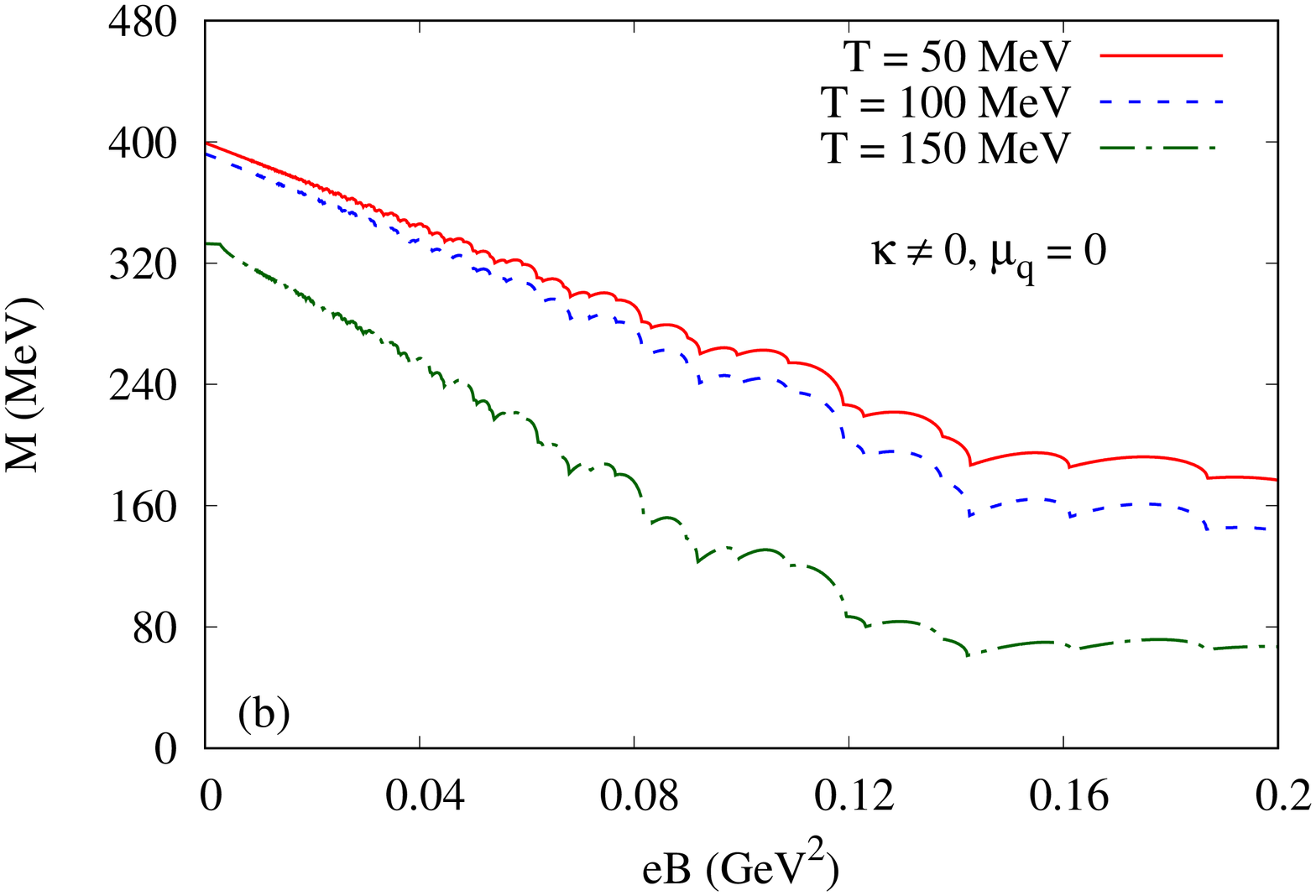}
 	\end{center}
 	\caption{Variation of constituent quark mass ($ M $) at $ \mu_q=0 $ (a) without considering AMM of quarks and (b) considering non-zero values of $ \kappa_f $ for temperatures of $ T = 50 , 100   $ and $ 150  $~MeV.  }
 	\label{M_vs_eB}
 \end{figure}

In Figs.~\ref{M_vs_eB}(a) and (b) we have plotted $ eB $-dependence of $ M $ at vanishing quark chemical potential 
with and without AMM of quarks for three different values of temperatures. We have not used any smoothing function as 
done in Refs~\cite{Sadooghi1,Sadooghi,Zhang} during numerical evaluation which leads to the oscillatory behaviour of $M$. 
These oscillations are related to the well known de Haas-van Alphen (dHvA) effect~\cite{Landau:1980mil} 
in the weak magnetic field regime and had also been observed 
in Refs.~\cite{Fayazbakhsh1,Fayazbakhsh2,Sadooghi,Sadooghi1,Ebert:1998gx,Inagaki:2004ih,Noronha:2007wg,Fukushima:2007fc,Orlovsky:2014kva}. 
It occurs whenever the Landau levels cross the quark Fermi surface. As can be noticed in Fig.~\ref{M_vs_eB}, the dHvA oscillations get 
smeared out with the increase in magnetic field (as LLL dominates) in agreement with Ref.~\cite{Sadooghi}.
As expected from Figs.~\ref{dep_cons_qmass}(a) and (b), for a particular temperature there is an overall 
increase of $ M $ with $ eB $ when AMM of quarks are not taken into consideration. On the other hand, inclusion of AMM 
leads to a reduction in $M$ with increasing $ eB $ which indicates the occurrence of IMC during the transition from 
broken to symmetry restored phase. 
\begin{figure}[h]
	\begin{center}
		\includegraphics[angle=-90, scale=0.3]{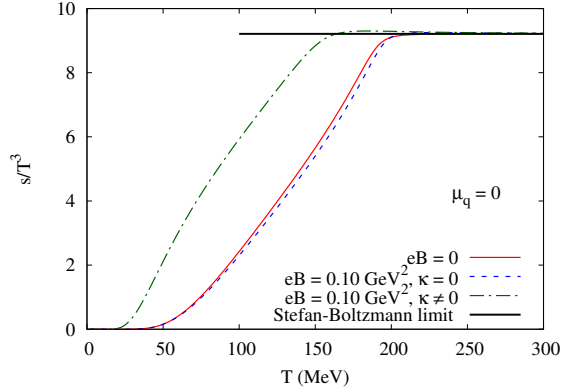}
	\end{center}
	\caption{Variation of scaled entropy density as function of temperature at $ \mu_q=0 $ for zero and non-zero values of $ eB $ with and without including AMM of quarks along with the Stefan-Boltzmann limit for the ideal gases.}
	\label{Fig.Entropy}
\end{figure}

Now we turn our attention to several thermodynamic observables defined in Sec.~\ref{TD_obs}. In Fig.~\ref{Fig.Entropy}, 
we have plotted the scaled entropy density ($s/T^3$) as functions of the temperature for three different conditions. 
In all cases, the scaled entropy density increases monotonically  with temperature and eventually saturates at the 
corresponding ideal gas limit. Since the transition to the high temperature phase, as shown in 
Figs.\ref{dep_cons_qmass}(a) and (b), is a rapid crossover the entropy varies continuously with increasing temperature. 
Note that, when we include AMM of quarks, the steep rise in $ s/T^3 $  starts at lower values of temperature compare to the case for vanishing AMM of quarks.
 \begin{figure}[h]
 	\begin{center}
 		\includegraphics[angle=-90, scale=0.3]{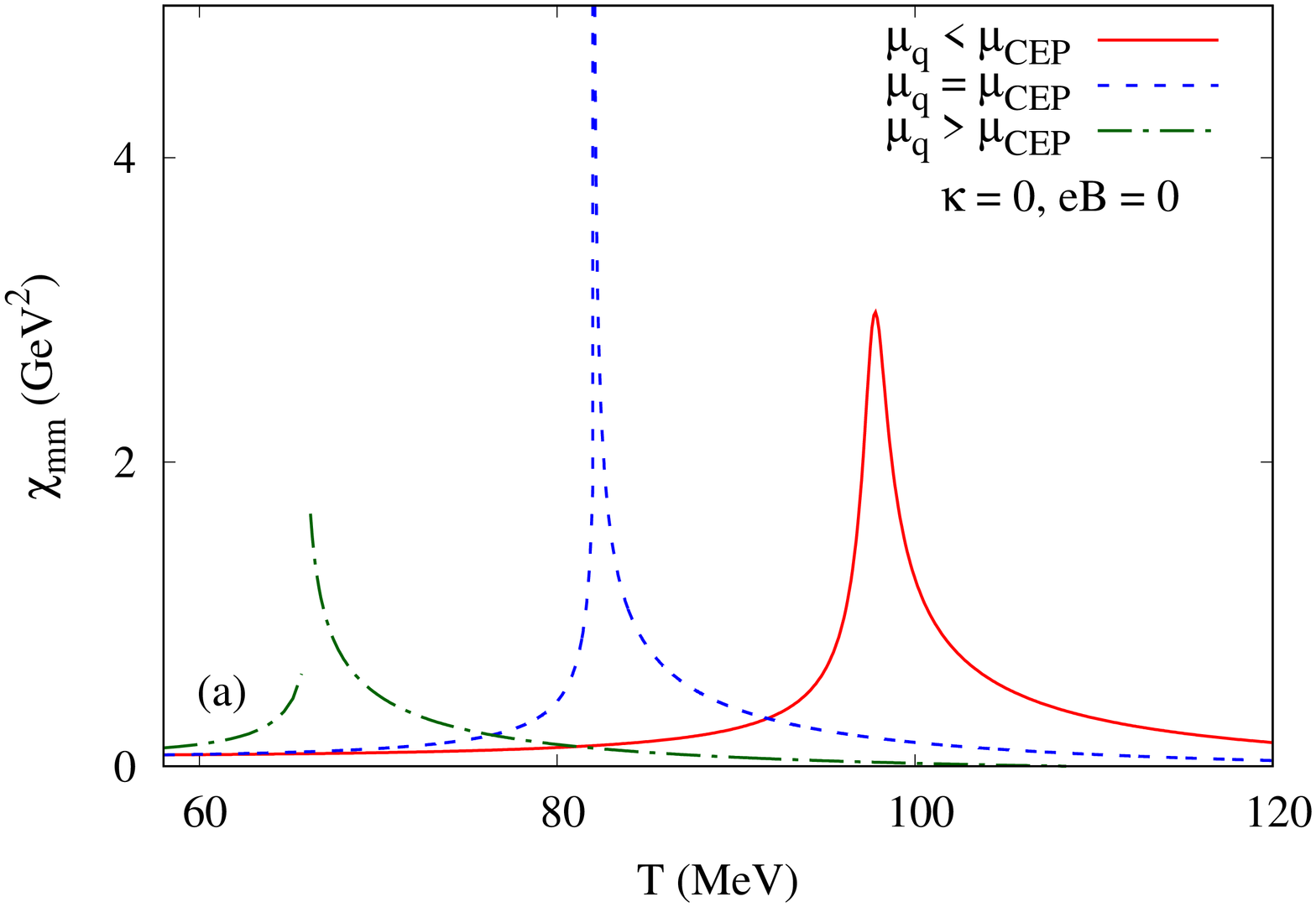}
 		\includegraphics[angle=-90, scale=0.3]{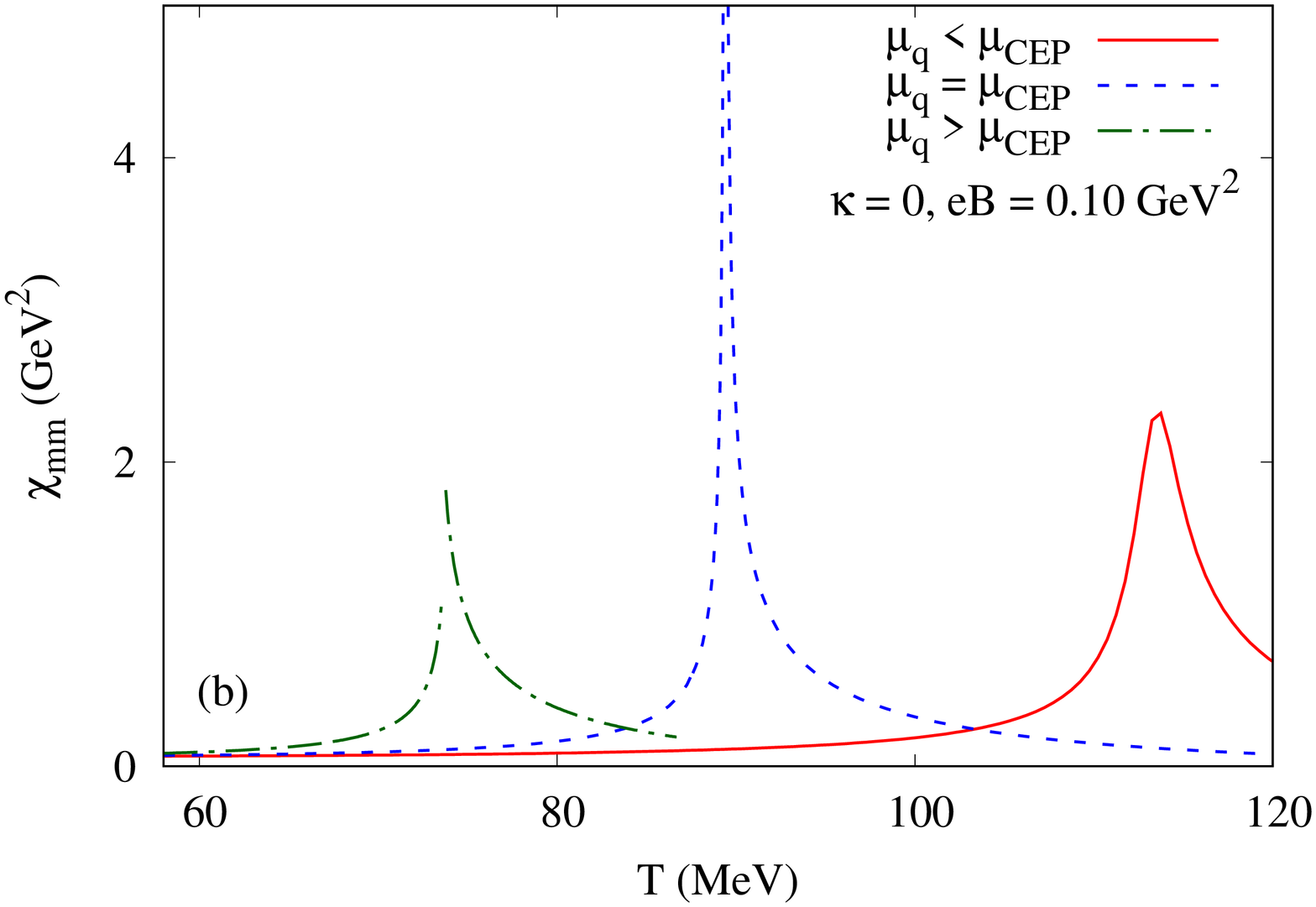} \\ 
 		\includegraphics[angle=-90, scale=0.3]{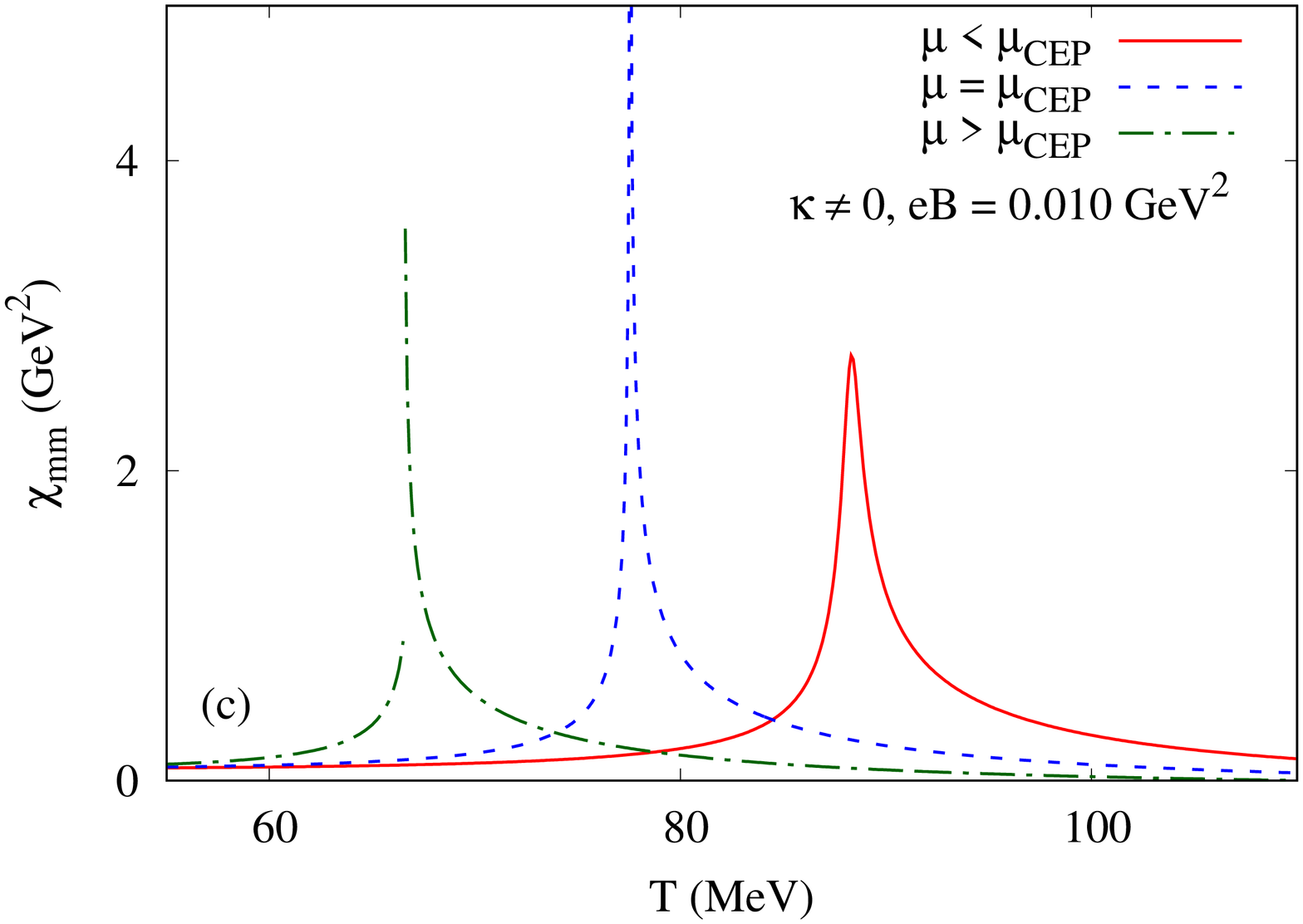}		
 		\includegraphics[angle=-90, scale=0.3]{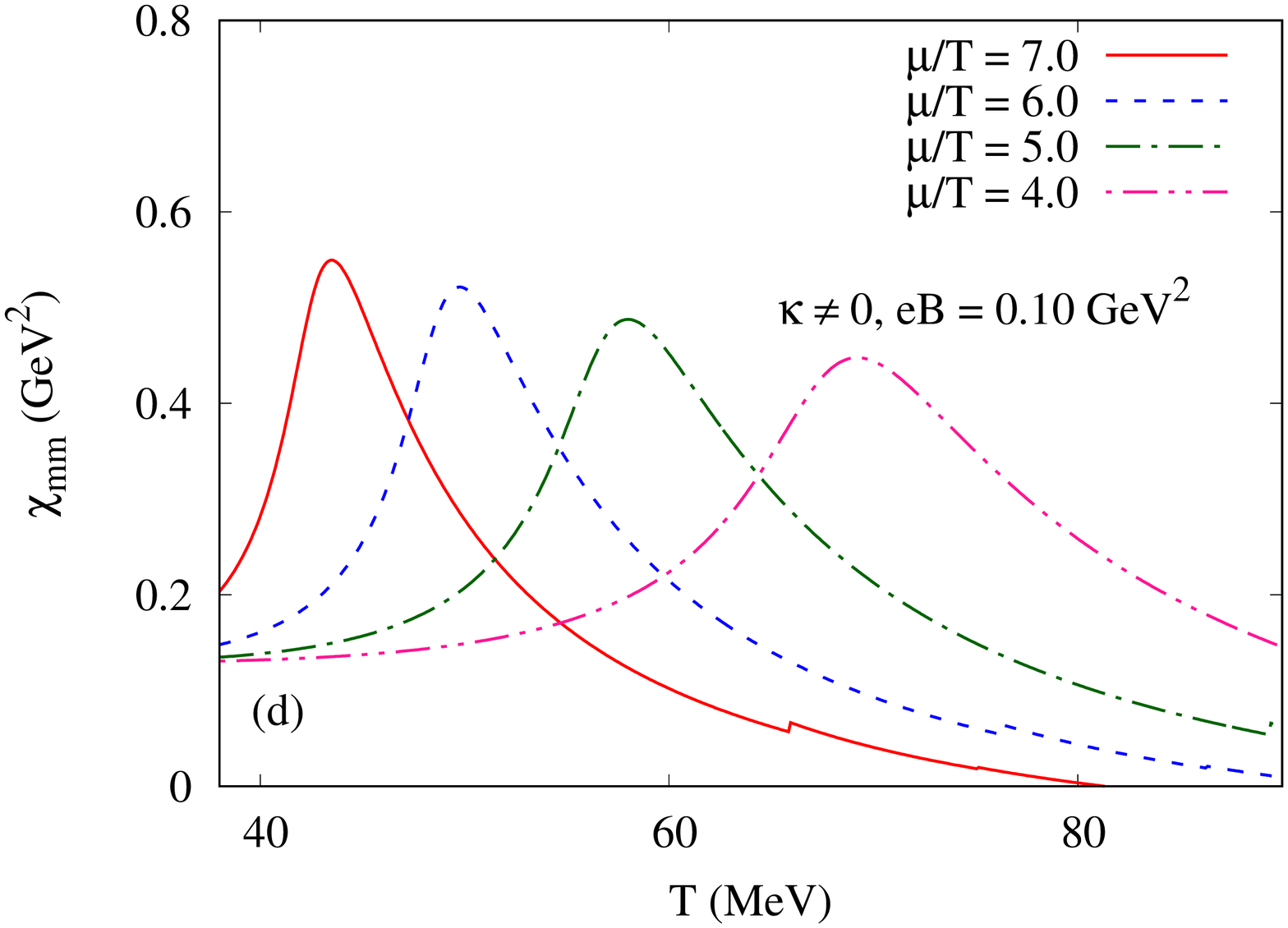}			
 	\end{center}
 	\caption{Variation of $ \chi_{mm} $ with temperature at different values of quark chemical potential for (a) $eB=0,\kappa_f =0  $ (b) $ eB=0.10~{\rm GeV}^2, \kappa_f =0 $, (c) $ eB=0.010~{\rm GeV}^2, \kappa_f \ne 0 $ and (d) $ eB=0.10~{\rm GeV}^2, \kappa_f \ne 0 $ }
  	\label{chiral_sus}
 \end{figure}

In Figs.~\ref{chiral_sus}(a) and (b) we have shown the variation of  chiral susceptibility($ \chi_{mm} $) as a function of $ T $ for $ eB=0, 0.10 \ {\rm GeV^2} $ at three different values of $ \mu_q $ without including the AMM of quarks. The behaviour is very similar in both cases. At $ \mu_q=\mu_\text{CEP} $,  $ \chi_{mm} $ diverges at $ T=T_\text{CEP} $ and we have a second order phase transition. This is known as the \textit{critical end point} (CEP).  For $ \mu_q < \mu_\text{CEP}  $, in the crossover region,  $ \chi_{mm} $ remains continuous for entire range of $  T $ and goes through a finite maximum at temperatures greater than $ T_\text{CEP} $. On the other hand, for $ \mu_q > \mu_\text{CEP}  $  we  have found a finite discontinuity at some temperature less than $ T_\text{CEP} $ and the transition is first order.  The only difference in Figs.~\ref{chiral_sus} (a) and (b) is the fact that for non zero background magnetic field $ T_\text{CEP} $ shifts towards the higher values of temperature. 
In Figs~\ref{chiral_sus}(c) and (d), we have included AMM of the quarks and studied the variation of $ \chi_{mm}  $ for $ eB=0.01 $ 
and $ 0.10 \ {\rm GeV^2} $ respectively. For the smaller value of magnetic field, a similar behaviour of $ \chi_{{mm}} $ 
(as Figs.~\ref{chiral_sus}(a) and (b)) for three different $ \mu_q $ around the CEP is observed; apart from the fact that $ T_\text{CEP} $, unlike the case of vanishing AMM, moves towards lower values of temperature. But, interestingly, $ \chi_{{mm}} $ in Fig~\ref{chiral_sus}(d) remains continuous for different values of $ \mu/T $ ratio and hence phase transition remains a crossover always which is also expected from Figs.~\ref{M_vs_T_diff_mu} and \ref{M_vs_mu}.
For the studied parameter set we have found that (a) for $ eB=0 $, 
CEP is located at $ \mu_\text{CEP} = 321~{\rm MeV}, T_\text{CEP} = 82~{\rm MeV} $, (b) for $ eB=0.10~{\rm GeV}^2 $ CEP is at $ \mu_\text{CEP} = 315~{\rm MeV}, 
T_\text{CEP} = 90~{\rm MeV}  $ and (c) for $ eB=0.010~{\rm GeV}^2 $ CEP is at $ \mu_\text{CEP} = 322~{\rm MeV}, T_\text{CEP} = 78~{\rm MeV}  $ .  

\begin{figure}[h]
	\begin{center}
		\includegraphics[angle=-90, scale=0.30]{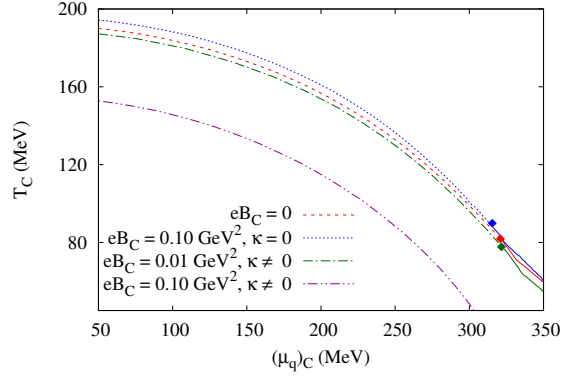}					
	\end{center}
	\caption{$T_C$-$ (\mu_q)_C$ phase diagram in NJL model for three different conditions. 
		The solid (dashed) lines denote the first-order (crossover) transition. 
		 The red, green and blue square points represent CEPs}
	\label{phase diagram}
\end{figure}

In Fig.~\ref{phase diagram} we have plotted the phase diagram associated with NJL model in the $T_C$-$ (\mu_q)_C$ plane 
for different external conditions. The essential features of the plots are similar except when we consider AMM of 
quarks at high background magnetic field. At higher temperature and low chemical potential 
the transition from chiral symmetry broken to restored phase is a crossover. On the other hand, at low temperature 
and high baryonic density the transition becomes first-order. Although, in presence of the magnetic field, the transition 
temperature for a fixed value of $ (\mu_q)_C $ increases and CEP moves towards the higher (lower) values of temperature 
(chemical potential of quarks) for vanishing AMM of quarks. The opposite effect is realized when AMM of quarks are taken 
into account. Interestingly, for high values of magnetic field in the later case, the transition temperature is found to go down significantly  and the transition remains a crossover for the entire range of $ T_C $ and $ (\mu_q)_C $ we have considered in the phase diagram. Note that the red, blue and green points in Fig.~\ref{phase diagram} represent the locations of the CEPs in three different cases.
\begin{figure}[h]
	\begin{center}
		\includegraphics[angle=-90, scale=0.30]{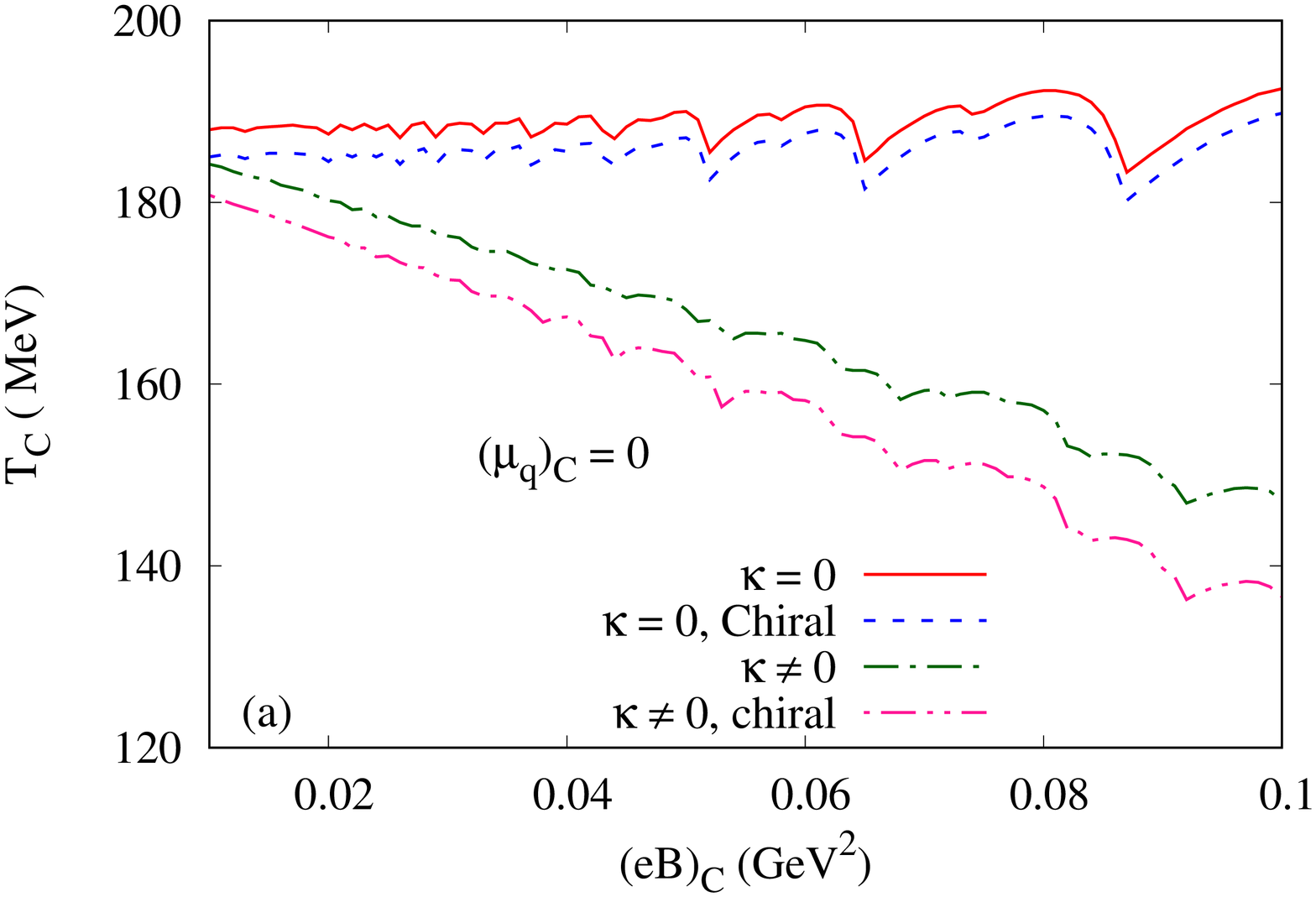}
		\includegraphics[angle=-90, scale=0.30]{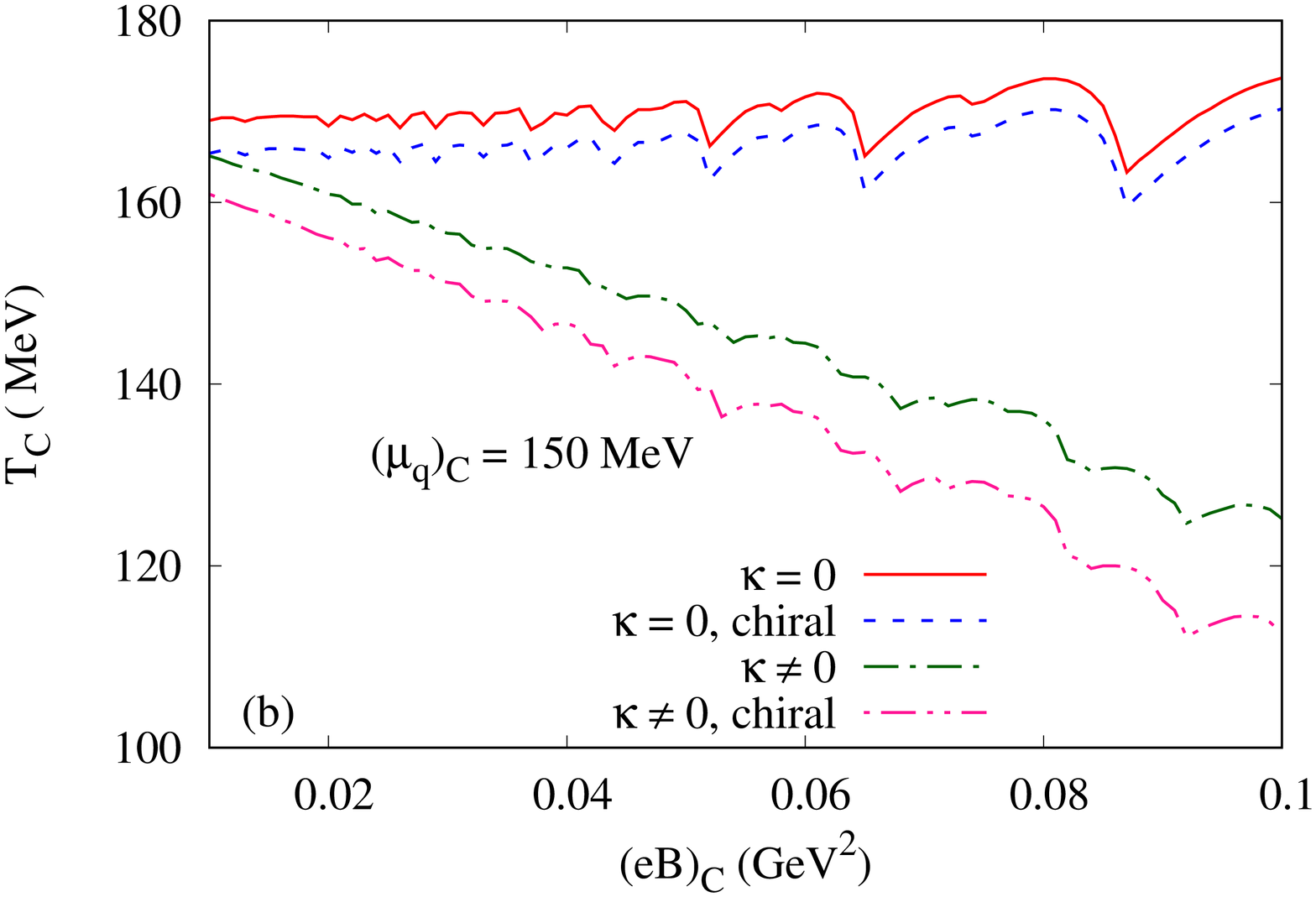}					
	\end{center}
	\caption{$T_C$-$ eB_C$ phase diagram in NJL model at (a) $ (\mu_q)_C=0 $ and (b) $ (\mu_q)_C=150 $ MeV along with 
		the corresponding chiral limits ($m=0$). }
	\label{Tc_vs_eB}
\end{figure}

In Figs.~\ref{Tc_vs_eB}(a)  and (b) we have plotted the variation of transition temperature ($T_c$ ) with the critical external magnetic 
field ($ eB_C $) at two different values of critical quark chemical potential ($ (\mu_q)_C =0$ and 150 MeV) with and 
without including AMM of quarks along with respective curves in the chiral limit ($m=0$). From Fig.~\ref{Tc_vs_eB}(a), it is 
evident that when the contributions of AMM of quarks are ignored, there is an overall increase in the transition temperature 
with increasing background magnetic field, again pointing towards the MC effect discussed earlier. On the contrary, when we 
include AMM, the transition temperature is reduced as $ (eB)_C $ increases, which leads to the IMC effect, 
which is obvious from Fig.~\ref{dep_cons_qmass}(d). In similar manner, from Fig.~\ref{Tc_vs_eB}(b) the MC and IMC for 
increasing  $ eB_C $ for zero and non-zero values of AMM can be noticed. Only difference is the fact that finite values of chemical potential results in a decrease of the magnitude of transition temperature. The overall behaviour of all 
these curves are same in their corresponding chiral limit except the fact that value of transition temperature decreases further.

\begin{figure}[h]
	\begin{center}
		\includegraphics[angle=-90, scale=0.30]{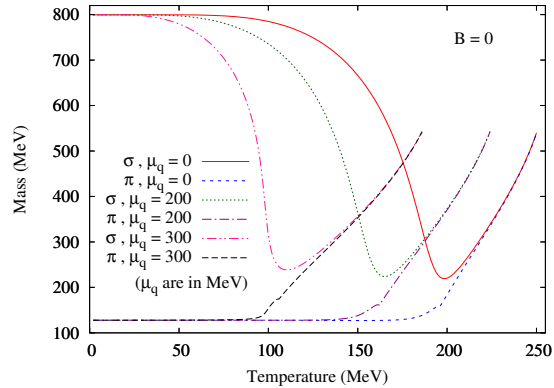}
	\end{center}
	\caption{Variation of masses of scalar ($\sigma$) and neutral pseudoscalar ($\pi^0$) mesons with temperature at 
		\textit{zero external magnetic field} and at three different values of quark chemical potential 
		($\mu_q=$ 0, 200 and 300 MeV).}
	\label{Fig.Meson.1}
\end{figure}

We now turn our attention to the mesonic properties in the NJL model. We will present results for the variation of masses 
of scalar meson $\sigma$ and neutral pseudoscalar meson $\pi^0$ with temperature, density and external magnetic field. 
Let us first consider the case of \textit{zero external magnetic field} for which the masses of $\sigma$ and $\pi^0$ are 
plotted as a function of temperature in Fig.~\ref{Fig.Meson.1} at three different values of quark chemical potential 
($\mu_q=$ 0, 200 and 300 MeV). In each of the three curves, one can notice that, the masses of $\sigma$ and $\pi^0$ 
start from their respective vacuum values at lower temperatures. With the increase in temperature, the $\sigma$ mass first 
decreases towards a minimum and then increases at the high temperature region. On the contrary, with the increase in temperature, 
the $\pi^0$ mass stays almost constant in the low temperature region and then starts increasing with temperature and 
merges with the mass curve of $\sigma$ meson owing to the restoration of chiral symmetry. The effect of increase in density is to 
shift the position of the minima of the $\sigma$ mass curve towards low temperature as well as to decrease the transition 
temperature ($T_c$) for the chiral symmetry restoration. Though, it is barely possible to precisely locate the $T_c$ from 
these curves, yet qualitatively the behaviour of $T_c$ with the change in density is consistent with Fig.~\ref{phase diagram}. 
\begin{figure}[h]
	\begin{center}
		\includegraphics[angle=-90, scale=0.30]{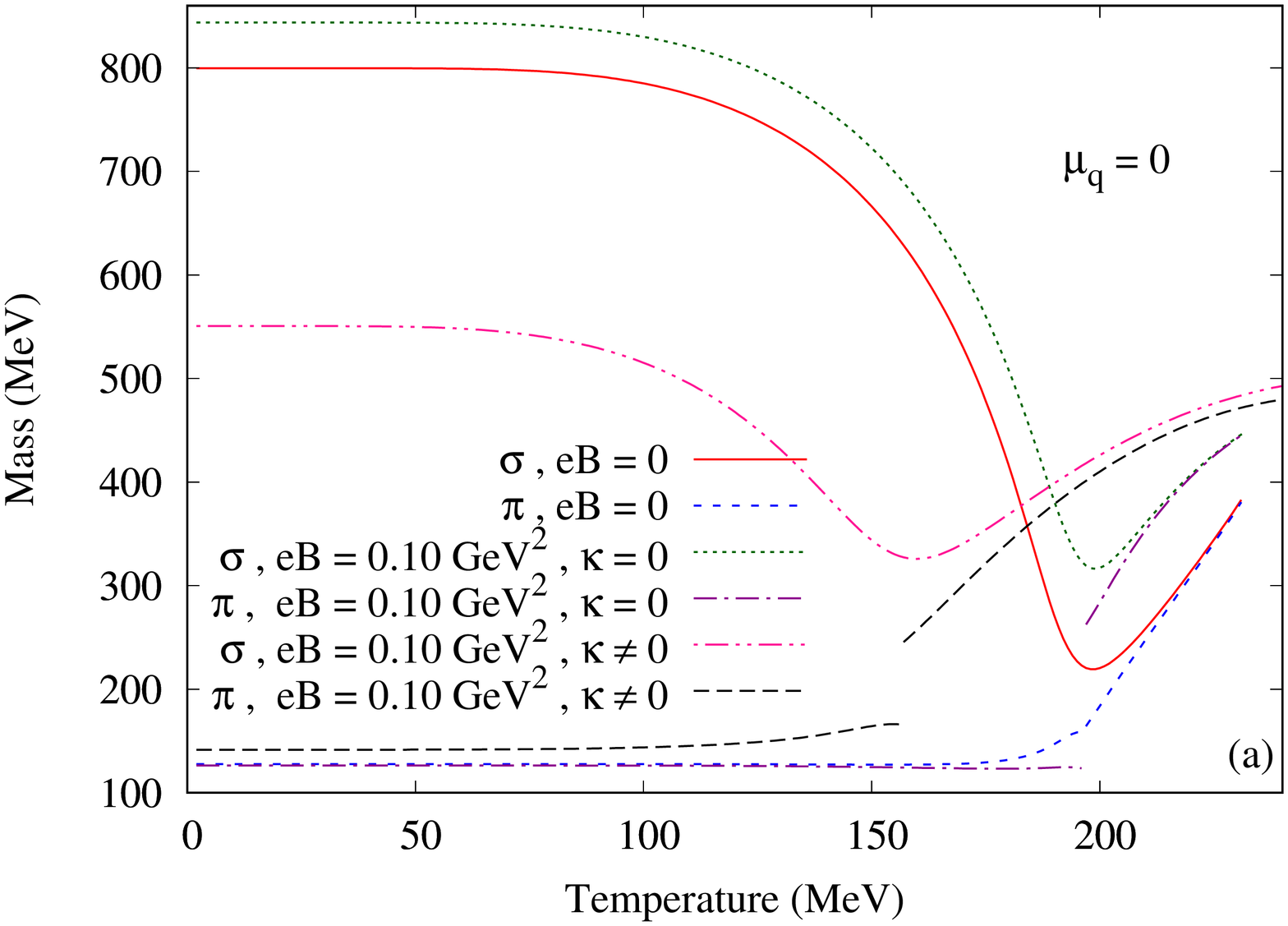} \includegraphics[angle=-90, scale=0.30]{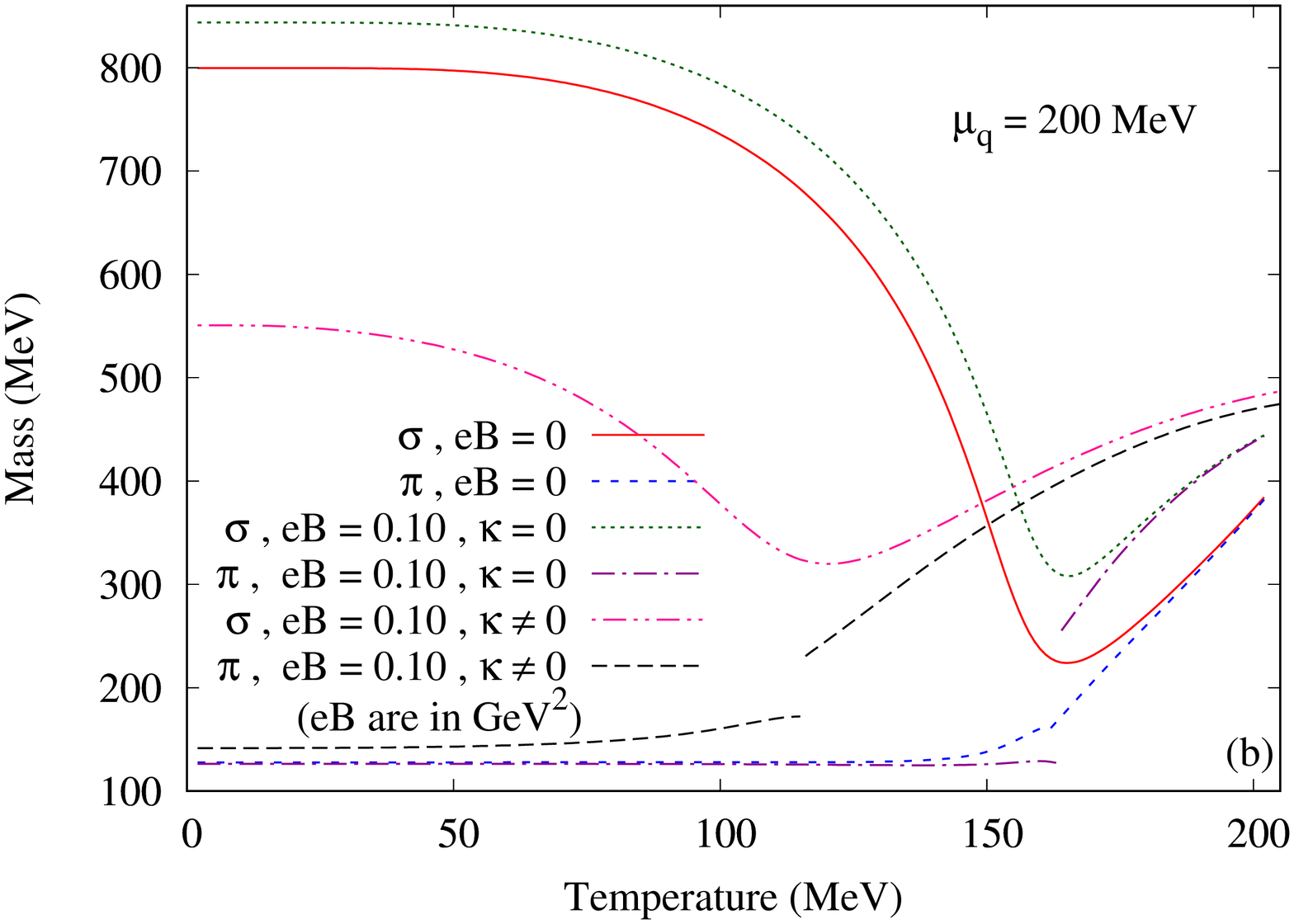}
	\end{center}
	\caption{Variation of masses of scalar ($\sigma$) and neutral pseudoscalar ($\pi^0$) mesons with temperature at two different 
		values of external magnetic field ($eB=$ 0 and 0.10 GeV$^2$) including and excluding the AMM of the quarks 
		for (a) $\mu_q=0$ and (b) $\mu_q=200$ MeV.}
	\label{Fig.Meson.2}
\end{figure}
\begin{figure}[h]
	\begin{center}
		\includegraphics[angle=-90, scale=0.30]{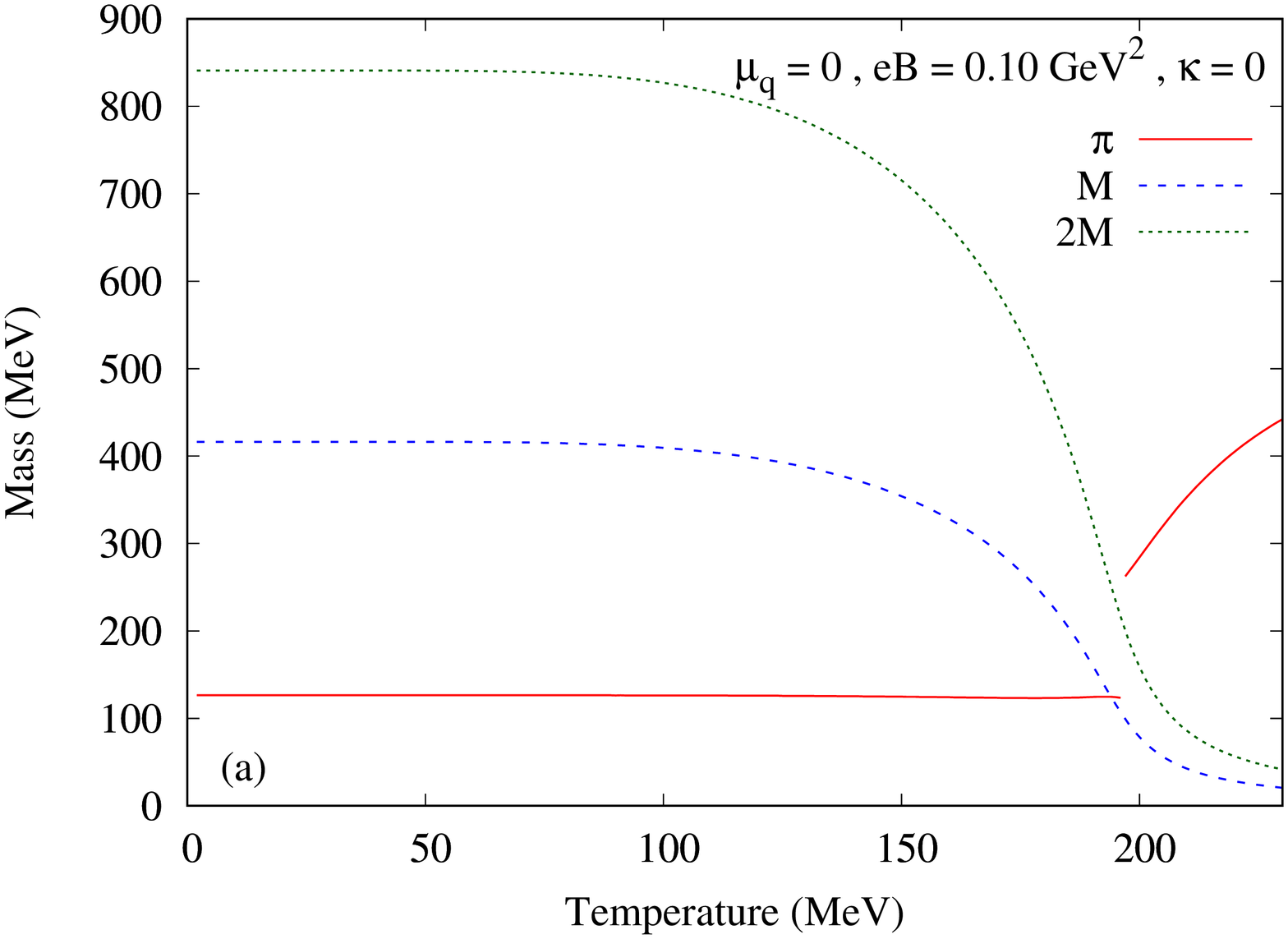} \includegraphics[angle=-90, scale=0.30]{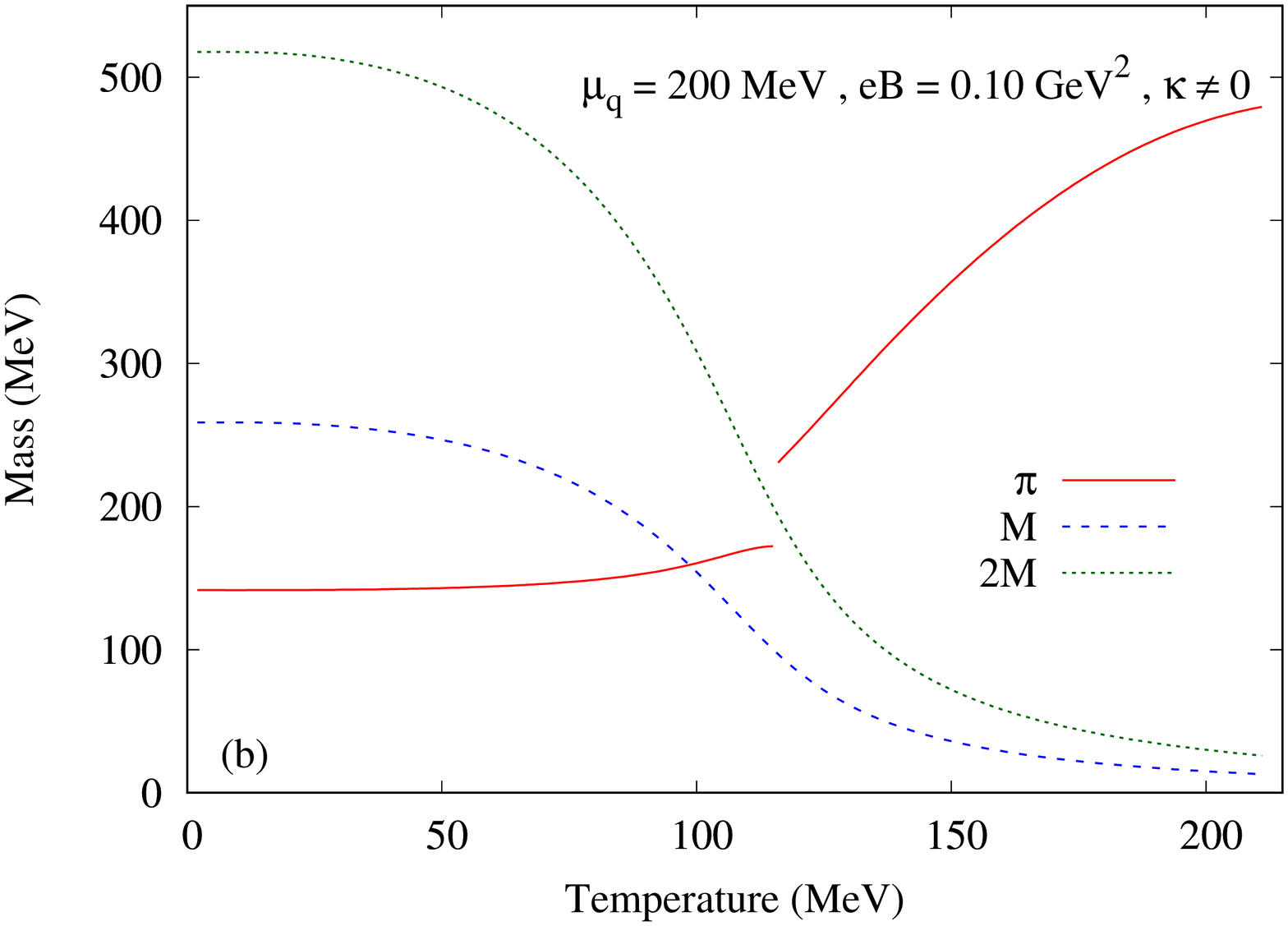} \\
		\includegraphics[angle=-90, scale=0.30]{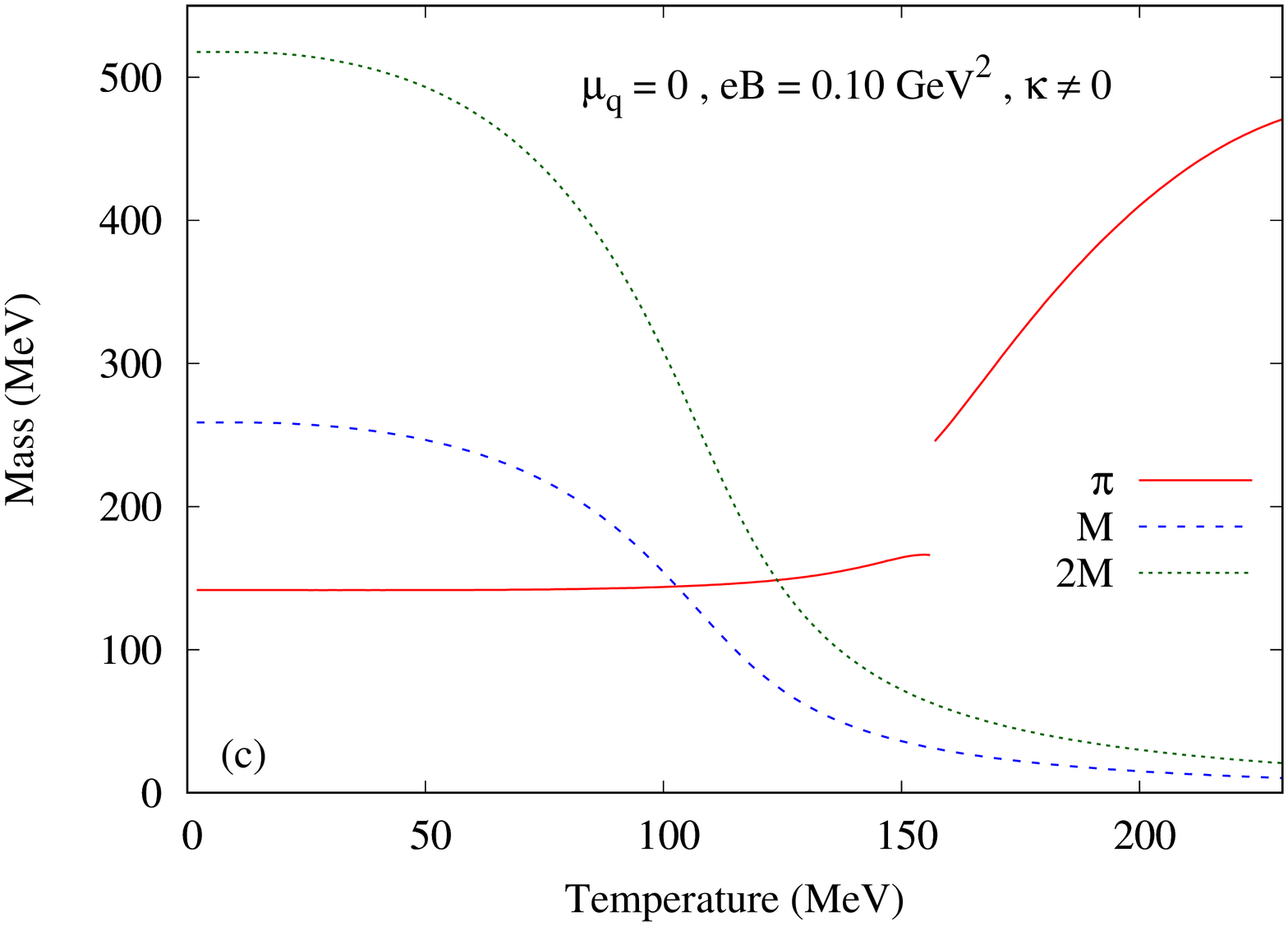} \includegraphics[angle=-90, scale=0.30]{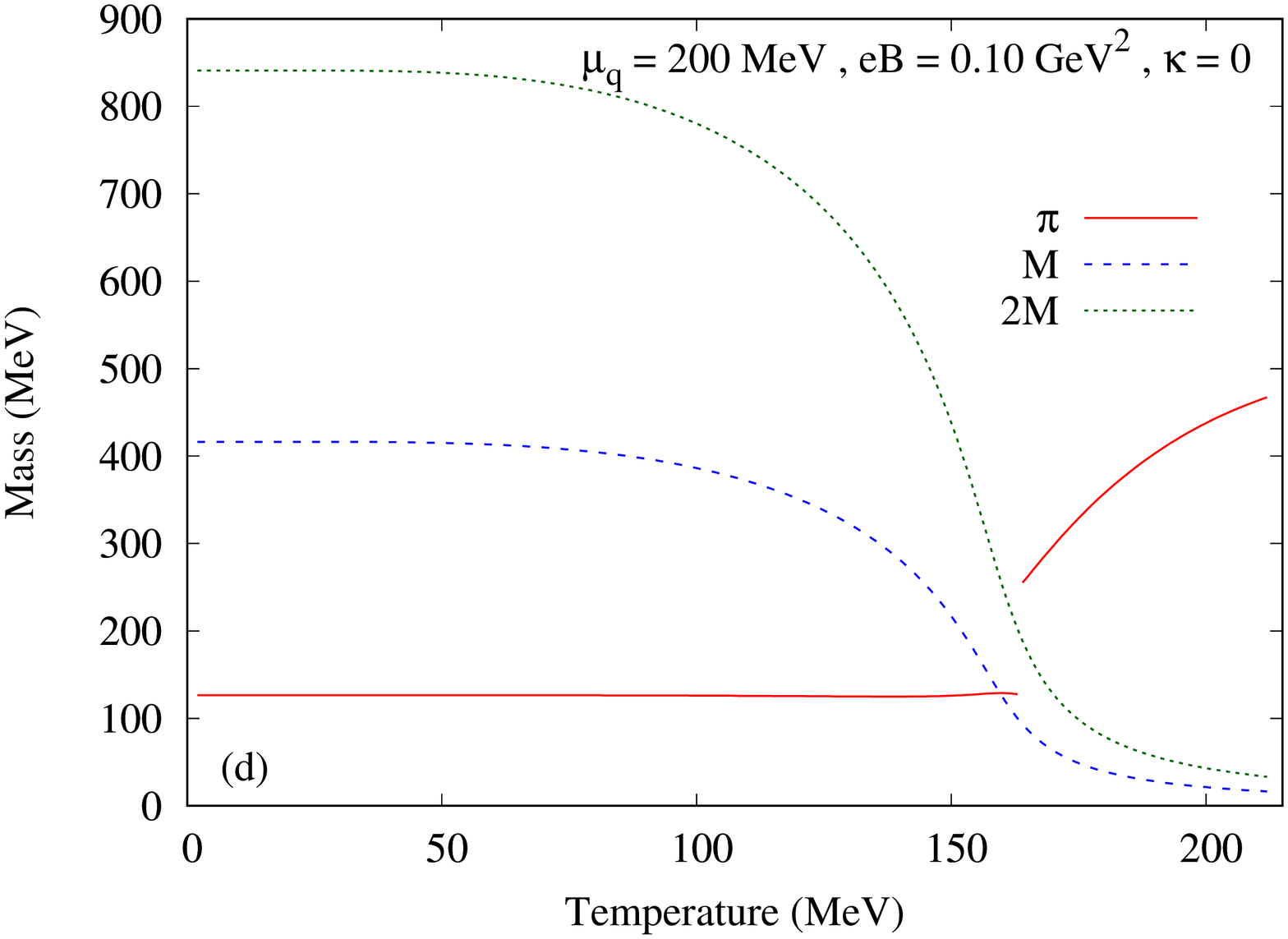}
	\end{center}
	\caption{Variation of $\pi^0$ mass, constituent quark mass and twice of the constituent quark mass with temperature 
		at $eB=0.10$ GeV$^2$ for (a) $\mu_q=0$ and $\kappa=0$, 
		(b) $\mu_q=200$ MeV and $\kappa=0$, (c) $\mu_q=0$ and $\kappa\ne0$ and (d) $\mu_q=200$ MeV and $\kappa\ne0$.}
	\label{Fig.Meson.3}
\end{figure}
\begin{figure}[h]
	\begin{center}
		\includegraphics[angle=-90, scale=0.320]{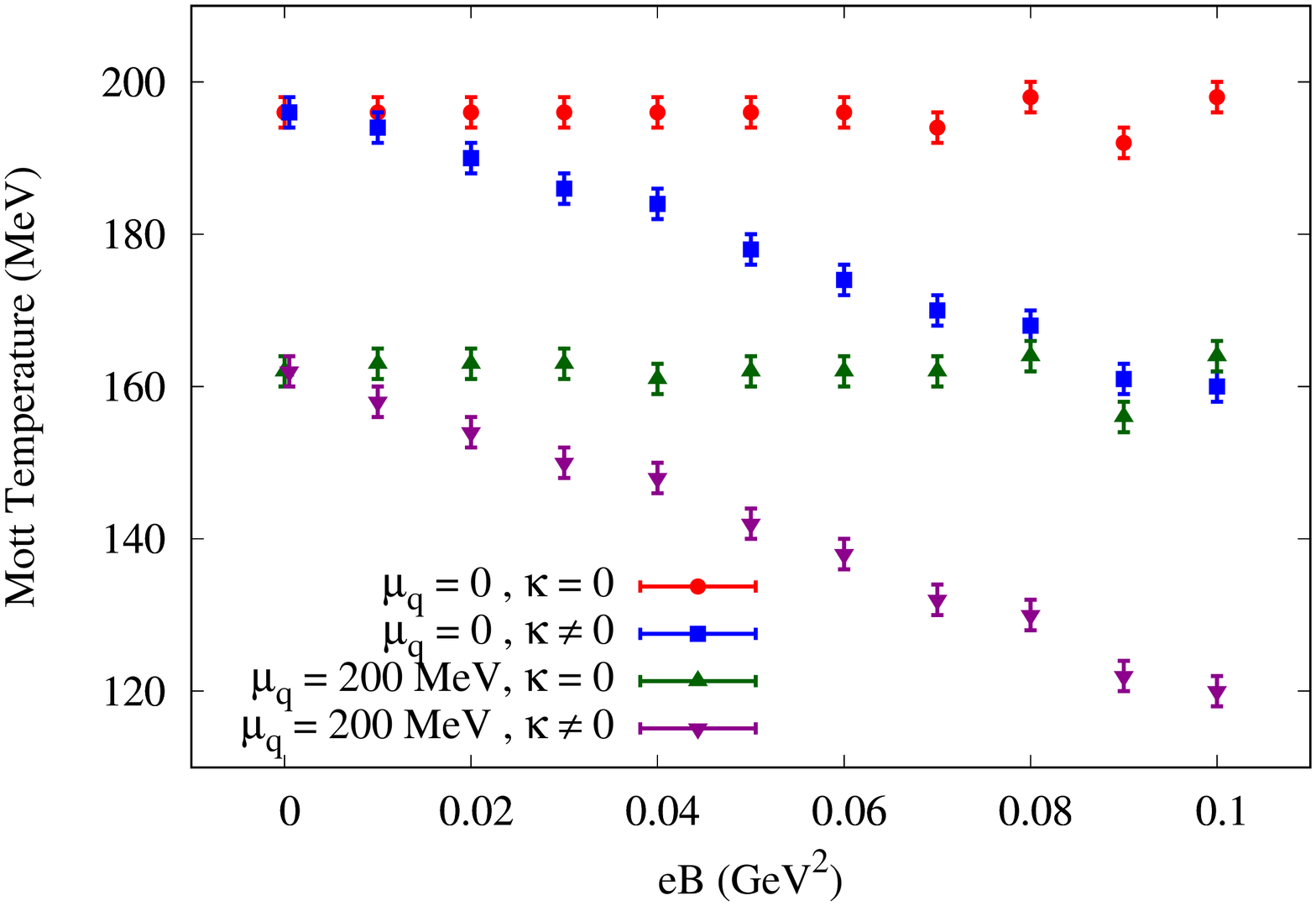}
	\end{center}
	\caption{Variation of Mott temperature with external magnetic field for two different values of quark chemical 
		potential ($\mu_q=0$ and 200 MeV respectively). Results including and excluding the AMM of quarks have been compared.}
	\label{Fig.Meson.4}
\end{figure}

Now, we switch on the external magnetic field. In Fig.~\ref{Fig.Meson.2}, the behaviour of masses of $\sigma$ and $\pi^0$ is 
shown as a function of temperature at two different values of external magnetic field ($eB=$ 0 and 0.10 GeV$^2$ respectively) 
including and excluding the AMM of the quarks. Figs.~\ref{Fig.Meson.2}(a) is obtained using quark 
chemical potential $\mu_q=0$ whereas Figs.~\ref{Fig.Meson.2}(b) corresponds $\mu_q=200$ MeV. 
As can be noticed from Figs.~\ref{Fig.Meson.2}(a) and (b), the mass of $\sigma$ increases with the increase in magnetic 
field when the AMM of quarks are not taken into account and the position of minima shows no significant shift over the 
temperature axis. On the contrary, while considering the AMM of quarks, 
the $\sigma$ mass decreases significantly with the increase in magnetic field in the low temperature region followed 
by a shift of the minima toward low temperature and the smearing of the same. 
The mass of $\pi^0$ is seen to decrease (increase) with the increase in magnetic field in the low (high) temperature region 
when the AMM of the quarks are turned off. However, turning on the AMM of the quarks, we find that the $\pi^0$ mass increases 
with the increase in magnetic field in all the temperature range. A sudden jump (discontinuity) of the $\pi^0$ mass at a 
particular temperature is also noticed for the non-zero external magnetic field values 
(in both the cases of $\kappa=0$ and $\kappa\ne0$).

Next, we look at the pion mass ($m_\pi$) in more detail and compare it with the constituent quark mass ($M$) in Fig.~\ref{Fig.Meson.3} 
where we have plotted $m_\pi$, $M$ and $2M$ as a function of temperature. 
In Figs.~\ref{Fig.Meson.3}(a) and (b), the AMM of the quarks are switched off (i.e. $\kappa=0$) and the quark 
chemical potential $\mu_q=0$ and 200 MeV are used; whereas Figs.~\ref{Fig.Meson.3}(c) and (d) 
correspond to $\kappa\ne0$ with $\mu_q=0$ and 200 MeV. 
Defining the Mott temperature ($T_\text{Mott}$) as 
the temperature beyond which $m_\pi\ge 2M$ i.e.
\begin{eqnarray}
m_\pi(T)\ge 2M(T) ~~ \text{if} ~~ T \ge T_\text{Mott}~,
\label{eq.mott}
\end{eqnarray}
one can notice from Fig.~\ref{Fig.Meson.3} that, $T_\text{Mott}$ decreases with the increase in quark chemical potential. 
Also, switching on the AMM of quarks, decreases the value of $T_\text{Mott}$ substantially with respect to the $\kappa=0$ case.

Few comments on the Mott dissociation in the context of NJL model are in order here. 
As discussed in Ref.~\cite{Hufner:1996pq}, the Mott transition in the context of solid state physics~\cite{MOTT:1968zz} has a close analogy with the confinement-deconfinement phase transition of QCD (hadrons to plasma of quarks and gluons). The Mott transition corresponds 
to the phase transition from a phase of bound states and constituents to a phase of constituents only. Whereas both Mott and 
confinement-deconfinement phase transitions involve the delocalization of the bound states into their constituents, the 
Mott transition is applicable for nonconfining systems as well; for example in the NJL model where there 
is no confinement.  
Here, with the increase in temperature and/or density, 
the dynamically generated constituent quark mass decreases owing to the partial restoration of chiral symmetry. 
One of the immediate impacts on the meson spectrum is the decrease of the continuum thresholds 
for $q\overline{q}$ scattering. This in turn lowers the binding energy for the pseudoscalar meson bound states (e.g. pions). 
The pions get dissociated while entering the continuum and thus become resonances having a finite lifetime. This phenomenon is also termed as 
the \textit{`Mott Effect'} or the \textit{`Mott Transition'} which 
occurs when the polarization function of the pion possesses an imaginary part or in other words the mass pole become complex. 
More details on this topic can be found in  Refs.~\cite{Hufner:1996pq,Dubinin:2016wvt,MaoWang,Mao_pi}.

This sudden jump of the pion mass at/or above the Mott temperature under external magnetic field is also observed 
and well studied in Refs.~\cite{MaoWang,Avancini}. The main reason being the dimensional reduction from (3+1)D to 
(1+1)D of the quark degrees of freedom due to which the number of accessible states for the $q\overline{q}$ resonant pair 
reduces. This in turn fails to guarantee some solution for the resonant $\pi^0$ state just above the $m_\pi \gtrsim 2M$. 
However, due to the Random Phase Approximation (RPA) leading to Eq.~\eqref{tran_meson_mass}, the external momenta in the RPA 
bubbles are both conserved and onshell. This makes the pion mass to have a sudden jump at/or above the Mott temperature to 
the lowest possible energy state accessible.

Finally, we present the variation of Mott temperature as a function of external magnetic field at two different 
values of quark chemical potential ($\mu_q =0$ and 200 MeV respectively) in Fig.~\ref{Fig.Meson.4}. The results including 
and excluding the AMM of quarks are also shown. As can be seen from the graph that, the Mott temperature decreases substantially 
with the increase in $\mu_q$ for both the cases of $\kappa=0$ and $\kappa\ne0$. For a particular value of $\mu_q$, 
$T_\text{Mott}$ is almost constant with the variation of the external magnetic field for $\kappa=0$. However, for $\kappa\ne0$, 
$T_\text{Mott}$ decreases with the increase in external magnetic field owing to the IMC.


\section{Summary \& Conclusion} \label{sec.summary}
In summary, using a field dependent three-momentum cut-off, we study the $2$-flavour 
NJL model at finite temperature and baryonic density 
in presence of arbitrary external magnetic field.
The constituent quark mass ($M$) is obtained by solving the \textit{gap equation} as a function of $T$, $ \mu_q $ and $ eB $ 
considering the AMM of quarks.  We have evaluated $ M $ as a function of $ T $ at vanishing baryonic density for different 
magnetic fields and it is found that the transition temperature from symmetry broken to restored phase increases with external 
magnetic field showing the enhancement of the quark anti-quark condensate, which can be identified as \textit{magnetic catalysis}. 
On the other hand, the opposite behaviour is observed when AMM of quarks is taken into consideration, indicating 
\textit{inverse magnetic catalysis}. Temperature variation of $ M $ is also studied for finite baryonic density by 
choosing two different representative values of $ \mu_q $ and it is found out that, even at $ \mu_q = 300 $ MeV, $M$ 
remains single valued through out the temperature range indicating a crossover transition but for $ \mu_q\simeq 330  $ MeV, 
it is observed that for certain range of $ T $, the \textit{gap equation} provides multiple solutions for $ M $, which indicates a first order transition. The same behaviour is also obtained while studying the $ \mu_q $  dependence of $ M $, 
the only difference is the fact that, multiple solutions for $ M $ are observed for lower values of temperature. The $ eB $ dependence of $ M $ is obtained at different $ T $ with and without considering AMM of quarks. 
In case of vanishing AMM, $  M $ becomes highly oscillatory function of $ eB $ but an overall increase in $ M $ with $ eB $ can be inferred. But inclusion of AMM results in a steady decrease in $ M $ with $ eB $. Critical behaviour of chiral susceptibility ($ \chi_{mm} $) has been examined in the 
vicinity of the phase transition. Finally the phase diagram of hot and dense magnetized quark matter, described by NJL 
model is obtained and for finite values of $ eB $, the CEP is found to shift towards higher temperature. 
On the contrary, when we include AMM of quarks CEP follows an opposite trend.  Interestingly at high $ eB $ for finite 
values of AMM, the transition remains crossover for larger range of $ T_C $ and $ (\mu_q)_C $. Thus, it can be inferred 
that in the presence of an external magnetic field, the AMM of quarks plays a crucial role in characterizing the properties of quark matter and its modification at finite temperature and density.

The masses of the scalar meson $\sigma$ and the neutral pseudo-scalar meson $\pi^0$ have been calculated at finite temperature, density 
and arbitrary external magnetic field using the RPA in the NJL model. For this, both the AMM of the quarks as well as infinite number of 
Landau levels of the quarks are considered in the calculation so that the results are valid for arbitrary strength of the external 
magnetic field. In this context, the Mott temperature corresponding to the transition from a bound to a resonant pionic state, has 
been calculated and its variation with external magnetic field and quark chemical potential is studied.

We find that, both the masses of $\sigma$ and $\pi^0$ reduces to their respective vacuum masses at $T\simeq0$ and $B\simeq0$. With the 
increase in temperature, $m_\sigma$ decreases while $m_{\pi}$ increases followed by a merging of their masses at high temperature owing 
to the partial restoration of the chiral symmetry. The external magnetic field affects the masses of these mesons in a non-trivial way; 
$m_{\pi}$ decreases (increases) with the increase in magnetic field in the low (high) temperature region 
when the AMM of the quarks are turned off. However, turning on the AMM of the quarks, $m_{\pi}$ increases 
with the increase in magnetic field in the temperature range considered. The $\pi^0$ mass suffers a sudden jump (discontinuity) 
at and above the Mott transition temperature for non-zero values external magnetic field. Finally, the $T_\text{Mott}$ is found to 
decrease significantly with the increases in density and external magnetic field when the AMM of the quarks are taken into consideration.


\appendix
\section{Derivatives of Distribution Functions}\label{nderivatives}
From Eq.~\eqref{energy} we get 
\begin{equation}
\dfrac{\partial E_{nfs}}{\partial M} = \dfrac{M}{E_{nfs}} \FB{ 1- \dfrac{s\kappa_fe_f B}{M_{nf}}}
\label{energy_mass_dreri}
\end{equation}
where $ M_{nf} = \sqrt{\MB{e_f B} (2n+1-s\xi_f) +M^2} $. 
Again differentiating Eq.~\eqref{distn_fn} w.r.t. $ M $ and using Eq.~\eqref{energy_mass_dreri} we can write 
\begin{equation}
\dfrac{\partial n^\pm}{\partial M} = - \beta \FB{n^\pm}^2 \exp\TB{\beta \FB{E_{nfs} \mp \mu}}\dfrac{\partial E_{nfs}}{\partial M}= -\beta n^\pm  \FB{1-n^\pm}\dfrac{M}{E_{nfs}} \FB{ 1- \dfrac{s\kappa_fe_f B}{M_{nf}}}.
\end{equation}
In a similar manner, the following expressions can be derived:
\begin{eqnarray}
\dfrac{\partial n^\pm }{\partial \mu } &=& \pm \beta n^\pm \FB{1-n^\pm}  \TB{1\mp \dfrac{M}{E_f} \FB{1- \dfrac{s\kappa_f e_f B}{M_{nf} } }\dfrac{\partial M}{\partial \mu } }\label{mu_derivative_n} \\
\dfrac{\partial n^\pm }{\partial T } &=&  \beta n^\pm \FB{1-n^\pm}  \TB{\dfrac{E_{nfs}\mp \mu  }{T}- \dfrac{M}{E_f} \FB{1- \dfrac{s\kappa_f e_f B}{M_{nf} } }\dfrac{\partial M}{\partial T } }\label{T_derivative_n}~.
\end{eqnarray}


\section{UV and AMM Blocking}\label{AMM_blocking}
In Sec.~\ref{deri_gap_eqn}, we have introduced two blocking factors while discussing the regularization scheme. 
To ensure the reality of UV cutoff $\Lambda_z $, we arrived at UV blocking condition expressed as
\begin{eqnarray}
\Lambda^2- \MB{e_f B} (2n+1-s\xi_f) -\FB{\kappa_fe_fB}^2 + 2sM_{nf} \kappa_fe_fB &\ge& 0 \label{cond_uv_block}~.
\end{eqnarray}
Let us first study the case when $ \kappa_f =0 $ for which the above condition becomes 
\begin{eqnarray}
\Lambda^2- \MB{e_f B} (2n+1-s\xi_f)  &\ge& 0 \nn \\
\implies n\le \dfrac{\Lambda^2}{2\MB{e_f B} } - \half (1-s\xi_f) \nn
\end{eqnarray}
which will give us a maximum value of the Landau level index, 
$ n_{\rm max} = \TB{ \dfrac{\Lambda^2}{2\MB{e_f B} } - \half (1-s\xi_f)} $ up to which the summation has to be performed. 
Here $ \TB{x} $ is the floor function which give greatest integer less than or equal to $ x $. 
Now for non-zero values of AMM, the UV blocking condition becomes non-linear in $n$ and thus 
it becomes difficult to find out the corresponding inequality for $ n $ analytically. 
So we have implemented the condition given in Eq.~\eqref{cond_uv_block} in our numerical calculations.

Now the AMM blocking condition is appearing to ensure the positivity of the Landau quantized transverse momentum of the 
quarks having an AMM. The AMM blocking condition is
\begin{eqnarray}
\MB{e_f B} (2n+1-s\xi_f) + \FB{\kappa_fe_fB}^2 - 2sM_{nf} \kappa_fe_fB\ge 0
\label{ammblockingcondition}
\end{eqnarray} 
which will restrict the lower values of the Landau level index $n$.

 
\section{Gap Equation in the Limit $B\rightarrow0$ }\label{Btendstozero}
In this appendix, we will show that, in the limit of vanishing external magnetic field, the analytic expression of the 
gap equation without external magnetic field is exactly reproduced. We will show the calculation for the temperature 
independent part containing the magnetic field dependent UV regulator for which a straight forward zero field limit is 
not obvious.

Let us consider the gap equation at $T=0$ and $B\ne0$ from Eq.~\eqref{mass_constituen_reg}:
\begin{eqnarray}
M = m + 4GN_c \sum_f \MB{e_fB }\sum_{n=0}^{\infty } \sum_{\{s\}} \int_{0}^{\Lambda_z} \dfrac{dp_z}{4\pi^2} 
\Theta( \Lambda^2 - \vec{p }_\perp^2) \ \Theta(\vec{p }_\perp^2)\dfrac{M}{E_{nfs}} \FB{1- \dfrac{s\kappa_fe_f B}{M_{nf}} }.
\end{eqnarray}
To take the limit $B\rightarrow0$ of the above equation, we first put all the terms containing the $\kappa_{f}e_fB$ equals to zero. 
Thus the above equation becomes
\begin{eqnarray}
M &=& m + 4GN_c \lim\limits_{B\rightarrow0} \sum_f \MB{e_fB }\sum_{n=0}^{\infty } (2-\delta_{n0}) \int_{0}^{\sqrt{\Lambda^2 -2n\MB{e_f B} }} \dfrac{dp_z}{4\pi^2} \Theta( \Lambda^2 -2n\MB{e_f B} ) \ \dfrac{M}{\sqrt{p_z^2 +2n\MB{e_f B}  +M^2}  }
\end{eqnarray}
where the sum over the spin index $s$ has been carried out. We can now analytically perform the $dp_z$ integration and get after 
some simplification
\begin{eqnarray}
M = m + \dfrac{MGN_c }{\pi^2} \lim\limits_{B\rightarrow0} \sum_f \MB{e_fB }\sum_{n=0}^{\infty } (2-\delta_{n0}) \Theta( \Lambda^2-2n\MB{e_f B})\tanh^{-1} \sqrt{\dfrac{\Lambda^2-2n\MB{e_f B}}{\Lambda^2 + M^2}}~.
\end{eqnarray}
Separating out the contribution of the LLL from the above equation, we get after a substitution of $ \tau_f = 2n\MB{e_f B} $
\begin{eqnarray}
 M = m + \dfrac{MGN_c }{\pi^2} \lim_{B\rightarrow0} \sum_f \MB{e_fB }\TB{ \tanh^{-1} 
 	\sqrt{\dfrac{\Lambda^2}{\Lambda^2 + M^2}} + 2 {\sum_{\tau_f=2|e_fB|}^{\infty}\!\!\!\!\!\!\!\!{}^{'}} 
 	\Theta( \Lambda^2-\tau_f)\tanh^{-1} \sqrt{\dfrac{\Lambda^2-\tau_f}{\Lambda^2 + M^2}}}\nn
\end{eqnarray}
where the primed summation $ \sum{}^{\prime } $ denotes an increment of $ 2e_f B $ of its index rather than $1$. 
Now as $ e_fB\rightarrow0 $, we can change the summation to an integration by doing the following substitution considering the 
continuum limit 
\begin{equation}
\sum_{\tau_f}{}^{'}\rightarrow  \dfrac{1}{2e_f B} \int_{2e_f B}^\infty d\tau_f.
\end{equation}
This leads to
\begin{eqnarray}
M &=& m + \dfrac{MGN_c }{\pi^2}\sum_f   \int_{0}^\infty  d\tau_f \Theta( \Lambda^2-\tau_f)\tanh^{-1}
\sqrt{\dfrac{\Lambda^2-\tau_f}{\Lambda^2 + M^2}}~.
\end{eqnarray}
Note that, the presence of the step function will restrict the upper limit of the $\tau_f$ integration. Performing the remaining 
$d\tau_f$ integral, we are left with
\begin{equation}\label{gap_zero}
M=m + \dfrac{GMN_fN_c }{\pi^2}\TB{  \Lambda \sqrt{\Lambda^2 + M^2 }- M^2\sinh^{-1} \FB{{\dfrac{\Lambda}{ M}}}}
\end{equation}
which is same as the vacuum term given in Ref.~\cite{Klevansky}. 


\section{Calculation of the Polarization Functions for $B\ne0$} \label{app.pola}
In this appendix, we will calculate the polarization functions in presence of external magnetic field and finite temperature. 
We first substitute Eq.~\eqref{eq.prop.SB} and \eqref{eq.prop.11B} into Eq.~\eqref{eq.repi.2} to write,
\begin{eqnarray}
\RE\overline{\overline{\Pi}}_a (q) &=& \RE \sum_f \sum_{s_k\in\SB{\pm1}}  \sum_{s_p\in\SB{\pm1}} 
\sum_{l=0}^\infty \sum_{n=0}^\infty i \kfourint{k} \mathcal{N}^a_{lns_ks_p}(k,q) \nn \\ 
&& \TB{ \dfrac{-1}{\vparasq{k} - \FB{M_l - s_k\kappa eB }^2 +i\epsilon } 
- 2\pi i \eta (k\cdot u)\delta \SB{\vparasq{k} -\FB{M_l - s_k\kappa eB }^2   }}  \nn \\ 
&& \TB{ \dfrac{-1}{\vparasq{p} - \FB{M_n - s_p\kappa eB }^2 +i\epsilon } - 2\pi i \eta (p\cdot u)\delta \SB{\vparasq{p} 
-\FB{M_n - s_p\kappa eB }^2   } } 
\label{eq.pola.1}
\end{eqnarray}
where the flavour indices have been suppressed inside the flavour sum for breviety and $\mathcal{N}^a_{lns_ks_p}(k,q)$ for the 
scalar ($\sigma$) and neutral pseudoscalar ($\pi^0$) channels are given by
\begin{eqnarray}
\mathcal{N}^{\sigma}_{lns_ks_p}(k,q) &=& N_c \Tr{ \D_{nfs_p} (q+k) \D_{lfs_k}(k) }
\FB{1-\delta^0_l \delta^{-1}_{s_k }}\FB{1-\delta^0_n \delta^{-1}_{s_p }} \\
\mathcal{N}^{\pi^0}_{lns_ks_p}(k,q) &=& -N_c \Tr{ \gamma^5 \D_{nfs_p} (q+k)\gamma^5 \D_{lfs_k}(k) }
\FB{1-\delta^0_l \delta^{-1}_{s_k }}\FB{1-\delta^0_n \delta^{-1}_{s_p }}.
\end{eqnarray}
Evaluating the traces over the Dirac matrices, the above equations become,
\begin{eqnarray}
\mathcal{N}^{a}_{lns_ks_p}(k,q) &=& j2N_ce^{-\alpha_k-\alpha_p}\frac{(-1)^{l+n}}{M_lM_n} 
\FB{1-\delta^0_l \delta^{-1}_{s_k }}\FB{1-\delta^0_n \delta^{-1}_{s_p }} \nn \\ &&
\Bigg[ 8L^1_{l-1}(2\alpha_k)L^1_{n-1}(2\alpha_p)(k_\perp\cdot p_\perp) \Big\{ s_ks_p(k_\parallel\cdot p_\parallel)
+js_ks_p(\kappa eB)^2 +jM_lM_n  -j \kappa eB(s_pM_l+s_kM_n) \Big\} \nn \\ &&
+ L_{l-1}(2\alpha_k)L_{n-1}(2\alpha_p)\Big\{j(k_\parallel\cdot p_\parallel)
(M_l-s_kM)(M_n-s_pM) \nn \\ && + \big\{s_k\kappa eBM-M_l(M-s_kM_l+\kappa eB)\big\}\big\{s_p\kappa eBM-M_n(M-s_pM_n+\kappa eB)\big\}\Big\}
\nn \\ && + L_{l}(2\alpha_k)L_{n}(2\alpha_p)\Big\{j(k_\parallel\cdot p_\parallel)
(M_l+s_kM)(M_n+s_pM) \nn \\ && + \big\{s_k\kappa eBM-M_l(M+s_kM_l-\kappa eB)\big\}\big\{s_p\kappa eBM-M_n(M+s_pM_n-\kappa eB)\big\}\Big\}
\Bigg] 
\end{eqnarray}
where $j=1$ for $a\equiv\sigma$ and $j=-1$ for $a\equiv\pi^0$.
Performing the $dk^0$ integral of Eq.~\eqref{eq.pola.1}, we get
\begin{eqnarray}
\RE\overline{\overline{\Pi}}_a (q) &=& \sum_f \sum_{s_k,s_p} \sum_{l,n} \int \dfrac{d^3 k}{(2\pi)^3}  
\mathcal{P} \TB{\dfrac{\mathcal{N}^a_{lns_ks_p}(k^0=-q^0+\omega^{ns_p}_{p})}{2 \omega^{ns_p}_{p} 
\SB{\FB{q^0 - \omega^{ns_p}_{p}}^2 - \FB{\omega^{ls_k}_{k}}^2 }} 
+\dfrac{\mathcal{N}^a_{lns_ks_p}(k^0=\omega^{ls_k}_{k})}{2 \omega^{ls_k}_{k} \SB{\FB{q^0 + \omega^{ls_k}_{k}}^2 - \FB{\omega^{ns_p}_{p}}^2 }} 
\right. \nn \\ && \hspace{1cm} \left. - \dfrac{\mathcal{N}^a_{lns_ks_p}(k^0=-\omega^{ls_k}_{k}) f^-(\omega^{ls_k}_{k}) }{2 \omega^{ls_k}_{k} 
\SB{\FB{q^0 - \omega^{ls_k}_{k}}^2 - \FB{\omega^{ns_p}_{p}}^2  }  } - \dfrac{\mathcal{N}^a_{lns_ks_p}(k^0=\omega^{ls_k}_{k}) f^+(\omega^{ls_k}_{k})}
{2 \omega^{ls_k}_{k} \SB{\FB{q^0 + \omega^{ls_k}_{k}}^2 - \FB{\omega^{ns_p}_{p}}^2  }  }   \right. \nn  \\ &&  \hspace{1cm} \left.
 -\dfrac{\mathcal{N}^a_{lns_ks_p}(k^0=-q^0-\omega^{ns_p}_{p}) f^-(\omega^{ns_p}_{p})   }{2 \omega^{ns_p}_{p} 
\SB{\FB{q^0 + \omega^{ns_p}_{p}}^2 - \FB{\omega^{ls_k}_{k}}^2  }  } - \dfrac{\mathcal{N}^a_{lns_ks_p}(k^0=-q^0 +\omega^{ns_p}_{p}) f^+(\omega^{ns_p}_{p})}
{2 \omega^{ns_p}_{p} \SB{\FB{q^0 - \omega^{ns_p}_{p}}^2 - \FB{\omega^{ls_k}_{k}}^2  }  } 
}
\label{eq.pola.2}
\end{eqnarray}
where $\omega^{ls_k}_k = \sqrt{k_z^2+(M_l-s_k\kappa eB)^2}$ in which the flavour index has been suppressed. 
Now considering $\vec{q}=\vec{0}$, the $d^2k_\perp$ integral in the above equation has been performed using the 
orthogonality of the Laguerre polynomials as
\begin{eqnarray}
	\RE\overline{\overline{\Pi}}_a (q) &=& \sum_f \sum_{s_k,s_p} \sum_{l,n} \int_{-\infty}^{+\infty} \dfrac{dk_z}{(2\pi)}  
	\mathcal{P} \TB{\dfrac{\tilde{\mathcal{N}}^a_{lns_ks_p}(k^0=-q^0+\omega^{ns_p}_{k})}{2 \omega^{ns_p}_{k} 
			\SB{\FB{q^0 - \omega^{ns_p}_{k}}^2 - \FB{\omega^{ls_k}_{k}}^2 }} 
		+\dfrac{\tilde{\mathcal{N}}^a_{lns_ks_p}(k^0=\omega^{ls_k}_{k})}{2 \omega^{ls_k}_{k} \SB{\FB{q^0 + \omega^{ls_k}_{k}}^2 - \FB{\omega^{ns_p}_{k}}^2 }} 
		\right. \nn \\ && \hspace{1cm} \left. - \dfrac{\tilde{\mathcal{N}}^a_{lns_ks_p}(k^0=-\omega^{ls_k}_{k}) f^-(\omega^{ls_k}_{k}) }{2 \omega^{ls_k}_{k} 
			\SB{\FB{q^0 - \omega^{ls_k}_{k}}^2 - \FB{\omega^{ns_p}_{k}}^2  }  } - \dfrac{\tilde{\mathcal{N}}^a_{lns_ks_p}(k^0=\omega^{ls_k}_{k}) f^+(\omega^{ls_k}_{k})}
		{2 \omega^{ls_k}_{k} \SB{\FB{q^0 + \omega^{ls_k}_{k}}^2 - \FB{\omega^{ns_p}_{k}}^2  }  }   \right. \nn  \\ &&  \hspace{1cm} \left.
		-\dfrac{\tilde{\mathcal{N}}^a_{lns_ks_p}(k^0=-q^0-\omega^{ns_p}_{k}) f^-(\omega^{ns_p}_{k})   }{2 \omega^{ns_p}_{k} 
			\SB{\FB{q^0 + \omega^{ns_p}_{k}}^2 - \FB{\omega^{ls_k}_{k}}^2  }  } - \dfrac{\tilde{\mathcal{N}}^a_{lns_ks_p}(k^0=-q^0 +\omega^{ns_p}_{k}) f^+(\omega^{ns_p}_{k})}
		{2 \omega^{ns_p}_{k} \SB{\FB{q^0 - \omega^{ns_p}_{k}}^2 - \FB{\omega^{ls_k}_{k}}^2  }  } 
	}
	\label{eq.pola.3}
\end{eqnarray}
where, 
\begin{eqnarray}
\tilde{\mathcal{N}}^{a}_{lns_ks_p}(k,q) &=& \delta_l^n j2N_c \frac{eB}{4\pi M_l^2} 
\FB{1-\delta^0_l \delta^{-1}_{s_k }}\FB{1-\delta^0_l \delta^{-1}_{s_p }} \nn \\ &&
\Bigg[ -4eBl \Big\{ s_ks_p(k_\parallel^2+k^0q^0)
+js_ks_p(\kappa eB)^2 +jM_l^2  -j \kappa eBM_l(s_p+s_k) \Big\} \nn \\ &&
+ \FB{1-\delta_l^0} \Big\{j(k_\parallel^2+k^0q^0)
(M_l-s_kM)(M_l-s_pM) \nn \\ && + \big\{s_k\kappa eBM-M_l(M-s_kM_l+\kappa eB)\big\}\big\{s_p\kappa eBM-M_l(M-s_pM_l+\kappa eB)\big\}\Big\}
\nn \\ && + j(k_\parallel^2+k^0q^0)
(M_l+s_kM)(M_l+s_pM) \nn \\ && + \big\{s_k\kappa eBM-M_l(M+s_kM_l-\kappa eB)\big\}\big\{s_p\kappa eBM-M_l(M+s_pM_l-\kappa eB)\big\}
\Bigg] ~. \label{eq.Ntilde}
\end{eqnarray}
The presence of Kronecker delta function in the above equation will eliminate one of the sums in Eq.~\eqref{eq.pola.3}. 
It is to be noted that, the temperature independent part of Eq.~\eqref{eq.pola.3} is ultra violate divergent which has to be 
properly regularized. Following the same regularization procedure as used in the gap equation (as discussed in Sec.~\ref{deri_gap_eqn}) we 
get, 
\begin{eqnarray}
\RE\overline{\overline{\Pi}}_a (q) &=& \sum_f \sum_{s_k,s_p} \sum_{l=0}^{\infty} \int_{0}^{\sqrt{\Lambda^2-\vec{k}_{\perp l}^2}} \dfrac{dk_z}{\pi}  
\Theta\FB{\vec{k}_{\perp l}^2} \Theta\FB{\vec{p}_{\perp l}^2} \Theta\FB{\Lambda^2-\vec{k}_{\perp l}^2} \Theta\FB{\Lambda^2-\vec{p}_{\perp l}^2} \nn \\ 
&& \mathcal{P} \TB{\dfrac{\tilde{\mathcal{N}}^a_{lls_ks_p}(k^0=-q^0+\omega^{ls_p}_{k})}{2 \omega^{ls_p}_{k} 
		\SB{\FB{q^0 - \omega^{ls_p}_{k}}^2 - \FB{\omega^{ls_k}_{k}}^2 }} 
	+\dfrac{\tilde{\mathcal{N}}^a_{lls_ks_p}(k^0=\omega^{ls_k}_{k})}{2 \omega^{ls_k}_{k} \SB{\FB{q^0 + \omega^{ls_k}_{k}}^2 - \FB{\omega^{ls_p}_{k}}^2 }} }
	   + \sum_f \sum_{s_k,s_p} \sum_{l=0}^{\infty} \int_{-\infty}^{+\infty} \dfrac{dk_z}{(2\pi)}  \nn \\ && 
	 \Theta\FB{\vec{k}_{\perp l}^2} \Theta\FB{\vec{p}_{\perp l}^2}
	 \mathcal{P} \TB{
	 - \dfrac{\tilde{\mathcal{N}}^a_{lls_ks_p}(k^0=-\omega^{ls_k}_{k}) f^-(\omega^{ls_k}_{k}) }{2 \omega^{ls_k}_{k} 
		\SB{\FB{q^0 - \omega^{ls_k}_{k}}^2 - \FB{\omega^{ls_p}_{k}}^2  }  } - \dfrac{\tilde{\mathcal{N}}^a_{lls_ks_p}(k^0=\omega^{ls_k}_{k}) f^+(\omega^{ls_k}_{k})}
	{2 \omega^{ls_k}_{k} \SB{\FB{q^0 + \omega^{ls_k}_{k}}^2 - \FB{\omega^{ls_p}_{k}}^2  }  }   \right. \nn  \\ &&  \hspace{0cm} \left.
	-\dfrac{\tilde{\mathcal{N}}^a_{lls_ks_p}(k^0=-q^0-\omega^{ls_p}_{k}) f^-(\omega^{ls_p}_{k})   }{2 \omega^{ls_p}_{k} 
		\SB{\FB{q^0 + \omega^{ls_p}_{k}}^2 - \FB{\omega^{ls_k}_{k}}^2  }  } - \dfrac{\tilde{\mathcal{N}}^a_{lls_ks_p}(k^0=-q^0 +\omega^{ls_p}_{k}) f^+(\omega^{ls_p}_{k})}
	{2 \omega^{ls_p}_{k} \SB{\FB{q^0 - \omega^{ls_p}_{k}}^2 - \FB{\omega^{ls_k}_{k}}^2  }  } 
}
\label{eq.pola.4}
\end{eqnarray}
where 
\begin{eqnarray}
\vec{k}_{\perp l}^2 &=& 2 leB +\FB{\kappa e B}^2 - 2s_kM_{l}(\kappa eB)  \\
\vec{p}_{\perp l}^2 &=& 2 leB +\FB{\kappa e B}^2 - 2s_pM_{l}(\kappa eB)~.
\end{eqnarray}
In Eq.~\eqref{eq.pola.4}, the step functions $\Theta\FB{\vec{k}_{\perp l}^2}$ and $\Theta\FB{\vec{p}_{\perp l}^2}$ represents 
the AMM blocking whereas $\Theta\FB{\Lambda^2-\vec{k}_{\perp l}^2}$ and $\Theta\FB{\Lambda^2-\vec{p}_{\perp l}^2}$ represents 
UV blocking of the loop quarks. These step functions will in turn restrict the minimum and maximum values of the summation 
index $l$ in Eq.~\eqref{eq.pola.4}.

\bibliographystyle{apsrev4-1}
\bibliography{nilanjan}

\begin{thebibliography}{96}%
\makeatletter
\providecommand \@ifxundefined [1]{%
 \@ifx{#1\undefined}
}%
\providecommand \@ifnum [1]{%
 \ifnum #1\expandafter \@firstoftwo
 \else \expandafter \@secondoftwo
 \fi
}%
\providecommand \@ifx [1]{%
 \ifx #1\expandafter \@firstoftwo
 \else \expandafter \@secondoftwo
 \fi
}%
\providecommand \natexlab [1]{#1}%
\providecommand \enquote  [1]{``#1''}%
\providecommand \bibnamefont  [1]{#1}%
\providecommand \bibfnamefont [1]{#1}%
\providecommand \citenamefont [1]{#1}%
\providecommand \href@noop [0]{\@secondoftwo}%
\providecommand \href [0]{\begingroup \@sanitize@url \@href}%
\providecommand \@href[1]{\@@startlink{#1}\@@href}%
\providecommand \@@href[1]{\endgroup#1\@@endlink}%
\providecommand \@sanitize@url [0]{\catcode `\\12\catcode `\$12\catcode
  `\&12\catcode `\#12\catcode `\^12\catcode `\_12\catcode `\%12\relax}%
\providecommand \@@startlink[1]{}%
\providecommand \@@endlink[0]{}%
\providecommand \url  [0]{\begingroup\@sanitize@url \@url }%
\providecommand \@url [1]{\endgroup\@href {#1}{\urlprefix }}%
\providecommand \urlprefix  [0]{URL }%
\providecommand \Eprint [0]{\href }%
\providecommand \doibase [0]{http://dx.doi.org/}%
\providecommand \selectlanguage [0]{\@gobble}%
\providecommand \bibinfo  [0]{\@secondoftwo}%
\providecommand \bibfield  [0]{\@secondoftwo}%
\providecommand \translation [1]{[#1]}%
\providecommand \BibitemOpen [0]{}%
\providecommand \bibitemStop [0]{}%
\providecommand \bibitemNoStop [0]{.\EOS\space}%
\providecommand \EOS [0]{\spacefactor3000\relax}%
\providecommand \BibitemShut  [1]{\csname bibitem#1\endcsname}%
\let\auto@bib@innerbib\@empty
\bibitem [{\citenamefont {Kharzeev}\ \emph {et~al.}(2013)\citenamefont
  {Kharzeev}, \citenamefont {Landsteiner}, \citenamefont {Schmitt},\ and\
  \citenamefont {Yee}}]{Lect_note}%
  \BibitemOpen
  \bibfield  {author} {\bibinfo {author} {\bibfnamefont {D.~E.}\ \bibnamefont
  {Kharzeev}}, \bibinfo {author} {\bibfnamefont {K.}~\bibnamefont
  {Landsteiner}}, \bibinfo {author} {\bibfnamefont {A.}~\bibnamefont
  {Schmitt}}, \ and\ \bibinfo {author} {\bibfnamefont {H.-U.}\ \bibnamefont
  {Yee}},\ }\href {\doibase 10.1007/978-3-642-37305-3_1} {\bibfield  {journal}
  {\bibinfo  {journal} {Lect. Notes Phys.}\ }\textbf {\bibinfo {volume}
  {871}},\ \bibinfo {pages} {1} (\bibinfo {year} {2013})},\ \Eprint
  {http://arxiv.org/abs/1211.6245} {arXiv:1211.6245 [hep-ph]} \BibitemShut
  {NoStop}%
\bibitem [{\citenamefont {Fukushima}\ \emph {et~al.}(2008)\citenamefont
  {Fukushima}, \citenamefont {Kharzeev},\ and\ \citenamefont
  {Warringa}}]{Fukushima}%
  \BibitemOpen
  \bibfield  {author} {\bibinfo {author} {\bibfnamefont {K.}~\bibnamefont
  {Fukushima}}, \bibinfo {author} {\bibfnamefont {D.~E.}\ \bibnamefont
  {Kharzeev}}, \ and\ \bibinfo {author} {\bibfnamefont {H.~J.}\ \bibnamefont
  {Warringa}},\ }\href {\doibase 10.1103/PhysRevD.78.074033} {\bibfield
  {journal} {\bibinfo  {journal} {Phys. Rev.}\ }\textbf {\bibinfo {volume}
  {D78}},\ \bibinfo {pages} {074033} (\bibinfo {year} {2008})},\ \Eprint
  {http://arxiv.org/abs/0808.3382} {arXiv:0808.3382 [hep-ph]} \BibitemShut
  {NoStop}%
\bibitem [{\citenamefont {Kharzeev}\ \emph {et~al.}(2008)\citenamefont
  {Kharzeev}, \citenamefont {McLerran},\ and\ \citenamefont
  {Warringa}}]{Kharzeev}%
  \BibitemOpen
  \bibfield  {author} {\bibinfo {author} {\bibfnamefont {D.~E.}\ \bibnamefont
  {Kharzeev}}, \bibinfo {author} {\bibfnamefont {L.~D.}\ \bibnamefont
  {McLerran}}, \ and\ \bibinfo {author} {\bibfnamefont {H.~J.}\ \bibnamefont
  {Warringa}},\ }\href {\doibase 10.1016/j.nuclphysa.2008.02.298} {\bibfield
  {journal} {\bibinfo  {journal} {Nucl. Phys.}\ }\textbf {\bibinfo {volume}
  {A803}},\ \bibinfo {pages} {227} (\bibinfo {year} {2008})},\ \Eprint
  {http://arxiv.org/abs/0711.0950} {arXiv:0711.0950 [hep-ph]} \BibitemShut
  {NoStop}%
\bibitem [{\citenamefont {Kharzeev}\ and\ \citenamefont
  {Warringa}(2009)}]{Kharzeev2}%
  \BibitemOpen
  \bibfield  {author} {\bibinfo {author} {\bibfnamefont {D.~E.}\ \bibnamefont
  {Kharzeev}}\ and\ \bibinfo {author} {\bibfnamefont {H.~J.}\ \bibnamefont
  {Warringa}},\ }\href {\doibase 10.1103/PhysRevD.80.034028} {\bibfield
  {journal} {\bibinfo  {journal} {Phys. Rev.}\ }\textbf {\bibinfo {volume}
  {D80}},\ \bibinfo {pages} {034028} (\bibinfo {year} {2009})},\ \Eprint
  {http://arxiv.org/abs/0907.5007} {arXiv:0907.5007 [hep-ph]} \BibitemShut
  {NoStop}%
\bibitem [{\citenamefont {Bali}\ \emph
  {et~al.}(2012{\natexlab{a}})\citenamefont {Bali}, \citenamefont {Bruckmann},
  \citenamefont {Endrodi}, \citenamefont {Fodor}, \citenamefont {Katz},
  \citenamefont {Krieg}, \citenamefont {Schafer},\ and\ \citenamefont
  {Szabo}}]{Bali}%
  \BibitemOpen
  \bibfield  {author} {\bibinfo {author} {\bibfnamefont {G.~S.}\ \bibnamefont
  {Bali}}, \bibinfo {author} {\bibfnamefont {F.}~\bibnamefont {Bruckmann}},
  \bibinfo {author} {\bibfnamefont {G.}~\bibnamefont {Endrodi}}, \bibinfo
  {author} {\bibfnamefont {Z.}~\bibnamefont {Fodor}}, \bibinfo {author}
  {\bibfnamefont {S.~D.}\ \bibnamefont {Katz}}, \bibinfo {author}
  {\bibfnamefont {S.}~\bibnamefont {Krieg}}, \bibinfo {author} {\bibfnamefont
  {A.}~\bibnamefont {Schafer}}, \ and\ \bibinfo {author} {\bibfnamefont
  {K.~K.}\ \bibnamefont {Szabo}},\ }\href {\doibase 10.1007/JHEP02(2012)044}
  {\bibfield  {journal} {\bibinfo  {journal} {JHEP}\ }\textbf {\bibinfo
  {volume} {02}},\ \bibinfo {pages} {044} (\bibinfo {year}
  {2012}{\natexlab{a}})},\ \Eprint {http://arxiv.org/abs/1111.4956}
  {arXiv:1111.4956 [hep-lat]} \BibitemShut {NoStop}%
\bibitem [{\citenamefont {Shovkovy}(2013)}]{Shovkovy}%
  \BibitemOpen
  \bibfield  {author} {\bibinfo {author} {\bibfnamefont {I.~A.}\ \bibnamefont
  {Shovkovy}},\ }\href {\doibase 10.1007/978-3-642-37305-3_2} {\bibfield
  {journal} {\bibinfo  {journal} {Lect. Notes Phys.}\ }\textbf {\bibinfo
  {volume} {871}},\ \bibinfo {pages} {13} (\bibinfo {year} {2013})},\ \Eprint
  {http://arxiv.org/abs/1207.5081} {arXiv:1207.5081 [hep-ph]} \BibitemShut
  {NoStop}%
\bibitem [{\citenamefont {Gusynin}\ \emph {et~al.}(1994)\citenamefont
  {Gusynin}, \citenamefont {Miransky},\ and\ \citenamefont
  {Shovkovy}}]{Gusynin1}%
  \BibitemOpen
  \bibfield  {author} {\bibinfo {author} {\bibfnamefont {V.~P.}\ \bibnamefont
  {Gusynin}}, \bibinfo {author} {\bibfnamefont {V.~A.}\ \bibnamefont
  {Miransky}}, \ and\ \bibinfo {author} {\bibfnamefont {I.~A.}\ \bibnamefont
  {Shovkovy}},\ }\href {\doibase 10.1103/PhysRevLett.76.1005,
  10.1103/PhysRevLett.73.3499} {\bibfield  {journal} {\bibinfo  {journal}
  {Phys. Rev. Lett.}\ }\textbf {\bibinfo {volume} {73}},\ \bibinfo {pages}
  {3499} (\bibinfo {year} {1994})},\ \bibinfo {note} {[Erratum: Phys. Rev.
  Lett.76,1005(1996)]},\ \Eprint {http://arxiv.org/abs/hep-ph/9405262}
  {arXiv:hep-ph/9405262 [hep-ph]} \BibitemShut {NoStop}%
\bibitem [{\citenamefont {Gusynin}\ \emph {et~al.}(1996)\citenamefont
  {Gusynin}, \citenamefont {Miransky},\ and\ \citenamefont
  {Shovkovy}}]{Gusynin2}%
  \BibitemOpen
  \bibfield  {author} {\bibinfo {author} {\bibfnamefont {V.~P.}\ \bibnamefont
  {Gusynin}}, \bibinfo {author} {\bibfnamefont {V.~A.}\ \bibnamefont
  {Miransky}}, \ and\ \bibinfo {author} {\bibfnamefont {I.~A.}\ \bibnamefont
  {Shovkovy}},\ }\href {\doibase 10.1016/0550-3213(96)00021-1} {\bibfield
  {journal} {\bibinfo  {journal} {Nucl. Phys.}\ }\textbf {\bibinfo {volume}
  {B462}},\ \bibinfo {pages} {249} (\bibinfo {year} {1996})},\ \Eprint
  {http://arxiv.org/abs/hep-ph/9509320} {arXiv:hep-ph/9509320 [hep-ph]}
  \BibitemShut {NoStop}%
\bibitem [{\citenamefont {Gusynin}\ \emph {et~al.}(1999)\citenamefont
  {Gusynin}, \citenamefont {Miransky},\ and\ \citenamefont
  {Shovkovy}}]{Gusynin3}%
  \BibitemOpen
  \bibfield  {author} {\bibinfo {author} {\bibfnamefont {V.~P.}\ \bibnamefont
  {Gusynin}}, \bibinfo {author} {\bibfnamefont {V.~A.}\ \bibnamefont
  {Miransky}}, \ and\ \bibinfo {author} {\bibfnamefont {I.~A.}\ \bibnamefont
  {Shovkovy}},\ }\href {\doibase 10.1016/S0550-3213(99)00573-8} {\bibfield
  {journal} {\bibinfo  {journal} {Nucl. Phys.}\ }\textbf {\bibinfo {volume}
  {B563}},\ \bibinfo {pages} {361} (\bibinfo {year} {1999})},\ \Eprint
  {http://arxiv.org/abs/hep-ph/9908320} {arXiv:hep-ph/9908320 [hep-ph]}
  \BibitemShut {NoStop}%
\bibitem [{\citenamefont {Preis}\ \emph {et~al.}(2011)\citenamefont {Preis},
  \citenamefont {Rebhan},\ and\ \citenamefont {Schmitt}}]{Preis}%
  \BibitemOpen
  \bibfield  {author} {\bibinfo {author} {\bibfnamefont {F.}~\bibnamefont
  {Preis}}, \bibinfo {author} {\bibfnamefont {A.}~\bibnamefont {Rebhan}}, \
  and\ \bibinfo {author} {\bibfnamefont {A.}~\bibnamefont {Schmitt}},\ }\href
  {\doibase 10.1007/JHEP03(2011)033} {\bibfield  {journal} {\bibinfo  {journal}
  {JHEP}\ }\textbf {\bibinfo {volume} {03}},\ \bibinfo {pages} {033} (\bibinfo
  {year} {2011})},\ \Eprint {http://arxiv.org/abs/1012.4785} {arXiv:1012.4785
  [hep-th]} \BibitemShut {NoStop}%
\bibitem [{\citenamefont {Preis}\ \emph {et~al.}(2013)\citenamefont {Preis},
  \citenamefont {Rebhan},\ and\ \citenamefont {Schmitt}}]{Preis2}%
  \BibitemOpen
  \bibfield  {author} {\bibinfo {author} {\bibfnamefont {F.}~\bibnamefont
  {Preis}}, \bibinfo {author} {\bibfnamefont {A.}~\bibnamefont {Rebhan}}, \
  and\ \bibinfo {author} {\bibfnamefont {A.}~\bibnamefont {Schmitt}},\ }\href
  {\doibase 10.1007/978-3-642-37305-3_3} {\bibfield  {journal} {\bibinfo
  {journal} {Lect. Notes Phys.}\ }\textbf {\bibinfo {volume} {871}},\ \bibinfo
  {pages} {51} (\bibinfo {year} {2013})},\ \Eprint
  {http://arxiv.org/abs/1208.0536} {arXiv:1208.0536 [hep-ph]} \BibitemShut
  {NoStop}%
\bibitem [{\citenamefont {Elmfors}\ \emph {et~al.}(1998)\citenamefont
  {Elmfors}, \citenamefont {Enqvist},\ and\ \citenamefont
  {Kainulainen}}]{Elmfors}%
  \BibitemOpen
  \bibfield  {author} {\bibinfo {author} {\bibfnamefont {P.}~\bibnamefont
  {Elmfors}}, \bibinfo {author} {\bibfnamefont {K.}~\bibnamefont {Enqvist}}, \
  and\ \bibinfo {author} {\bibfnamefont {K.}~\bibnamefont {Kainulainen}},\
  }\href {\doibase 10.1016/S0370-2693(98)01117-4} {\bibfield  {journal}
  {\bibinfo  {journal} {Phys. Lett.}\ }\textbf {\bibinfo {volume} {B440}},\
  \bibinfo {pages} {269} (\bibinfo {year} {1998})},\ \Eprint
  {http://arxiv.org/abs/hep-ph/9806403} {arXiv:hep-ph/9806403 [hep-ph]}
  \BibitemShut {NoStop}%
\bibitem [{\citenamefont {Skalozub}\ and\ \citenamefont
  {Bordag}(2000)}]{Skalozub}%
  \BibitemOpen
  \bibfield  {author} {\bibinfo {author} {\bibfnamefont {V.}~\bibnamefont
  {Skalozub}}\ and\ \bibinfo {author} {\bibfnamefont {M.}~\bibnamefont
  {Bordag}},\ }\href {\doibase 10.1142/S0217751X00000161} {\bibfield  {journal}
  {\bibinfo  {journal} {Int. J. Mod. Phys.}\ }\textbf {\bibinfo {volume}
  {A15}},\ \bibinfo {pages} {349} (\bibinfo {year} {2000})},\ \Eprint
  {http://arxiv.org/abs/hep-ph/9904333} {arXiv:hep-ph/9904333 [hep-ph]}
  \BibitemShut {NoStop}%
\bibitem [{\citenamefont {Sadooghi}\ and\ \citenamefont
  {Anaraki}(2008)}]{Sadooghi_ew}%
  \BibitemOpen
  \bibfield  {author} {\bibinfo {author} {\bibfnamefont {N.}~\bibnamefont
  {Sadooghi}}\ and\ \bibinfo {author} {\bibfnamefont {K.~S.}\ \bibnamefont
  {Anaraki}},\ }\href {\doibase 10.1103/PhysRevD.78.125019} {\bibfield
  {journal} {\bibinfo  {journal} {Phys. Rev.}\ }\textbf {\bibinfo {volume}
  {D78}},\ \bibinfo {pages} {125019} (\bibinfo {year} {2008})},\ \Eprint
  {http://arxiv.org/abs/0805.0078} {arXiv:0805.0078 [hep-ph]} \BibitemShut
  {NoStop}%
\bibitem [{\citenamefont {Navarro}\ \emph {et~al.}(2010)\citenamefont
  {Navarro}, \citenamefont {Sanchez}, \citenamefont {Tejeda-Yeomans},
  \citenamefont {Ayala},\ and\ \citenamefont {Piccinelli}}]{Navarro}%
  \BibitemOpen
  \bibfield  {author} {\bibinfo {author} {\bibfnamefont {J.}~\bibnamefont
  {Navarro}}, \bibinfo {author} {\bibfnamefont {A.}~\bibnamefont {Sanchez}},
  \bibinfo {author} {\bibfnamefont {M.~E.}\ \bibnamefont {Tejeda-Yeomans}},
  \bibinfo {author} {\bibfnamefont {A.}~\bibnamefont {Ayala}}, \ and\ \bibinfo
  {author} {\bibfnamefont {G.}~\bibnamefont {Piccinelli}},\ }\href {\doibase
  10.1103/PhysRevD.82.123007} {\bibfield  {journal} {\bibinfo  {journal} {Phys.
  Rev.}\ }\textbf {\bibinfo {volume} {D82}},\ \bibinfo {pages} {123007}
  (\bibinfo {year} {2010})},\ \Eprint {http://arxiv.org/abs/1007.4208}
  {arXiv:1007.4208 [hep-ph]} \BibitemShut {NoStop}%
\bibitem [{\citenamefont {Fayazbakhsh}\ and\ \citenamefont
  {Sadooghi}(2010)}]{Fayazbakhsh1}%
  \BibitemOpen
  \bibfield  {author} {\bibinfo {author} {\bibfnamefont {S.}~\bibnamefont
  {Fayazbakhsh}}\ and\ \bibinfo {author} {\bibfnamefont {N.}~\bibnamefont
  {Sadooghi}},\ }\href {\doibase 10.1103/PhysRevD.82.045010} {\bibfield
  {journal} {\bibinfo  {journal} {Phys. Rev.}\ }\textbf {\bibinfo {volume}
  {D82}},\ \bibinfo {pages} {045010} (\bibinfo {year} {2010})},\ \Eprint
  {http://arxiv.org/abs/1005.5022} {arXiv:1005.5022 [hep-ph]} \BibitemShut
  {NoStop}%
\bibitem [{\citenamefont {Fayazbakhsh}\ and\ \citenamefont
  {Sadooghi}(2011)}]{Fayazbakhsh2}%
  \BibitemOpen
  \bibfield  {author} {\bibinfo {author} {\bibfnamefont {S.}~\bibnamefont
  {Fayazbakhsh}}\ and\ \bibinfo {author} {\bibfnamefont {N.}~\bibnamefont
  {Sadooghi}},\ }\href {\doibase 10.1103/PhysRevD.83.025026} {\bibfield
  {journal} {\bibinfo  {journal} {Phys. Rev.}\ }\textbf {\bibinfo {volume}
  {D83}},\ \bibinfo {pages} {025026} (\bibinfo {year} {2011})},\ \Eprint
  {http://arxiv.org/abs/1009.6125} {arXiv:1009.6125 [hep-ph]} \BibitemShut
  {NoStop}%
\bibitem [{\citenamefont {Skokov}(2012)}]{Skokov}%
  \BibitemOpen
  \bibfield  {author} {\bibinfo {author} {\bibfnamefont {V.}~\bibnamefont
  {Skokov}},\ }\href {\doibase 10.1103/PhysRevD.85.034026} {\bibfield
  {journal} {\bibinfo  {journal} {Phys. Rev.}\ }\textbf {\bibinfo {volume}
  {D85}},\ \bibinfo {pages} {034026} (\bibinfo {year} {2012})},\ \Eprint
  {http://arxiv.org/abs/1112.5137} {arXiv:1112.5137 [hep-ph]} \BibitemShut
  {NoStop}%
\bibitem [{\citenamefont {Fukushima}\ and\ \citenamefont
  {Pawlowski}(2012)}]{Fukushima2}%
  \BibitemOpen
  \bibfield  {author} {\bibinfo {author} {\bibfnamefont {K.}~\bibnamefont
  {Fukushima}}\ and\ \bibinfo {author} {\bibfnamefont {J.~M.}\ \bibnamefont
  {Pawlowski}},\ }\href {\doibase 10.1103/PhysRevD.86.076013} {\bibfield
  {journal} {\bibinfo  {journal} {Phys. Rev.}\ }\textbf {\bibinfo {volume}
  {D86}},\ \bibinfo {pages} {076013} (\bibinfo {year} {2012})},\ \Eprint
  {http://arxiv.org/abs/1203.4330} {arXiv:1203.4330 [hep-ph]} \BibitemShut
  {NoStop}%
\bibitem [{\citenamefont {Chernodub}(2011)}]{Chernodub1}%
  \BibitemOpen
  \bibfield  {author} {\bibinfo {author} {\bibfnamefont {M.~N.}\ \bibnamefont
  {Chernodub}},\ }\href {\doibase 10.1103/PhysRevLett.106.142003} {\bibfield
  {journal} {\bibinfo  {journal} {Phys. Rev. Lett.}\ }\textbf {\bibinfo
  {volume} {106}},\ \bibinfo {pages} {142003} (\bibinfo {year} {2011})},\
  \Eprint {http://arxiv.org/abs/1101.0117} {arXiv:1101.0117 [hep-ph]}
  \BibitemShut {NoStop}%
\bibitem [{\citenamefont {Chernodub}\ \emph {et~al.}(2012)\citenamefont
  {Chernodub}, \citenamefont {Van~Doorsselaere},\ and\ \citenamefont
  {Verschelde}}]{Chernodub2}%
  \BibitemOpen
  \bibfield  {author} {\bibinfo {author} {\bibfnamefont {M.~N.}\ \bibnamefont
  {Chernodub}}, \bibinfo {author} {\bibfnamefont {J.}~\bibnamefont
  {Van~Doorsselaere}}, \ and\ \bibinfo {author} {\bibfnamefont
  {H.}~\bibnamefont {Verschelde}},\ }\href {\doibase
  10.1103/PhysRevD.85.045002} {\bibfield  {journal} {\bibinfo  {journal} {Phys.
  Rev.}\ }\textbf {\bibinfo {volume} {D85}},\ \bibinfo {pages} {045002}
  (\bibinfo {year} {2012})},\ \Eprint {http://arxiv.org/abs/1111.4401}
  {arXiv:1111.4401 [hep-ph]} \BibitemShut {NoStop}%
\bibitem [{\citenamefont {Skokov}\ \emph {et~al.}(2009)\citenamefont {Skokov},
  \citenamefont {Illarionov},\ and\ \citenamefont {Toneev}}]{Skokov2}%
  \BibitemOpen
  \bibfield  {author} {\bibinfo {author} {\bibfnamefont {V.}~\bibnamefont
  {Skokov}}, \bibinfo {author} {\bibfnamefont {A.~{\relax Yu}.}\ \bibnamefont
  {Illarionov}}, \ and\ \bibinfo {author} {\bibfnamefont {V.}~\bibnamefont
  {Toneev}},\ }\href {\doibase 10.1142/S0217751X09047570} {\bibfield  {journal}
  {\bibinfo  {journal} {Int. J. Mod. Phys.}\ }\textbf {\bibinfo {volume}
  {A24}},\ \bibinfo {pages} {5925} (\bibinfo {year} {2009})},\ \Eprint
  {http://arxiv.org/abs/0907.1396} {arXiv:0907.1396 [nucl-th]} \BibitemShut
  {NoStop}%
\bibitem [{\citenamefont {Gursoy}\ \emph {et~al.}(2014)\citenamefont {Gursoy},
  \citenamefont {Kharzeev},\ and\ \citenamefont {Rajagopal}}]{Gursoy}%
  \BibitemOpen
  \bibfield  {author} {\bibinfo {author} {\bibfnamefont {U.}~\bibnamefont
  {Gursoy}}, \bibinfo {author} {\bibfnamefont {D.}~\bibnamefont {Kharzeev}}, \
  and\ \bibinfo {author} {\bibfnamefont {K.}~\bibnamefont {Rajagopal}},\ }\href
  {\doibase 10.1103/PhysRevC.89.054905} {\bibfield  {journal} {\bibinfo
  {journal} {Phys. Rev.}\ }\textbf {\bibinfo {volume} {C89}},\ \bibinfo {pages}
  {054905} (\bibinfo {year} {2014})},\ \Eprint {http://arxiv.org/abs/1401.3805}
  {arXiv:1401.3805 [hep-ph]} \BibitemShut {NoStop}%
\bibitem [{\citenamefont {Vachaspati}(1991)}]{Vachaspati}%
  \BibitemOpen
  \bibfield  {author} {\bibinfo {author} {\bibfnamefont {T.}~\bibnamefont
  {Vachaspati}},\ }\href {\doibase 10.1016/0370-2693(91)90051-Q} {\bibfield
  {journal} {\bibinfo  {journal} {Phys. Lett.}\ }\textbf {\bibinfo {volume}
  {B265}},\ \bibinfo {pages} {258} (\bibinfo {year} {1991})}\BibitemShut
  {NoStop}%
\bibitem [{\citenamefont {Campanelli}(2013)}]{Campanelli}%
  \BibitemOpen
  \bibfield  {author} {\bibinfo {author} {\bibfnamefont {L.}~\bibnamefont
  {Campanelli}},\ }\href {\doibase 10.1103/PhysRevLett.111.061301} {\bibfield
  {journal} {\bibinfo  {journal} {Phys. Rev. Lett.}\ }\textbf {\bibinfo
  {volume} {111}},\ \bibinfo {pages} {061301} (\bibinfo {year} {2013})},\
  \Eprint {http://arxiv.org/abs/1304.6534} {arXiv:1304.6534 [astro-ph.CO]}
  \BibitemShut {NoStop}%
\bibitem [{\citenamefont {Duncan}\ and\ \citenamefont
  {Thompson}(1992)}]{Duncan}%
  \BibitemOpen
  \bibfield  {author} {\bibinfo {author} {\bibfnamefont {R.~C.}\ \bibnamefont
  {Duncan}}\ and\ \bibinfo {author} {\bibfnamefont {C.}~\bibnamefont
  {Thompson}},\ }\href {\doibase 10.1086/186413} {\bibfield  {journal}
  {\bibinfo  {journal} {Astrophys. J.}\ }\textbf {\bibinfo {volume} {392}},\
  \bibinfo {pages} {L9} (\bibinfo {year} {1992})}\BibitemShut {NoStop}%
\bibitem [{\citenamefont {Thompson}\ and\ \citenamefont
  {Duncan}(1993)}]{Duncan2}%
  \BibitemOpen
  \bibfield  {author} {\bibinfo {author} {\bibfnamefont {C.}~\bibnamefont
  {Thompson}}\ and\ \bibinfo {author} {\bibfnamefont {R.~C.}\ \bibnamefont
  {Duncan}},\ }\href {\doibase 10.1086/172580} {\bibfield  {journal} {\bibinfo
  {journal} {Astrophys. J.}\ }\textbf {\bibinfo {volume} {408}},\ \bibinfo
  {pages} {194} (\bibinfo {year} {1993})}\BibitemShut {NoStop}%
\bibitem [{\citenamefont {Lai}\ and\ \citenamefont {Shapiro}(1991)}]{Lai}%
  \BibitemOpen
  \bibfield  {author} {\bibinfo {author} {\bibfnamefont {D.}~\bibnamefont
  {Lai}}\ and\ \bibinfo {author} {\bibfnamefont {S.~L.}\ \bibnamefont
  {Shapiro}},\ }\href@noop {} {\bibfield  {journal} {\bibinfo  {journal}
  {Astrophys. J.}\ }\textbf {\bibinfo {volume} {383}},\ \bibinfo {pages} {745}
  (\bibinfo {year} {1991})}\BibitemShut {NoStop}%
\bibitem [{\citenamefont {Novoselov}\ \emph {et~al.}(2005)\citenamefont
  {Novoselov}, \citenamefont {Geim}, \citenamefont {Morozov}, \citenamefont
  {Jiang}, \citenamefont {Katsnelson}, \citenamefont {Grigorieva},
  \citenamefont {Dubonos},\ and\ \citenamefont {Firsov}}]{Novoselov}%
  \BibitemOpen
  \bibfield  {author} {\bibinfo {author} {\bibfnamefont {K.~S.}\ \bibnamefont
  {Novoselov}}, \bibinfo {author} {\bibfnamefont {A.~K.}\ \bibnamefont {Geim}},
  \bibinfo {author} {\bibfnamefont {S.~V.}\ \bibnamefont {Morozov}}, \bibinfo
  {author} {\bibfnamefont {D.}~\bibnamefont {Jiang}}, \bibinfo {author}
  {\bibfnamefont {M.~I.}\ \bibnamefont {Katsnelson}}, \bibinfo {author}
  {\bibfnamefont {I.~V.}\ \bibnamefont {Grigorieva}}, \bibinfo {author}
  {\bibfnamefont {S.~V.}\ \bibnamefont {Dubonos}}, \ and\ \bibinfo {author}
  {\bibfnamefont {A.~A.}\ \bibnamefont {Firsov}},\ }\href {\doibase
  10.1038/nature04233} {\bibfield  {journal} {\bibinfo  {journal} {Nature}\
  }\textbf {\bibinfo {volume} {438}},\ \bibinfo {pages} {197} (\bibinfo {year}
  {2005})},\ \Eprint {http://arxiv.org/abs/cond-mat/0509330}
  {arXiv:cond-mat/0509330 [cond-mat.mes-hall]} \BibitemShut {NoStop}%
\bibitem [{\citenamefont {Zhang}\ \emph {et~al.}(2005)\citenamefont {Zhang},
  \citenamefont {Tan}, \citenamefont {Stormer},\ and\ \citenamefont
  {Kim}}]{Zhang_cond}%
  \BibitemOpen
  \bibfield  {author} {\bibinfo {author} {\bibfnamefont {Y.}~\bibnamefont
  {Zhang}}, \bibinfo {author} {\bibfnamefont {Y.-W.}\ \bibnamefont {Tan}},
  \bibinfo {author} {\bibfnamefont {H.~L.}\ \bibnamefont {Stormer}}, \ and\
  \bibinfo {author} {\bibfnamefont {P.}~\bibnamefont {Kim}},\ }\href {\doibase
  10.1038/nature04235} {\bibfield  {journal} {\bibinfo  {journal} {Nature}\
  }\textbf {\bibinfo {volume} {438}},\ \bibinfo {pages} {201} (\bibinfo {year}
  {2005})}\BibitemShut {NoStop}%
\bibitem [{\citenamefont {de~Forcrand}\ and\ \citenamefont
  {Philipsen}(2007)}]{Forcrand1}%
  \BibitemOpen
  \bibfield  {author} {\bibinfo {author} {\bibfnamefont {P.}~\bibnamefont
  {de~Forcrand}}\ and\ \bibinfo {author} {\bibfnamefont {O.}~\bibnamefont
  {Philipsen}},\ }\href {\doibase 10.1088/1126-6708/2007/01/077} {\bibfield
  {journal} {\bibinfo  {journal} {JHEP}\ }\textbf {\bibinfo {volume} {01}},\
  \bibinfo {pages} {077} (\bibinfo {year} {2007})},\ \Eprint
  {http://arxiv.org/abs/hep-lat/0607017} {arXiv:hep-lat/0607017 [hep-lat]}
  \BibitemShut {NoStop}%
\bibitem [{\citenamefont {de~Forcrand}\ and\ \citenamefont
  {Philipsen}(2008{\natexlab{a}})}]{Forcrand2}%
  \BibitemOpen
  \bibfield  {author} {\bibinfo {author} {\bibfnamefont {P.}~\bibnamefont
  {de~Forcrand}}\ and\ \bibinfo {author} {\bibfnamefont {O.}~\bibnamefont
  {Philipsen}},\ }\href {\doibase 10.1088/1126-6708/2008/11/012} {\bibfield
  {journal} {\bibinfo  {journal} {JHEP}\ }\textbf {\bibinfo {volume} {11}},\
  \bibinfo {pages} {012} (\bibinfo {year} {2008}{\natexlab{a}})},\ \Eprint
  {http://arxiv.org/abs/0808.1096} {arXiv:0808.1096 [hep-lat]} \BibitemShut
  {NoStop}%
\bibitem [{\citenamefont {de~Forcrand}\ and\ \citenamefont
  {Philipsen}(2008{\natexlab{b}})}]{Forcrand3}%
  \BibitemOpen
  \bibfield  {author} {\bibinfo {author} {\bibfnamefont {P.}~\bibnamefont
  {de~Forcrand}}\ and\ \bibinfo {author} {\bibfnamefont {O.}~\bibnamefont
  {Philipsen}},\ }\bibfield  {booktitle} {\emph {\bibinfo {booktitle}
  {{Proceedings, 26th International Symposium on Lattice field theory (Lattice
  2008): Williamsburg, USA, July 14-19, 2008}}},\ }\href {\doibase
  10.22323/1.066.0208} {\bibfield  {journal} {\bibinfo  {journal} {PoS}\
  }\textbf {\bibinfo {volume} {LATTICE2008}},\ \bibinfo {pages} {208} (\bibinfo
  {year} {2008}{\natexlab{b}})},\ \Eprint {http://arxiv.org/abs/0811.3858}
  {arXiv:0811.3858 [hep-lat]} \BibitemShut {NoStop}%
\bibitem [{\citenamefont {Brandt}\ \emph {et~al.}(2016)\citenamefont {Brandt},
  \citenamefont {Bali}, \citenamefont {Endrödi},\ and\ \citenamefont
  {Glässle}}]{Bali_lat}%
  \BibitemOpen
  \bibfield  {author} {\bibinfo {author} {\bibfnamefont {B.~B.}\ \bibnamefont
  {Brandt}}, \bibinfo {author} {\bibfnamefont {G.}~\bibnamefont {Bali}},
  \bibinfo {author} {\bibfnamefont {G.}~\bibnamefont {Endrödi}}, \ and\
  \bibinfo {author} {\bibfnamefont {B.}~\bibnamefont {Glässle}},\ }\bibfield
  {booktitle} {\emph {\bibinfo {booktitle} {{Proceedings, 33rd International
  Symposium on Lattice Field Theory (Lattice 2015): Kobe, Japan, July 14-18,
  2015}}},\ }\href {\doibase 10.22323/1.251.0265} {\bibfield  {journal}
  {\bibinfo  {journal} {PoS}\ }\textbf {\bibinfo {volume} {LATTICE2015}},\
  \bibinfo {pages} {265} (\bibinfo {year} {2016})},\ \Eprint
  {http://arxiv.org/abs/1510.03899} {arXiv:1510.03899 [hep-lat]} \BibitemShut
  {NoStop}%
\bibitem [{\citenamefont {Luschevskaya}\ and\ \citenamefont
  {Larina}(2014)}]{Luschevskaya}%
  \BibitemOpen
  \bibfield  {author} {\bibinfo {author} {\bibfnamefont {E.~V.}\ \bibnamefont
  {Luschevskaya}}\ and\ \bibinfo {author} {\bibfnamefont {O.~V.}\ \bibnamefont
  {Larina}},\ }\href {\doibase 10.1016/j.nuclphysb.2014.04.003} {\bibfield
  {journal} {\bibinfo  {journal} {Nucl. Phys.}\ }\textbf {\bibinfo {volume}
  {B884}},\ \bibinfo {pages} {1} (\bibinfo {year} {2014})},\ \Eprint
  {http://arxiv.org/abs/1203.5699} {arXiv:1203.5699 [hep-lat]} \BibitemShut
  {NoStop}%
\bibitem [{\citenamefont {Aoki}\ \emph
  {et~al.}(2006{\natexlab{a}})\citenamefont {Aoki}, \citenamefont {Endrodi},
  \citenamefont {Fodor}, \citenamefont {Katz},\ and\ \citenamefont
  {Szabo}}]{Aoki}%
  \BibitemOpen
  \bibfield  {author} {\bibinfo {author} {\bibfnamefont {Y.}~\bibnamefont
  {Aoki}}, \bibinfo {author} {\bibfnamefont {G.}~\bibnamefont {Endrodi}},
  \bibinfo {author} {\bibfnamefont {Z.}~\bibnamefont {Fodor}}, \bibinfo
  {author} {\bibfnamefont {S.~D.}\ \bibnamefont {Katz}}, \ and\ \bibinfo
  {author} {\bibfnamefont {K.~K.}\ \bibnamefont {Szabo}},\ }\href {\doibase
  10.1038/nature05120} {\bibfield  {journal} {\bibinfo  {journal} {Nature}\
  }\textbf {\bibinfo {volume} {443}},\ \bibinfo {pages} {675} (\bibinfo {year}
  {2006}{\natexlab{a}})},\ \Eprint {http://arxiv.org/abs/hep-lat/0611014}
  {arXiv:hep-lat/0611014 [hep-lat]} \BibitemShut {NoStop}%
\bibitem [{\citenamefont {Aoki}\ \emph
  {et~al.}(2006{\natexlab{b}})\citenamefont {Aoki}, \citenamefont {Endrodi},
  \citenamefont {Fodor}, \citenamefont {Katz},\ and\ \citenamefont
  {Szabo}}]{Aoki2}%
  \BibitemOpen
  \bibfield  {author} {\bibinfo {author} {\bibfnamefont {Y.}~\bibnamefont
  {Aoki}}, \bibinfo {author} {\bibfnamefont {G.}~\bibnamefont {Endrodi}},
  \bibinfo {author} {\bibfnamefont {Z.}~\bibnamefont {Fodor}}, \bibinfo
  {author} {\bibfnamefont {S.~D.}\ \bibnamefont {Katz}}, \ and\ \bibinfo
  {author} {\bibfnamefont {K.~K.}\ \bibnamefont {Szabo}},\ }\href {\doibase
  10.1038/nature05120} {\bibfield  {journal} {\bibinfo  {journal} {Nature}\
  }\textbf {\bibinfo {volume} {443}},\ \bibinfo {pages} {675} (\bibinfo {year}
  {2006}{\natexlab{b}})},\ \Eprint {http://arxiv.org/abs/hep-lat/0611014}
  {arXiv:hep-lat/0611014 [hep-lat]} \BibitemShut {NoStop}%
\bibitem [{\citenamefont {Nambu}\ and\ \citenamefont
  {Jona-Lasinio}(1961{\natexlab{a}})}]{Nambu1}%
  \BibitemOpen
  \bibfield  {author} {\bibinfo {author} {\bibfnamefont {Y.}~\bibnamefont
  {Nambu}}\ and\ \bibinfo {author} {\bibfnamefont {G.}~\bibnamefont
  {Jona-Lasinio}},\ }\href {\doibase 10.1103/PhysRev.124.246} {\bibfield
  {journal} {\bibinfo  {journal} {Phys. Rev.}\ }\textbf {\bibinfo {volume}
  {124}},\ \bibinfo {pages} {246} (\bibinfo {year} {1961}{\natexlab{a}})},\
  \bibinfo {note} {[,141(1961)]}\BibitemShut {NoStop}%
\bibitem [{\citenamefont {Nambu}\ and\ \citenamefont
  {Jona-Lasinio}(1961{\natexlab{b}})}]{Nambu2}%
  \BibitemOpen
  \bibfield  {author} {\bibinfo {author} {\bibfnamefont {Y.}~\bibnamefont
  {Nambu}}\ and\ \bibinfo {author} {\bibfnamefont {G.}~\bibnamefont
  {Jona-Lasinio}},\ }\href {\doibase 10.1103/PhysRev.122.345} {\bibfield
  {journal} {\bibinfo  {journal} {Phys. Rev.}\ }\textbf {\bibinfo {volume}
  {122}},\ \bibinfo {pages} {345} (\bibinfo {year} {1961}{\natexlab{b}})},\
  \bibinfo {note} {[,127(1961)]}\BibitemShut {NoStop}%
\bibitem [{\citenamefont {Klevansky}(1992)}]{Klevansky}%
  \BibitemOpen
  \bibfield  {author} {\bibinfo {author} {\bibfnamefont {S.~P.}\ \bibnamefont
  {Klevansky}},\ }\href {\doibase 10.1103/RevModPhys.64.649} {\bibfield
  {journal} {\bibinfo  {journal} {Rev. Mod. Phys.}\ }\textbf {\bibinfo {volume}
  {64}},\ \bibinfo {pages} {649} (\bibinfo {year} {1992})}\BibitemShut
  {NoStop}%
\bibitem [{\citenamefont {Hatsuda}\ and\ \citenamefont
  {Kunihiro}(1994)}]{Hatsuda1}%
  \BibitemOpen
  \bibfield  {author} {\bibinfo {author} {\bibfnamefont {T.}~\bibnamefont
  {Hatsuda}}\ and\ \bibinfo {author} {\bibfnamefont {T.}~\bibnamefont
  {Kunihiro}},\ }\href {\doibase 10.1016/0370-1573(94)90022-1} {\bibfield
  {journal} {\bibinfo  {journal} {Phys. Rept.}\ }\textbf {\bibinfo {volume}
  {247}},\ \bibinfo {pages} {221} (\bibinfo {year} {1994})},\ \Eprint
  {http://arxiv.org/abs/hep-ph/9401310} {arXiv:hep-ph/9401310 [hep-ph]}
  \BibitemShut {NoStop}%
\bibitem [{\citenamefont {Vogl}\ and\ \citenamefont {Weise}(1991)}]{Vogl}%
  \BibitemOpen
  \bibfield  {author} {\bibinfo {author} {\bibfnamefont {U.}~\bibnamefont
  {Vogl}}\ and\ \bibinfo {author} {\bibfnamefont {W.}~\bibnamefont {Weise}},\
  }\href {\doibase 10.1016/0146-6410(91)90005-9} {\bibfield  {journal}
  {\bibinfo  {journal} {Prog. Part. Nucl. Phys.}\ }\textbf {\bibinfo {volume}
  {27}},\ \bibinfo {pages} {195} (\bibinfo {year} {1991})}\BibitemShut
  {NoStop}%
\bibitem [{\citenamefont {Buballa}(2005)}]{Buballa}%
  \BibitemOpen
  \bibfield  {author} {\bibinfo {author} {\bibfnamefont {M.}~\bibnamefont
  {Buballa}},\ }\href {\doibase 10.1016/j.physrep.2004.11.004} {\bibfield
  {journal} {\bibinfo  {journal} {Phys. Rept.}\ }\textbf {\bibinfo {volume}
  {407}},\ \bibinfo {pages} {205} (\bibinfo {year} {2005})},\ \Eprint
  {http://arxiv.org/abs/hep-ph/0402234} {arXiv:hep-ph/0402234 [hep-ph]}
  \BibitemShut {NoStop}%
\bibitem [{\citenamefont {Reinders}\ \emph {et~al.}(1985)\citenamefont
  {Reinders}, \citenamefont {Rubinstein},\ and\ \citenamefont
  {Yazaki}}]{Reinder}%
  \BibitemOpen
  \bibfield  {author} {\bibinfo {author} {\bibfnamefont {L.~J.}\ \bibnamefont
  {Reinders}}, \bibinfo {author} {\bibfnamefont {H.}~\bibnamefont
  {Rubinstein}}, \ and\ \bibinfo {author} {\bibfnamefont {S.}~\bibnamefont
  {Yazaki}},\ }\href {\doibase 10.1016/0370-1573(85)90065-1} {\bibfield
  {journal} {\bibinfo  {journal} {Phys. Rept.}\ }\textbf {\bibinfo {volume}
  {127}},\ \bibinfo {pages} {1} (\bibinfo {year} {1985})}\BibitemShut {NoStop}%
\bibitem [{\citenamefont {Klevansky}\ and\ \citenamefont
  {Lemmer}(1989)}]{Klevansky2}%
  \BibitemOpen
  \bibfield  {author} {\bibinfo {author} {\bibfnamefont {S.~P.}\ \bibnamefont
  {Klevansky}}\ and\ \bibinfo {author} {\bibfnamefont {R.~H.}\ \bibnamefont
  {Lemmer}},\ }\href {\doibase 10.1103/PhysRevD.39.3478} {\bibfield  {journal}
  {\bibinfo  {journal} {Phys. Rev.}\ }\textbf {\bibinfo {volume} {D39}},\
  \bibinfo {pages} {3478} (\bibinfo {year} {1989})}\BibitemShut {NoStop}%
\bibitem [{\citenamefont {Mao}(2016)}]{Mao}%
  \BibitemOpen
  \bibfield  {author} {\bibinfo {author} {\bibfnamefont {S.}~\bibnamefont
  {Mao}},\ }\href {\doibase 10.1016/j.physletb.2016.05.018} {\bibfield
  {journal} {\bibinfo  {journal} {Phys. Lett.}\ }\textbf {\bibinfo {volume}
  {B758}},\ \bibinfo {pages} {195} (\bibinfo {year} {2016})},\ \Eprint
  {http://arxiv.org/abs/1602.06503} {arXiv:1602.06503 [hep-ph]} \BibitemShut
  {NoStop}%
\bibitem [{\citenamefont {Fayazbakhsh}\ \emph {et~al.}(2012)\citenamefont
  {Fayazbakhsh}, \citenamefont {Sadeghian},\ and\ \citenamefont
  {Sadooghi}}]{Sadooghi1}%
  \BibitemOpen
  \bibfield  {author} {\bibinfo {author} {\bibfnamefont {S.}~\bibnamefont
  {Fayazbakhsh}}, \bibinfo {author} {\bibfnamefont {S.}~\bibnamefont
  {Sadeghian}}, \ and\ \bibinfo {author} {\bibfnamefont {N.}~\bibnamefont
  {Sadooghi}},\ }\href {\doibase 10.1103/PhysRevD.86.085042} {\bibfield
  {journal} {\bibinfo  {journal} {Phys. Rev.}\ }\textbf {\bibinfo {volume}
  {D86}},\ \bibinfo {pages} {085042} (\bibinfo {year} {2012})},\ \Eprint
  {http://arxiv.org/abs/1206.6051} {arXiv:1206.6051 [hep-ph]} \BibitemShut
  {NoStop}%
\bibitem [{\citenamefont {Ruggieri}\ \emph {et~al.}(2013)\citenamefont
  {Ruggieri}, \citenamefont {Tachibana},\ and\ \citenamefont
  {Greco}}]{Ruggieri}%
  \BibitemOpen
  \bibfield  {author} {\bibinfo {author} {\bibfnamefont {M.}~\bibnamefont
  {Ruggieri}}, \bibinfo {author} {\bibfnamefont {M.}~\bibnamefont {Tachibana}},
  \ and\ \bibinfo {author} {\bibfnamefont {V.}~\bibnamefont {Greco}},\ }\href
  {\doibase 10.1007/JHEP07(2013)165} {\bibfield  {journal} {\bibinfo  {journal}
  {JHEP}\ }\textbf {\bibinfo {volume} {07}},\ \bibinfo {pages} {165} (\bibinfo
  {year} {2013})},\ \Eprint {http://arxiv.org/abs/1305.0137} {arXiv:1305.0137
  [hep-ph]} \BibitemShut {NoStop}%
\bibitem [{\citenamefont {Bali}\ \emph {et~al.}(2014)\citenamefont {Bali},
  \citenamefont {Bruckmann}, \citenamefont {Endrödi}, \citenamefont {Katz},\
  and\ \citenamefont {Schäfer}}]{Bali_lat2}%
  \BibitemOpen
  \bibfield  {author} {\bibinfo {author} {\bibfnamefont {G.~S.}\ \bibnamefont
  {Bali}}, \bibinfo {author} {\bibfnamefont {F.}~\bibnamefont {Bruckmann}},
  \bibinfo {author} {\bibfnamefont {G.}~\bibnamefont {Endrödi}}, \bibinfo
  {author} {\bibfnamefont {S.~D.}\ \bibnamefont {Katz}}, \ and\ \bibinfo
  {author} {\bibfnamefont {A.}~\bibnamefont {Schäfer}},\ }\href {\doibase
  10.1007/JHEP08(2014)177} {\bibfield  {journal} {\bibinfo  {journal} {JHEP}\
  }\textbf {\bibinfo {volume} {08}},\ \bibinfo {pages} {177} (\bibinfo {year}
  {2014})},\ \Eprint {http://arxiv.org/abs/1406.0269} {arXiv:1406.0269
  [hep-lat]} \BibitemShut {NoStop}%
\bibitem [{\citenamefont {Bali}\ \emph
  {et~al.}(2012{\natexlab{b}})\citenamefont {Bali}, \citenamefont {Bruckmann},
  \citenamefont {Endrodi}, \citenamefont {Fodor}, \citenamefont {Katz},\ and\
  \citenamefont {Schafer}}]{Bali_lat3}%
  \BibitemOpen
  \bibfield  {author} {\bibinfo {author} {\bibfnamefont {G.~S.}\ \bibnamefont
  {Bali}}, \bibinfo {author} {\bibfnamefont {F.}~\bibnamefont {Bruckmann}},
  \bibinfo {author} {\bibfnamefont {G.}~\bibnamefont {Endrodi}}, \bibinfo
  {author} {\bibfnamefont {Z.}~\bibnamefont {Fodor}}, \bibinfo {author}
  {\bibfnamefont {S.~D.}\ \bibnamefont {Katz}}, \ and\ \bibinfo {author}
  {\bibfnamefont {A.}~\bibnamefont {Schafer}},\ }\href {\doibase
  10.1103/PhysRevD.86.071502} {\bibfield  {journal} {\bibinfo  {journal} {Phys.
  Rev.}\ }\textbf {\bibinfo {volume} {D86}},\ \bibinfo {pages} {071502}
  (\bibinfo {year} {2012}{\natexlab{b}})},\ \Eprint
  {http://arxiv.org/abs/1206.4205} {arXiv:1206.4205 [hep-lat]} \BibitemShut
  {NoStop}%
\bibitem [{\citenamefont {Bornyakov}\ \emph {et~al.}(2014)\citenamefont
  {Bornyakov}, \citenamefont {Buividovich}, \citenamefont {Cundy},
  \citenamefont {Kochetkov},\ and\ \citenamefont {Schäfer}}]{Bornyakov}%
  \BibitemOpen
  \bibfield  {author} {\bibinfo {author} {\bibfnamefont {V.~G.}\ \bibnamefont
  {Bornyakov}}, \bibinfo {author} {\bibfnamefont {P.~V.}\ \bibnamefont
  {Buividovich}}, \bibinfo {author} {\bibfnamefont {N.}~\bibnamefont {Cundy}},
  \bibinfo {author} {\bibfnamefont {O.~A.}\ \bibnamefont {Kochetkov}}, \ and\
  \bibinfo {author} {\bibfnamefont {A.}~\bibnamefont {Schäfer}},\ }\href
  {\doibase 10.1103/PhysRevD.90.034501} {\bibfield  {journal} {\bibinfo
  {journal} {Phys. Rev.}\ }\textbf {\bibinfo {volume} {D90}},\ \bibinfo {pages}
  {034501} (\bibinfo {year} {2014})},\ \Eprint {http://arxiv.org/abs/1312.5628}
  {arXiv:1312.5628 [hep-lat]} \BibitemShut {NoStop}%
\bibitem [{\citenamefont {Ruggieri}\ \emph {et~al.}(2014)\citenamefont
  {Ruggieri}, \citenamefont {Oliva}, \citenamefont {Castorina}, \citenamefont
  {Gatto},\ and\ \citenamefont {Greco}}]{Ruggieri2}%
  \BibitemOpen
  \bibfield  {author} {\bibinfo {author} {\bibfnamefont {M.}~\bibnamefont
  {Ruggieri}}, \bibinfo {author} {\bibfnamefont {L.}~\bibnamefont {Oliva}},
  \bibinfo {author} {\bibfnamefont {P.}~\bibnamefont {Castorina}}, \bibinfo
  {author} {\bibfnamefont {R.}~\bibnamefont {Gatto}}, \ and\ \bibinfo {author}
  {\bibfnamefont {V.}~\bibnamefont {Greco}},\ }\href {\doibase
  10.1016/j.physletb.2014.05.073} {\bibfield  {journal} {\bibinfo  {journal}
  {Phys. Lett.}\ }\textbf {\bibinfo {volume} {B734}},\ \bibinfo {pages} {255}
  (\bibinfo {year} {2014})},\ \Eprint {http://arxiv.org/abs/1402.0737}
  {arXiv:1402.0737 [hep-ph]} \BibitemShut {NoStop}%
\bibitem [{\citenamefont {Andersen}\ \emph {et~al.}(2016)\citenamefont
  {Andersen}, \citenamefont {Naylor},\ and\ \citenamefont
  {Tranberg}}]{Andersen}%
  \BibitemOpen
  \bibfield  {author} {\bibinfo {author} {\bibfnamefont {J.~O.}\ \bibnamefont
  {Andersen}}, \bibinfo {author} {\bibfnamefont {W.~R.}\ \bibnamefont
  {Naylor}}, \ and\ \bibinfo {author} {\bibfnamefont {A.}~\bibnamefont
  {Tranberg}},\ }\href {\doibase 10.1103/RevModPhys.88.025001} {\bibfield
  {journal} {\bibinfo  {journal} {Rev. Mod. Phys.}\ }\textbf {\bibinfo {volume}
  {88}},\ \bibinfo {pages} {025001} (\bibinfo {year} {2016})},\ \Eprint
  {http://arxiv.org/abs/1411.7176} {arXiv:1411.7176 [hep-ph]} \BibitemShut
  {NoStop}%
\bibitem [{\citenamefont {Ayala}\ \emph {et~al.}(2016)\citenamefont {Ayala},
  \citenamefont {Dominguez}, \citenamefont {Hernandez}, \citenamefont {Loewe},\
  and\ \citenamefont {Zamora}}]{Ayala2}%
  \BibitemOpen
  \bibfield  {author} {\bibinfo {author} {\bibfnamefont {A.}~\bibnamefont
  {Ayala}}, \bibinfo {author} {\bibfnamefont {C.~A.}\ \bibnamefont
  {Dominguez}}, \bibinfo {author} {\bibfnamefont {L.~A.}\ \bibnamefont
  {Hernandez}}, \bibinfo {author} {\bibfnamefont {M.}~\bibnamefont {Loewe}}, \
  and\ \bibinfo {author} {\bibfnamefont {R.}~\bibnamefont {Zamora}},\ }\href
  {\doibase 10.1016/j.physletb.2016.05.058} {\bibfield  {journal} {\bibinfo
  {journal} {Phys. Lett.}\ }\textbf {\bibinfo {volume} {B759}},\ \bibinfo
  {pages} {99} (\bibinfo {year} {2016})},\ \Eprint
  {http://arxiv.org/abs/1510.09134} {arXiv:1510.09134 [hep-ph]} \BibitemShut
  {NoStop}%
\bibitem [{\citenamefont {Strickland}\ \emph {et~al.}(2012)\citenamefont
  {Strickland}, \citenamefont {Dexheimer},\ and\ \citenamefont
  {Menezes}}]{Strickland}%
  \BibitemOpen
  \bibfield  {author} {\bibinfo {author} {\bibfnamefont {M.}~\bibnamefont
  {Strickland}}, \bibinfo {author} {\bibfnamefont {V.}~\bibnamefont
  {Dexheimer}}, \ and\ \bibinfo {author} {\bibfnamefont {D.~P.}\ \bibnamefont
  {Menezes}},\ }\href {\doibase 10.1103/PhysRevD.86.125032} {\bibfield
  {journal} {\bibinfo  {journal} {Phys. Rev.}\ }\textbf {\bibinfo {volume}
  {D86}},\ \bibinfo {pages} {125032} (\bibinfo {year} {2012})},\ \Eprint
  {http://arxiv.org/abs/1209.3276} {arXiv:1209.3276 [nucl-th]} \BibitemShut
  {NoStop}%
\bibitem [{\citenamefont {Mukherjee}\ \emph {et~al.}(2018)\citenamefont
  {Mukherjee}, \citenamefont {Ghosh}, \citenamefont {Mandal}, \citenamefont
  {Sarkar},\ and\ \citenamefont {Roy}}]{Arghya}%
  \BibitemOpen
  \bibfield  {author} {\bibinfo {author} {\bibfnamefont {A.}~\bibnamefont
  {Mukherjee}}, \bibinfo {author} {\bibfnamefont {S.}~\bibnamefont {Ghosh}},
  \bibinfo {author} {\bibfnamefont {M.}~\bibnamefont {Mandal}}, \bibinfo
  {author} {\bibfnamefont {S.}~\bibnamefont {Sarkar}}, \ and\ \bibinfo {author}
  {\bibfnamefont {P.}~\bibnamefont {Roy}},\ }\href {\doibase
  10.1103/PhysRevD.98.056024} {\bibfield  {journal} {\bibinfo  {journal} {Phys.
  Rev.}\ }\textbf {\bibinfo {volume} {D98}},\ \bibinfo {pages} {056024}
  (\bibinfo {year} {2018})},\ \Eprint {http://arxiv.org/abs/1809.07028}
  {arXiv:1809.07028 [hep-ph]} \BibitemShut {NoStop}%
\bibitem [{\citenamefont {Bicudo}\ \emph {et~al.}(1999)\citenamefont {Bicudo},
  \citenamefont {Ribeiro},\ and\ \citenamefont {Fernandes}}]{Pedro}%
  \BibitemOpen
  \bibfield  {author} {\bibinfo {author} {\bibfnamefont {P.~J.~A.}\
  \bibnamefont {Bicudo}}, \bibinfo {author} {\bibfnamefont {J.~E. F.~T.}\
  \bibnamefont {Ribeiro}}, \ and\ \bibinfo {author} {\bibfnamefont
  {R.}~\bibnamefont {Fernandes}},\ }\href {\doibase 10.1103/PhysRevC.59.1107}
  {\bibfield  {journal} {\bibinfo  {journal} {Phys. Rev.}\ }\textbf {\bibinfo
  {volume} {C59}},\ \bibinfo {pages} {1107} (\bibinfo {year} {1999})},\ \Eprint
  {http://arxiv.org/abs/hep-ph/9806243} {arXiv:hep-ph/9806243 [hep-ph]}
  \BibitemShut {NoStop}%
\bibitem [{\citenamefont {Chang}\ \emph {et~al.}(2011)\citenamefont {Chang},
  \citenamefont {Liu},\ and\ \citenamefont {Roberts}}]{Chang}%
  \BibitemOpen
  \bibfield  {author} {\bibinfo {author} {\bibfnamefont {L.}~\bibnamefont
  {Chang}}, \bibinfo {author} {\bibfnamefont {Y.-X.}\ \bibnamefont {Liu}}, \
  and\ \bibinfo {author} {\bibfnamefont {C.~D.}\ \bibnamefont {Roberts}},\
  }\href {\doibase 10.1103/PhysRevLett.106.072001} {\bibfield  {journal}
  {\bibinfo  {journal} {Phys. Rev. Lett.}\ }\textbf {\bibinfo {volume} {106}},\
  \bibinfo {pages} {072001} (\bibinfo {year} {2011})},\ \Eprint
  {http://arxiv.org/abs/1009.3458} {arXiv:1009.3458 [nucl-th]} \BibitemShut
  {NoStop}%
\bibitem [{\citenamefont {Fayazbakhsh}\ and\ \citenamefont
  {Sadooghi}(2014)}]{Sadooghi}%
  \BibitemOpen
  \bibfield  {author} {\bibinfo {author} {\bibfnamefont {S.}~\bibnamefont
  {Fayazbakhsh}}\ and\ \bibinfo {author} {\bibfnamefont {N.}~\bibnamefont
  {Sadooghi}},\ }\href {\doibase 10.1103/PhysRevD.90.105030} {\bibfield
  {journal} {\bibinfo  {journal} {Phys. Rev.}\ }\textbf {\bibinfo {volume}
  {D90}},\ \bibinfo {pages} {105030} (\bibinfo {year} {2014})},\ \Eprint
  {http://arxiv.org/abs/1408.5457} {arXiv:1408.5457 [hep-ph]} \BibitemShut
  {NoStop}%
\bibitem [{\citenamefont {Hatsuda}\ and\ \citenamefont
  {Kunihiro}(1985)}]{HatsudaPRL}%
  \BibitemOpen
  \bibfield  {author} {\bibinfo {author} {\bibfnamefont {T.}~\bibnamefont
  {Hatsuda}}\ and\ \bibinfo {author} {\bibfnamefont {T.}~\bibnamefont
  {Kunihiro}},\ }\href {\doibase 10.1103/PhysRevLett.55.158} {\bibfield
  {journal} {\bibinfo  {journal} {Phys. Rev. Lett.}\ }\textbf {\bibinfo
  {volume} {55}},\ \bibinfo {pages} {158} (\bibinfo {year} {1985})}\BibitemShut
  {NoStop}%
\bibitem [{\citenamefont {Shuryak}(1990)}]{Shuryak}%
  \BibitemOpen
  \bibfield  {author} {\bibinfo {author} {\bibfnamefont {E.~V.}\ \bibnamefont
  {Shuryak}},\ }\bibfield  {booktitle} {\emph {\bibinfo {booktitle} {{Workshop
  on Heavy Ion Physics at the AGS Upton, New York, March 3-4, 1990}}},\ }\href
  {\doibase 10.1103/PhysRevD.42.1764} {\bibfield  {journal} {\bibinfo
  {journal} {Phys. Rev.}\ }\textbf {\bibinfo {volume} {D42}},\ \bibinfo {pages}
  {1764} (\bibinfo {year} {1990})}\BibitemShut {NoStop}%
\bibitem [{\citenamefont {Ayala}\ \emph {et~al.}(2002)\citenamefont {Ayala},
  \citenamefont {Amore},\ and\ \citenamefont {Aranda}}]{Ayala}%
  \BibitemOpen
  \bibfield  {author} {\bibinfo {author} {\bibfnamefont {A.}~\bibnamefont
  {Ayala}}, \bibinfo {author} {\bibfnamefont {P.}~\bibnamefont {Amore}}, \ and\
  \bibinfo {author} {\bibfnamefont {A.}~\bibnamefont {Aranda}},\ }\href
  {\doibase 10.1103/PhysRevC.66.045205} {\bibfield  {journal} {\bibinfo
  {journal} {Phys. Rev.}\ }\textbf {\bibinfo {volume} {C66}},\ \bibinfo {pages}
  {045205} (\bibinfo {year} {2002})},\ \Eprint
  {http://arxiv.org/abs/hep-ph/0207081} {arXiv:hep-ph/0207081 [hep-ph]}
  \BibitemShut {NoStop}%
\bibitem [{\citenamefont {Inagaki}\ \emph {et~al.}(2008)\citenamefont
  {Inagaki}, \citenamefont {Kimura},\ and\ \citenamefont
  {Kvinikhidze}}]{Inagaki}%
  \BibitemOpen
  \bibfield  {author} {\bibinfo {author} {\bibfnamefont {T.}~\bibnamefont
  {Inagaki}}, \bibinfo {author} {\bibfnamefont {D.}~\bibnamefont {Kimura}}, \
  and\ \bibinfo {author} {\bibfnamefont {A.}~\bibnamefont {Kvinikhidze}},\
  }\href {\doibase 10.1103/PhysRevD.77.116004} {\bibfield  {journal} {\bibinfo
  {journal} {Phys. Rev.}\ }\textbf {\bibinfo {volume} {D77}},\ \bibinfo {pages}
  {116004} (\bibinfo {year} {2008})},\ \Eprint {http://arxiv.org/abs/0712.1336}
  {arXiv:0712.1336 [hep-ph]} \BibitemShut {NoStop}%
\bibitem [{\citenamefont {Hansen}\ \emph {et~al.}(2007)\citenamefont {Hansen},
  \citenamefont {Alberico}, \citenamefont {Beraudo}, \citenamefont {Molinari},
  \citenamefont {Nardi},\ and\ \citenamefont {Ratti}}]{Hansen}%
  \BibitemOpen
  \bibfield  {author} {\bibinfo {author} {\bibfnamefont {H.}~\bibnamefont
  {Hansen}}, \bibinfo {author} {\bibfnamefont {W.~M.}\ \bibnamefont
  {Alberico}}, \bibinfo {author} {\bibfnamefont {A.}~\bibnamefont {Beraudo}},
  \bibinfo {author} {\bibfnamefont {A.}~\bibnamefont {Molinari}}, \bibinfo
  {author} {\bibfnamefont {M.}~\bibnamefont {Nardi}}, \ and\ \bibinfo {author}
  {\bibfnamefont {C.}~\bibnamefont {Ratti}},\ }\href {\doibase
  10.1103/PhysRevD.75.065004} {\bibfield  {journal} {\bibinfo  {journal} {Phys.
  Rev.}\ }\textbf {\bibinfo {volume} {D75}},\ \bibinfo {pages} {065004}
  (\bibinfo {year} {2007})},\ \Eprint {http://arxiv.org/abs/hep-ph/0609116}
  {arXiv:hep-ph/0609116 [hep-ph]} \BibitemShut {NoStop}%
\bibitem [{\citenamefont {Zhang}\ \emph {et~al.}(2016)\citenamefont {Zhang},
  \citenamefont {Fu},\ and\ \citenamefont {Liu}}]{Zhang}%
  \BibitemOpen
  \bibfield  {author} {\bibinfo {author} {\bibfnamefont {R.}~\bibnamefont
  {Zhang}}, \bibinfo {author} {\bibfnamefont {W.-j.}\ \bibnamefont {Fu}}, \
  and\ \bibinfo {author} {\bibfnamefont {Y.-x.}\ \bibnamefont {Liu}},\ }\href
  {\doibase 10.1140/epjc/s10052-016-4123-8} {\bibfield  {journal} {\bibinfo
  {journal} {Eur. Phys. J.}\ }\textbf {\bibinfo {volume} {C76}},\ \bibinfo
  {pages} {307} (\bibinfo {year} {2016})},\ \Eprint
  {http://arxiv.org/abs/1604.08888} {arXiv:1604.08888 [hep-ph]} \BibitemShut
  {NoStop}%
\bibitem [{\citenamefont {Mao}(2019)}]{Mao_pi}%
  \BibitemOpen
  \bibfield  {author} {\bibinfo {author} {\bibfnamefont {S.}~\bibnamefont
  {Mao}},\ }\href {\doibase 10.1103/PhysRevD.99.056005} {\bibfield  {journal}
  {\bibinfo  {journal} {Phys. Rev.}\ }\textbf {\bibinfo {volume} {D99}},\
  \bibinfo {pages} {056005} (\bibinfo {year} {2019})},\ \Eprint
  {http://arxiv.org/abs/1808.10242} {arXiv:1808.10242 [nucl-th]} \BibitemShut
  {NoStop}%
\bibitem [{\citenamefont {Avancini}\ \emph
  {et~al.}(2019{\natexlab{a}})\citenamefont {Avancini}, \citenamefont
  {Farias},\ and\ \citenamefont {Tavares}}]{Avancini}%
  \BibitemOpen
  \bibfield  {author} {\bibinfo {author} {\bibfnamefont {S.~S.}\ \bibnamefont
  {Avancini}}, \bibinfo {author} {\bibfnamefont {R.~L.~S.}\ \bibnamefont
  {Farias}}, \ and\ \bibinfo {author} {\bibfnamefont {W.~R.}\ \bibnamefont
  {Tavares}},\ }\href {\doibase 10.1103/PhysRevD.99.056009} {\bibfield
  {journal} {\bibinfo  {journal} {Phys. Rev.}\ }\textbf {\bibinfo {volume}
  {D99}},\ \bibinfo {pages} {056009} (\bibinfo {year} {2019}{\natexlab{a}})},\
  \Eprint {http://arxiv.org/abs/1812.00945} {arXiv:1812.00945 [hep-ph]}
  \BibitemShut {NoStop}%
\bibitem [{\citenamefont {Pushkina}\ \emph {et~al.}(2005)\citenamefont
  {Pushkina}, \citenamefont {de~Forcrand}, \citenamefont {Garcia~Perez},
  \citenamefont {Kim}, \citenamefont {Matsufuru}, \citenamefont {Nakamura},
  \citenamefont {Stamatescu}, \citenamefont {Takaishi},\ and\ \citenamefont
  {Umeda}}]{Pushkina}%
  \BibitemOpen
  \bibfield  {author} {\bibinfo {author} {\bibfnamefont {I.}~\bibnamefont
  {Pushkina}}, \bibinfo {author} {\bibfnamefont {P.}~\bibnamefont
  {de~Forcrand}}, \bibinfo {author} {\bibfnamefont {M.}~\bibnamefont
  {Garcia~Perez}}, \bibinfo {author} {\bibfnamefont {S.}~\bibnamefont {Kim}},
  \bibinfo {author} {\bibfnamefont {H.}~\bibnamefont {Matsufuru}}, \bibinfo
  {author} {\bibfnamefont {A.}~\bibnamefont {Nakamura}}, \bibinfo {author}
  {\bibfnamefont {I.-O.}\ \bibnamefont {Stamatescu}}, \bibinfo {author}
  {\bibfnamefont {T.}~\bibnamefont {Takaishi}}, \ and\ \bibinfo {author}
  {\bibfnamefont {T.}~\bibnamefont {Umeda}} (\bibinfo {collaboration}
  {QCD-TARO}),\ }\href {\doibase 10.1016/j.physletb.2005.01.006} {\bibfield
  {journal} {\bibinfo  {journal} {Phys. Lett.}\ }\textbf {\bibinfo {volume}
  {B609}},\ \bibinfo {pages} {265} (\bibinfo {year} {2005})},\ \Eprint
  {http://arxiv.org/abs/hep-lat/0410017} {arXiv:hep-lat/0410017 [hep-lat]}
  \BibitemShut {NoStop}%
\bibitem [{\citenamefont {Wissel}\ \emph {et~al.}(2006)\citenamefont {Wissel},
  \citenamefont {Laermann}, \citenamefont {Shcheredin}, \citenamefont {Datta},\
  and\ \citenamefont {Karsch}}]{Wissel}%
  \BibitemOpen
  \bibfield  {author} {\bibinfo {author} {\bibfnamefont {S.}~\bibnamefont
  {Wissel}}, \bibinfo {author} {\bibfnamefont {E.}~\bibnamefont {Laermann}},
  \bibinfo {author} {\bibfnamefont {S.}~\bibnamefont {Shcheredin}}, \bibinfo
  {author} {\bibfnamefont {S.}~\bibnamefont {Datta}}, \ and\ \bibinfo {author}
  {\bibfnamefont {F.}~\bibnamefont {Karsch}},\ }\href {\doibase
  10.22323/1.020.0164} {\bibfield  {journal} {\bibinfo  {journal} {PoS}\
  }\textbf {\bibinfo {volume} {LAT2005}},\ \bibinfo {pages} {164} (\bibinfo
  {year} {2006})},\ \Eprint {http://arxiv.org/abs/hep-lat/0510031}
  {arXiv:hep-lat/0510031 [hep-lat]} \BibitemShut {NoStop}%
\bibitem [{\citenamefont {Hansson}\ and\ \citenamefont
  {Zahed}(1992)}]{Hansson}%
  \BibitemOpen
  \bibfield  {author} {\bibinfo {author} {\bibfnamefont {T.~H.}\ \bibnamefont
  {Hansson}}\ and\ \bibinfo {author} {\bibfnamefont {I.}~\bibnamefont
  {Zahed}},\ }\href {\doibase 10.1016/0550-3213(92)90353-D} {\bibfield
  {journal} {\bibinfo  {journal} {Nucl. Phys.}\ }\textbf {\bibinfo {volume}
  {B374}},\ \bibinfo {pages} {277} (\bibinfo {year} {1992})}\BibitemShut
  {NoStop}%
\bibitem [{\citenamefont {Laine}\ and\ \citenamefont
  {Vepsalainen}(2004)}]{Laine}%
  \BibitemOpen
  \bibfield  {author} {\bibinfo {author} {\bibfnamefont {M.}~\bibnamefont
  {Laine}}\ and\ \bibinfo {author} {\bibfnamefont {M.}~\bibnamefont
  {Vepsalainen}},\ }\href {\doibase 10.1088/1126-6708/2004/02/004} {\bibfield
  {journal} {\bibinfo  {journal} {JHEP}\ }\textbf {\bibinfo {volume} {02}},\
  \bibinfo {pages} {004} (\bibinfo {year} {2004})},\ \Eprint
  {http://arxiv.org/abs/hep-ph/0311268} {arXiv:hep-ph/0311268 [hep-ph]}
  \BibitemShut {NoStop}%
\bibitem [{\citenamefont {Alberico}\ \emph {et~al.}(2005)\citenamefont
  {Alberico}, \citenamefont {Beraudo},\ and\ \citenamefont
  {Molinari}}]{Alberico}%
  \BibitemOpen
  \bibfield  {author} {\bibinfo {author} {\bibfnamefont {W.~M.}\ \bibnamefont
  {Alberico}}, \bibinfo {author} {\bibfnamefont {A.}~\bibnamefont {Beraudo}}, \
  and\ \bibinfo {author} {\bibfnamefont {A.}~\bibnamefont {Molinari}},\ }\href
  {\doibase 10.1016/j.nuclphysa.2004.12.070} {\bibfield  {journal} {\bibinfo
  {journal} {Nucl. Phys.}\ }\textbf {\bibinfo {volume} {A750}},\ \bibinfo
  {pages} {359} (\bibinfo {year} {2005})},\ \Eprint
  {http://arxiv.org/abs/hep-ph/0411346} {arXiv:hep-ph/0411346 [hep-ph]}
  \BibitemShut {NoStop}%
\bibitem [{\citenamefont {Alberico}\ \emph {et~al.}(2006)\citenamefont
  {Alberico}, \citenamefont {Beraudo}, \citenamefont {Czerski},\ and\
  \citenamefont {Molinari}}]{Alberico2}%
  \BibitemOpen
  \bibfield  {author} {\bibinfo {author} {\bibfnamefont {W.~M.}\ \bibnamefont
  {Alberico}}, \bibinfo {author} {\bibfnamefont {A.}~\bibnamefont {Beraudo}},
  \bibinfo {author} {\bibfnamefont {P.}~\bibnamefont {Czerski}}, \ and\
  \bibinfo {author} {\bibfnamefont {A.}~\bibnamefont {Molinari}},\ }\href
  {\doibase 10.1016/j.nuclphysa.2006.06.006} {\bibfield  {journal} {\bibinfo
  {journal} {Nucl. Phys.}\ }\textbf {\bibinfo {volume} {A775}},\ \bibinfo
  {pages} {188} (\bibinfo {year} {2006})},\ \Eprint
  {http://arxiv.org/abs/hep-ph/0605060} {arXiv:hep-ph/0605060 [hep-ph]}
  \BibitemShut {NoStop}%
\bibitem [{\citenamefont {Karsch}\ \emph {et~al.}(2001)\citenamefont {Karsch},
  \citenamefont {Mustafa},\ and\ \citenamefont {Thoma}}]{Karsch}%
  \BibitemOpen
  \bibfield  {author} {\bibinfo {author} {\bibfnamefont {F.}~\bibnamefont
  {Karsch}}, \bibinfo {author} {\bibfnamefont {M.~G.}\ \bibnamefont {Mustafa}},
  \ and\ \bibinfo {author} {\bibfnamefont {M.~H.}\ \bibnamefont {Thoma}},\
  }\href {\doibase 10.1016/S0370-2693(00)01322-8} {\bibfield  {journal}
  {\bibinfo  {journal} {Phys. Lett.}\ }\textbf {\bibinfo {volume} {B497}},\
  \bibinfo {pages} {249} (\bibinfo {year} {2001})},\ \Eprint
  {http://arxiv.org/abs/hep-ph/0007093} {arXiv:hep-ph/0007093 [hep-ph]}
  \BibitemShut {NoStop}%
\bibitem [{\citenamefont {Morimoto}\ \emph {et~al.}(2018)\citenamefont
  {Morimoto}, \citenamefont {Tsue}, \citenamefont {da~Providencia},
  \citenamefont {Providencia},\ and\ \citenamefont
  {Yamamura}}]{Morimoto:2018pzk}%
  \BibitemOpen
  \bibfield  {author} {\bibinfo {author} {\bibfnamefont {M.}~\bibnamefont
  {Morimoto}}, \bibinfo {author} {\bibfnamefont {Y.}~\bibnamefont {Tsue}},
  \bibinfo {author} {\bibfnamefont {J.}~\bibnamefont {da~Providencia}},
  \bibinfo {author} {\bibfnamefont {C.}~\bibnamefont {Providencia}}, \ and\
  \bibinfo {author} {\bibfnamefont {M.}~\bibnamefont {Yamamura}},\ }\href
  {\doibase 10.1142/S0218301318500283} {\bibfield  {journal} {\bibinfo
  {journal} {Int. J. Mod. Phys.}\ }\textbf {\bibinfo {volume} {E27}},\ \bibinfo
  {pages} {1850028} (\bibinfo {year} {2018})},\ \Eprint
  {http://arxiv.org/abs/1801.03633} {arXiv:1801.03633 [hep-ph]} \BibitemShut
  {NoStop}%
\bibitem [{\citenamefont {Volkov}(1993)}]{Volkov}%
  \BibitemOpen
  \bibfield  {author} {\bibinfo {author} {\bibfnamefont {M.~K.}\ \bibnamefont
  {Volkov}},\ }\href@noop {} {\bibfield  {journal} {\bibinfo  {journal} {Phys.
  Part. Nucl.}\ }\textbf {\bibinfo {volume} {24}},\ \bibinfo {pages} {35}
  (\bibinfo {year} {1993})}\BibitemShut {NoStop}%
\bibitem [{\citenamefont {Menezes}\ \emph {et~al.}(2009)\citenamefont
  {Menezes}, \citenamefont {Benghi~Pinto}, \citenamefont {Avancini},
  \citenamefont {Perez~Martinez},\ and\ \citenamefont
  {Providencia}}]{Menenzes}%
  \BibitemOpen
  \bibfield  {author} {\bibinfo {author} {\bibfnamefont {D.~P.}\ \bibnamefont
  {Menezes}}, \bibinfo {author} {\bibfnamefont {M.}~\bibnamefont
  {Benghi~Pinto}}, \bibinfo {author} {\bibfnamefont {S.~S.}\ \bibnamefont
  {Avancini}}, \bibinfo {author} {\bibfnamefont {A.}~\bibnamefont
  {Perez~Martinez}}, \ and\ \bibinfo {author} {\bibfnamefont {C.}~\bibnamefont
  {Providencia}},\ }\href {\doibase 10.1103/PhysRevC.79.035807} {\bibfield
  {journal} {\bibinfo  {journal} {Phys. Rev.}\ }\textbf {\bibinfo {volume}
  {C79}},\ \bibinfo {pages} {035807} (\bibinfo {year} {2009})},\ \Eprint
  {http://arxiv.org/abs/0811.3361} {arXiv:0811.3361 [nucl-th]} \BibitemShut
  {NoStop}%
\bibitem [{\citenamefont {Avancini}\ \emph
  {et~al.}(2019{\natexlab{b}})\citenamefont {Avancini}, \citenamefont {Farias},
  \citenamefont {Scoccola},\ and\ \citenamefont {Tavares}}]{Avancini:2019wed}%
  \BibitemOpen
  \bibfield  {author} {\bibinfo {author} {\bibfnamefont {S.~S.}\ \bibnamefont
  {Avancini}}, \bibinfo {author} {\bibfnamefont {R.~L.~S.}\ \bibnamefont
  {Farias}}, \bibinfo {author} {\bibfnamefont {N.~N.}\ \bibnamefont
  {Scoccola}}, \ and\ \bibinfo {author} {\bibfnamefont {W.~R.}\ \bibnamefont
  {Tavares}},\ }\href@noop {} {\  (\bibinfo {year} {2019}{\natexlab{b}})},\
  \Eprint {http://arxiv.org/abs/1904.02730} {arXiv:1904.02730 [hep-ph]}
  \BibitemShut {NoStop}%
\bibitem [{\citenamefont {Mallik}\ and\ \citenamefont {Sarkar}(2016)}]{Sourav}%
  \BibitemOpen
  \bibfield  {author} {\bibinfo {author} {\bibfnamefont {S.}~\bibnamefont
  {Mallik}}\ and\ \bibinfo {author} {\bibfnamefont {S.}~\bibnamefont
  {Sarkar}},\ }\href {\doibase 10.1017/9781316535585} {\emph {\bibinfo {title}
  {{Hadrons at Finite Temperature}}}}\ (\bibinfo  {publisher} {Cambridge
  University Press},\ \bibinfo {address} {Cambridge},\ \bibinfo {year}
  {2016})\BibitemShut {NoStop}%
\bibitem [{\citenamefont {Bellac}(2011)}]{Bellac:2011kqa}%
  \BibitemOpen
  \bibfield  {author} {\bibinfo {author} {\bibfnamefont {M.~L.}\ \bibnamefont
  {Bellac}},\ }\href {\doibase 10.1017/CBO9780511721700} {\emph {\bibinfo
  {title} {{Thermal Field Theory}}}},\ Cambridge Monographs on Mathematical
  Physics\ (\bibinfo  {publisher} {Cambridge University Press},\ \bibinfo
  {year} {2011})\BibitemShut {NoStop}%
\bibitem [{\citenamefont {Schwinger}(1951)}]{Schwinger:1951nm}%
  \BibitemOpen
  \bibfield  {author} {\bibinfo {author} {\bibfnamefont {J.~S.}\ \bibnamefont
  {Schwinger}},\ }\href {\doibase 10.1103/PhysRev.82.664} {\bibfield  {journal}
  {\bibinfo  {journal} {Phys. Rev.}\ }\textbf {\bibinfo {volume} {82}},\
  \bibinfo {pages} {664} (\bibinfo {year} {1951})},\ \bibinfo {note}
  {[,116(1951)]}\BibitemShut {NoStop}%
\bibitem [{\citenamefont {D'Olivo}\ \emph {et~al.}(2003)\citenamefont
  {D'Olivo}, \citenamefont {Nieves},\ and\ \citenamefont {Sahu}}]{Olivio}%
  \BibitemOpen
  \bibfield  {author} {\bibinfo {author} {\bibfnamefont {J.~C.}\ \bibnamefont
  {D'Olivo}}, \bibinfo {author} {\bibfnamefont {J.~F.}\ \bibnamefont {Nieves}},
  \ and\ \bibinfo {author} {\bibfnamefont {S.}~\bibnamefont {Sahu}},\ }\href
  {\doibase 10.1103/PhysRevD.67.025018} {\bibfield  {journal} {\bibinfo
  {journal} {Phys. Rev.}\ }\textbf {\bibinfo {volume} {D67}},\ \bibinfo {pages}
  {025018} (\bibinfo {year} {2003})},\ \Eprint
  {http://arxiv.org/abs/hep-ph/0208146} {arXiv:hep-ph/0208146 [hep-ph]}
  \BibitemShut {NoStop}%
\bibitem [{\citenamefont {Ghosh}\ \emph {et~al.}(2019)\citenamefont {Ghosh},
  \citenamefont {Mukherjee}, \citenamefont {Roy},\ and\ \citenamefont
  {Sarkar}}]{Ghosh:2019fet}%
  \BibitemOpen
  \bibfield  {author} {\bibinfo {author} {\bibfnamefont {S.}~\bibnamefont
  {Ghosh}}, \bibinfo {author} {\bibfnamefont {A.}~\bibnamefont {Mukherjee}},
  \bibinfo {author} {\bibfnamefont {P.}~\bibnamefont {Roy}}, \ and\ \bibinfo
  {author} {\bibfnamefont {S.}~\bibnamefont {Sarkar}},\ }\href@noop {} {\
  (\bibinfo {year} {2019})},\ \Eprint {http://arxiv.org/abs/1901.02290}
  {arXiv:1901.02290 [hep-ph]} \BibitemShut {NoStop}%
\bibitem [{\citenamefont {Aguirre}(2017)}]{Aguirre}%
  \BibitemOpen
  \bibfield  {author} {\bibinfo {author} {\bibfnamefont {R.}~\bibnamefont
  {Aguirre}},\ }\href {\doibase 10.1103/PhysRevD.95.074029} {\bibfield
  {journal} {\bibinfo  {journal} {Phys. Rev.}\ }\textbf {\bibinfo {volume}
  {D95}},\ \bibinfo {pages} {074029} (\bibinfo {year} {2017})},\ \Eprint
  {http://arxiv.org/abs/1612.00327} {arXiv:1612.00327 [nucl-th]} \BibitemShut
  {NoStop}%
\bibitem [{\citenamefont {Sasaki}\ \emph {et~al.}(2008)\citenamefont {Sasaki},
  \citenamefont {Friman},\ and\ \citenamefont {Redlich}}]{Sasaki}%
  \BibitemOpen
  \bibfield  {author} {\bibinfo {author} {\bibfnamefont {C.}~\bibnamefont
  {Sasaki}}, \bibinfo {author} {\bibfnamefont {B.}~\bibnamefont {Friman}}, \
  and\ \bibinfo {author} {\bibfnamefont {K.}~\bibnamefont {Redlich}},\ }\href
  {\doibase 10.1103/PhysRevD.77.034024} {\bibfield  {journal} {\bibinfo
  {journal} {Phys. Rev.}\ }\textbf {\bibinfo {volume} {D77}},\ \bibinfo {pages}
  {034024} (\bibinfo {year} {2008})},\ \Eprint {http://arxiv.org/abs/0712.2761}
  {arXiv:0712.2761 [hep-ph]} \BibitemShut {NoStop}%
\bibitem [{\citenamefont {Inagaki}\ \emph
  {et~al.}(2004{\natexlab{a}})\citenamefont {Inagaki}, \citenamefont {Kimura},\
  and\ \citenamefont {Murata}}]{Inagaki2}%
  \BibitemOpen
  \bibfield  {author} {\bibinfo {author} {\bibfnamefont {T.}~\bibnamefont
  {Inagaki}}, \bibinfo {author} {\bibfnamefont {D.}~\bibnamefont {Kimura}}, \
  and\ \bibinfo {author} {\bibfnamefont {T.}~\bibnamefont {Murata}},\ }\href
  {\doibase 10.1143/PTP.111.371} {\bibfield  {journal} {\bibinfo  {journal}
  {Prog. Theor. Phys.}\ }\textbf {\bibinfo {volume} {111}},\ \bibinfo {pages}
  {371} (\bibinfo {year} {2004}{\natexlab{a}})},\ \Eprint
  {http://arxiv.org/abs/hep-ph/0312005} {arXiv:hep-ph/0312005 [hep-ph]}
  \BibitemShut {NoStop}%
\bibitem [{\citenamefont {Landau}\ and\ \citenamefont
  {Lifshitz}(1980)}]{Landau:1980mil}%
  \BibitemOpen
  \bibfield  {author} {\bibinfo {author} {\bibfnamefont {L.~D.}\ \bibnamefont
  {Landau}}\ and\ \bibinfo {author} {\bibfnamefont {E.~M.}\ \bibnamefont
  {Lifshitz}},\ }\href@noop {} {\emph {\bibinfo {title} {{Statistical Physics,
  Part 1}}}},\ \bibinfo {series} {Course of Theoretical Physics}, Vol.~\bibinfo
  {volume} {5}\ (\bibinfo  {publisher} {Butterworth-Heinemann},\ \bibinfo
  {address} {Oxford},\ \bibinfo {year} {1980})\BibitemShut {NoStop}%
\bibitem [{\citenamefont {Ebert}\ and\ \citenamefont
  {Vshivtsev}(1998)}]{Ebert:1998gx}%
  \BibitemOpen
  \bibfield  {author} {\bibinfo {author} {\bibfnamefont {D.}~\bibnamefont
  {Ebert}}\ and\ \bibinfo {author} {\bibfnamefont {A.~S.}\ \bibnamefont
  {Vshivtsev}},\ }\href@noop {} {\  (\bibinfo {year} {1998})},\ \Eprint
  {http://arxiv.org/abs/hep-ph/9806421} {arXiv:hep-ph/9806421 [hep-ph]}
  \BibitemShut {NoStop}%
\bibitem [{\citenamefont {Inagaki}\ \emph
  {et~al.}(2004{\natexlab{b}})\citenamefont {Inagaki}, \citenamefont {Kimura},\
  and\ \citenamefont {Murata}}]{Inagaki:2004ih}%
  \BibitemOpen
  \bibfield  {author} {\bibinfo {author} {\bibfnamefont {T.}~\bibnamefont
  {Inagaki}}, \bibinfo {author} {\bibfnamefont {D.}~\bibnamefont {Kimura}}, \
  and\ \bibinfo {author} {\bibfnamefont {T.}~\bibnamefont {Murata}},\
  }\bibfield  {booktitle} {\emph {\bibinfo {booktitle} {{Finite density QCD.
  Proceedings, International Workshop, Nara, Japan, July 10-12, 2003}}},\
  }\href {\doibase 10.1143/PTPS.153.321} {\bibfield  {journal} {\bibinfo
  {journal} {Prog. Theor. Phys. Suppl.}\ }\textbf {\bibinfo {volume} {153}},\
  \bibinfo {pages} {321} (\bibinfo {year} {2004}{\natexlab{b}})},\ \Eprint
  {http://arxiv.org/abs/hep-ph/0404219} {arXiv:hep-ph/0404219 [hep-ph]}
  \BibitemShut {NoStop}%
\bibitem [{\citenamefont {Noronha}\ and\ \citenamefont
  {Shovkovy}(2007)}]{Noronha:2007wg}%
  \BibitemOpen
  \bibfield  {author} {\bibinfo {author} {\bibfnamefont {J.~L.}\ \bibnamefont
  {Noronha}}\ and\ \bibinfo {author} {\bibfnamefont {I.~A.}\ \bibnamefont
  {Shovkovy}},\ }\href {\doibase 10.1103/PhysRevD.76.105030,
  10.1103/PhysRevD.86.049901} {\bibfield  {journal} {\bibinfo  {journal} {Phys.
  Rev.}\ }\textbf {\bibinfo {volume} {D76}},\ \bibinfo {pages} {105030}
  (\bibinfo {year} {2007})},\ \bibinfo {note} {[Erratum: Phys.
  Rev.D86,049901(2012)]},\ \Eprint {http://arxiv.org/abs/0708.0307}
  {arXiv:0708.0307 [hep-ph]} \BibitemShut {NoStop}%
\bibitem [{\citenamefont {Fukushima}\ and\ \citenamefont
  {Warringa}(2008)}]{Fukushima:2007fc}%
  \BibitemOpen
  \bibfield  {author} {\bibinfo {author} {\bibfnamefont {K.}~\bibnamefont
  {Fukushima}}\ and\ \bibinfo {author} {\bibfnamefont {H.~J.}\ \bibnamefont
  {Warringa}},\ }\href {\doibase 10.1103/PhysRevLett.100.032007} {\bibfield
  {journal} {\bibinfo  {journal} {Phys. Rev. Lett.}\ }\textbf {\bibinfo
  {volume} {100}},\ \bibinfo {pages} {032007} (\bibinfo {year} {2008})},\
  \Eprint {http://arxiv.org/abs/0707.3785} {arXiv:0707.3785 [hep-ph]}
  \BibitemShut {NoStop}%
\bibitem [{\citenamefont {Orlovsky}\ and\ \citenamefont
  {Simonov}(2015)}]{Orlovsky:2014kva}%
  \BibitemOpen
  \bibfield  {author} {\bibinfo {author} {\bibfnamefont {V.~D.}\ \bibnamefont
  {Orlovsky}}\ and\ \bibinfo {author} {\bibfnamefont {{\relax Yu}.~A.}\
  \bibnamefont {Simonov}},\ }\href {\doibase 10.1142/S0217751X15500608}
  {\bibfield  {journal} {\bibinfo  {journal} {Int. J. Mod. Phys.}\ }\textbf
  {\bibinfo {volume} {A30}},\ \bibinfo {pages} {1550060} (\bibinfo {year}
  {2015})},\ \Eprint {http://arxiv.org/abs/1406.1056} {arXiv:1406.1056
  [hep-ph]} \BibitemShut {NoStop}%
\bibitem [{\citenamefont {Hufner}\ \emph {et~al.}(1996)\citenamefont {Hufner},
  \citenamefont {Klevansky},\ and\ \citenamefont {Rehberg}}]{Hufner:1996pq}%
  \BibitemOpen
  \bibfield  {author} {\bibinfo {author} {\bibfnamefont {J.}~\bibnamefont
  {Hufner}}, \bibinfo {author} {\bibfnamefont {S.~P.}\ \bibnamefont
  {Klevansky}}, \ and\ \bibinfo {author} {\bibfnamefont {P.}~\bibnamefont
  {Rehberg}},\ }\href {\doibase 10.1016/0375-9474(96)00206-0} {\bibfield
  {journal} {\bibinfo  {journal} {Nucl. Phys.}\ }\textbf {\bibinfo {volume}
  {A606}},\ \bibinfo {pages} {260} (\bibinfo {year} {1996})}\BibitemShut
  {NoStop}%
\bibitem [{\citenamefont {MOTT}(1968)}]{MOTT:1968zz}%
  \BibitemOpen
  \bibfield  {author} {\bibinfo {author} {\bibfnamefont {N.~F.}\ \bibnamefont
  {MOTT}},\ }\href {\doibase 10.1103/RevModPhys.40.677} {\bibfield  {journal}
  {\bibinfo  {journal} {Rev. Mod. Phys.}\ }\textbf {\bibinfo {volume} {40}},\
  \bibinfo {pages} {677} (\bibinfo {year} {1968})}\BibitemShut {NoStop}%
\bibitem [{\citenamefont {Blaschke}\ \emph {et~al.}(2017)\citenamefont
  {Blaschke}, \citenamefont {Dubinin}, \citenamefont {Radzhabov},\ and\
  \citenamefont {Wergieluk}}]{Dubinin:2016wvt}%
  \BibitemOpen
  \bibfield  {author} {\bibinfo {author} {\bibfnamefont {D.}~\bibnamefont
  {Blaschke}}, \bibinfo {author} {\bibfnamefont {A.}~\bibnamefont {Dubinin}},
  \bibinfo {author} {\bibfnamefont {A.}~\bibnamefont {Radzhabov}}, \ and\
  \bibinfo {author} {\bibfnamefont {A.}~\bibnamefont {Wergieluk}},\ }\href
  {\doibase 10.1103/PhysRevD.96.094008} {\bibfield  {journal} {\bibinfo
  {journal} {Phys. Rev.}\ }\textbf {\bibinfo {volume} {D96}},\ \bibinfo {pages}
  {094008} (\bibinfo {year} {2017})},\ \Eprint
  {http://arxiv.org/abs/1608.05383} {arXiv:1608.05383 [hep-ph]} \BibitemShut
  {NoStop}%
\bibitem [{\citenamefont {Mao}\ and\ \citenamefont {Wang}(2017)}]{MaoWang}%
  \BibitemOpen
  \bibfield  {author} {\bibinfo {author} {\bibfnamefont {S.}~\bibnamefont
  {Mao}}\ and\ \bibinfo {author} {\bibfnamefont {Y.}~\bibnamefont {Wang}},\
  }\href {\doibase 10.1103/PhysRevD.96.034004} {\bibfield  {journal} {\bibinfo
  {journal} {Phys. Rev.}\ }\textbf {\bibinfo {volume} {D96}},\ \bibinfo {pages}
  {034004} (\bibinfo {year} {2017})},\ \Eprint
  {http://arxiv.org/abs/1702.04868} {arXiv:1702.04868 [hep-ph]} \BibitemShut
  {NoStop}%
\end{thebibliography}%

\end{document}